\documentclass[twocolumn]{aastex62}

\shorttitle{HOPS-108}
\shortauthors{Tobin et al.}

\newcommand{\thco}{\mbox{$^{13}$CO}}
\newcommand{\twco}{\mbox{$^{12}$CO}}

\newcommand{\kms}{\mbox{km s$^{-1}$}}

\newcommand{\lsun}{\mbox{L$_{\sun}$}}
\newcommand{\msun}{\mbox{M$_{\sun}$}}

\newcommand{\lbol}{\mbox{L$_{bol}$}}
\begin{document}

\title{The VLA/ALMA Nascent Disk and Multiplicity (VANDAM) Survey of Orion Protostars I. Identifying and Characterizing
the Protostellar Content of the OMC2-FIR4 and OMC2-FIR3 Regions}
\author{John J. Tobin}
\affiliation{National Radio Astronomy Observatory, 520 Edgemont Rd., Charlottesville, VA 22903, USA}
\author{S. Thomas Megeath}
\affiliation{Department of Physics and Astronomy, University of Toledo, Toledo, OH 43560}
\author{Merel van 't Hoff}
\affiliation{Leiden Observatory, Leiden University, P.O. Box 9513, 2300-RA Leiden, The Netherlands}
\author{Ana Karla D{\'i}az-Rodr{\'i}guez }
\affiliation{Instituto de Astrof\'{\i}sica de Andaluc\'{\i}a, CSIC, Glorieta de la Astronom\'{\i}a
s/n, E-18008 Granada, Spain}
\author{Nickalas Reynolds}
\affiliation{Homer L. Dodge Department of Physics and Astronomy, University of Oklahoma, 440 W. Brooks Street, Norman, OK 73019, USA}
\author{Mayra Osorio}
\affiliation{Instituto de Astrof\'{\i}sica de Andaluc\'{\i}a, CSIC, Glorieta de la Astronom\'{\i}a
s/n, E-18008 Granada, Spain}
\author{Guillem Anglada}
\affiliation{Instituto de Astrof\'{\i}sica de Andaluc\'{\i}a, CSIC, Glorieta de la Astronom\'{\i}a
s/n, E-18008 Granada, Spain}
\author[0000-0001-9800-6248]{Elise Furlan}
\affiliation{IPAC, Mail Code 314-6, Caltech, 1200 E. California Blvd., 
Pasadena,
CA 91125, USA}
\author{Nicole Karnath}
\affiliation{Department of Physics and Astronomy, University of Toledo, Toledo, OH 43560}
\author{Stella S. R. Offner}
\affiliation{The University of Texas at Austin, 2500 Speedway, Austin, TX USA}
\author{Patrick Sheehan}
\affiliation{National Radio Astronomy Observatory, 520 Edgemont Rd., Charlottesville, VA 22903, USA}
\author{Sarah I. Sadavoy}
\affiliation{Harvard-Smithsonian Center for Astrophysics, 60 Garden St, MS 78, Cambridge, MA 02138}
\author{Amelia M. Stutz}
\affiliation{Departmento de Astronom\'{i}a, Universidad de Concepci\'{o}n,
Casilla 160-C, Concepci\'{o}n, Chile}
\affiliation{Max-Planck-Institute for Astronomy, K\"onigstuhl 17, 69117 Heidelberg, Germany}
\author{William J. Fischer}
\affiliation{Space Telescope Science Institute, Baltimore, MD, USA}
\author{Mihkel Kama}
\affiliation{Institute of Astronomy, Madingley Road, Cambridge CB3 OHA, UK}
\author{Magnus Persson}
\affiliation{Chalmers University of Technology, Department of Space, Earth and Environment, Sweden}
\author{James Di Francesco}
\affiliation{Herzberg Astronomy and Astrophysics Programs, National Research Council of Canada, 5071 West Saanich Road, Victoria BC V9E 2E7, Canada}
\author{Leslie W. Looney}
\affiliation{Department of Astronomy, University of Illinois, Urbana, IL 61801}
\author{Dan M. Watson}
\affiliation{Department of Physics and Astronomy, University of Rochester, Rochester, NY 14627}
\author{Zhi-Yun Li}
\affiliation{Department of Astronomy, University of Virginia, Charlottesville, VA 22903}
\author{Ian Stephens}
\affiliation{Harvard-Smithsonian Center for Astrophysics, 60 Garden St, MS 78, Cambridge, MA 02138}
\author{Claire J. Chandler}
\affiliation{National Radio Astronomy Observatory, P.O. Box O, Socorro, NM 87801}
\author{Erin Cox}
\affiliation{Center for Interdisciplinary Exploration and Research in Astrophysics (CIERA) and Department of Physics and Astronomy, Northwestern University, Evanston, IL 60208, USA}
\author{Michael M. Dunham}
\affiliation{Department of Physics, State University of New York Fredonia, Fredonia, New York 14063, USA}
\affiliation{Harvard-Smithsonian Center for Astrophysics, 60 Garden St, MS 78, Cambridge, MA 02138}
\author{Kaitlin Kratter}
\affiliation{University of Arizona, Steward Observatory, Tucson, AZ 85721}
\author{Marina Kounkel}
\affiliation{Department of Physics and Astronomy, Western Washington University, 516 High St., Bellingham, WA 98225, USA}
\author{Brian Mazur}
\affiliation{Department of Physics and Astronomy, University of Toledo, Toledo, OH 43560}
\author{Nadia M. Murillo}
\affiliation{Leiden Observatory, Leiden University, P.O. Box 9513, 2300-RA Leiden, The Netherlands}
\author{Lisa Patel}
\affiliation{Homer L. Dodge Department of Physics and Astronomy, University of Oklahoma, 440 W. Brooks Street, Norman, OK 73019, USA}
\author{Laura Perez}
\affiliation{Departamento de Astronom\'ia, Universidad de Chile, Camino El Observatorio 1515, Las Condes, Santiago, Chile}
\author{Dominique Segura-Cox}
\affiliation{Max-Planck-Institut f{\"u}r extraterrestrische Physik, Giessenbachstrasse 1, D-85748 Garching, Germany}
\author{Rajeeb Sharma}
\affiliation{Homer L. Dodge Department of Physics and Astronomy, University of Oklahoma, 440 W. Brooks Street, Norman, OK 73019, USA}
\author{{\L}ukasz Tychoniec}
\affiliation{Leiden Observatory, Leiden University, P.O. Box 9513, 2300-RA Leiden, The Netherlands}
\author{Friedrich Wyrowski}
\affiliation{Max-Planck-Institut f\"ur Radioastronomie, Auf dem H\"ugel 69, 53121, Bonn, Germany}

\begin{abstract}

We present ALMA (0.87~mm) and VLA (9~mm) observations toward OMC2-FIR4 and OMC2-FIR3
within the Orion integral-shaped filament that are thought to be
the nearest regions of intermediate mass star formation.
We characterize the continuum sources within these regions 
on $\sim$40~AU (0\farcs1) scales
and associated molecular line emission at a factor of $\sim$30
better resolution than previous observations at similar wavelengths. We identify six compact 
continuum sources within OMC2-FIR4, four in OMC2-FIR3, and one additional
source just outside OMC2-FIR4. This continuum emission is tracing the 
inner envelope and/or disk emission on less than 100~AU scales. 
HOPS-108 is the only protostar in OMC2-FIR4 that exhibits
emission from high-excitation transitions of complex organic molecules
(e.g., methanol and other lines) coincident with the continuum emission.
HOPS-370 in OMC2-FIR3 with L~$\sim$~360~\lsun, also exhibits emission
from high-excitation methanol and other lines.
The methanol emission
toward these two protostars is indicative of temperatures high enough to thermally evaporate 
methanol from icy dust grains; overall these protostars 
have characteristics
similar to hot corinos. We do not identify a clear outflow 
from HOPS-108 in \twco, but find evidence of interaction between the outflow/jet 
from HOPS-370 and the OMC2-FIR4 
region. The multitude of observational constraints 
indicate that HOPS-108 is likely a low to intermediate-mass protostar
in its main mass accretion phase 
and it is the most luminous protostar in OMC2-FIR4. 
The high resolution data 
presented here are essential for disentangling
the embedded protostars from their 
surrounding dusty environments and characterizing them.

\end{abstract}

\section{Introduction}
The formation of intermediate to high-mass protostars has yet to be fully
characterized observationally \citep[e.g.,][]{tan2014}. The uncertainty is, 
in part, because
intermediate and high-mass stars are significantly more rare than low-mass stars.
Furthermore,
many examples of intermediate to high-mass protostars are at distances greater than
1~kpc \citep{cyganowski2017,motte2017}, and they are typically more deeply embedded
than low-mass protostars making their characterization challenging 
\citep[e.g., Orion BN-KL;][]{gezari1998,debuizer2012,ginsburg2018}.
This typically large distance makes the identification and characterization
of intermediate-mass protostars difficult, especially because multiplicity increases with
stellar mass \citep[e.g.,][]{vankempen2012,duchene2013,moe2017}. For the sake
of discussion in this paper, we refer to stars with M$_*$~$<$~2~\msun\ as low-mass, 
2~\msun~$\leq$~M$_*$~$<$8~\msun\ intermediate-mass, and M$_*$~$>$8~\msun\ as high-mass. And
with respect to the protostellar phase, a protostar that is expected 
to ultimately form a low, intermediate, 
or high-mass star is referred to a low, intermediate, or high-mass protostar.

The Integral-Shaped Filament (ISF) within the Orion A molecular cloud at a distance
of $\sim$400~pc \citep{kounkel2017} harbors
 several attractive intermediate-mass protostar candidates. The ISF
comprises Orion Molecular Clouds (OMC) 1, 2, and 3, where OMC1 begins south of
the Orion Nebula Cluster (ONC) and OMC2 and OMC3 are located north of the ONC; just north of 
OMC3 is NGC 1977 \citep{peterson2008}. In particular,
the regions identified by \citet{mezger1990} as OMC2-FIR3 and FIR4 are often looked to
as candidate intermediate-to-high mass protostars and/or 
proto-clusters \citep{shimajiri2008, fontani2015,ceccarelli2014}. 
The total gas masses of OMC2-FIR4 and OMC2-FIR3 have been estimated to be $\sim$33~\msun\
and $\sim$17~\msun, respectively, from their 850~\micron\ continuum emission maps \citep{nutter2007}.

The protostars within these regions,
however, are expected to be lower-mass than the known high-mass protostars
in the BN-KL region, of which source I was recently measured to 
have a protostar mass of $\sim$15~\msun\ from its disk rotation \citep{ginsburg2018}. 
It has been difficult, however, to accurately measure the 
multi-wavelength emission from individual protostars in the
OMC2 and 3 regions due to the high protostellar density, especially at wavelengths longer
than 24~\micron\ where most of the luminosity of a protostar is emitted \citep{dunham2008,furlan2016}. 

The ISF of Orion has been the 
target of photometric studies with the \textit{Herschel Space Observatory}
as part of the \textit{Herschel} Orion Protostar Survey (HOPS) \citep{furlan2016}. 
Within this survey, the protostars associated with OMC2-FIR3 and FIR4 
 were resolved in the mid-to-far-infrared by 
\citet{furlan2014} and \citet{adams2012} from 3.6~\micron\ to 100~\micron. 
They found that HOPS-370, associated with 
FIR3, has a high bolometric luminosity (\lbol $\sim$360~L$_{\sun}$) indicative of
at least an intermediate-mass protostar. However, 
the nature of FIR4 was less clear, having previously been suggested to be a 
high mass protostar, which is in conflict with the observed luminosity 
of the most closely associated protostar, HOPS-108. The luminosity of 
HOPS-108 in the mid-to-far-infrared (L$_{bol}$ $\sim$37~L$_{\sun}$) is lower than HOPS-370 at 
wavelengths $<$100~\micron. This indicates that the HOPS-108 protostar could be less
massive (or accreting less rapidly) than HOPS-370, despite residing within the more
massive FIR4 core. \lbol\ can both over- and underestimate the
total internal luminosity of a protostellar system, 
however, due to inclination, obscuration, and 
some light escaping through the outflow cavities \citep{whitney2003a}. 
Furthermore, at wavelengths between 160~\micron\ to 0.87~mm the emission from HOPS-108 
could not be separated from that of the
FIR4 core, the peak of which is displaced $\sim$4\farcs5 (1800 AU) from HOPS-108.
\citet{furlan2014} fit a modified blackbody to the emission from the FIR4 core 
between 160~\micron\ and 0.87~mm and found a temperature of 22 K and 
luminosity of 137~\lsun; a substantial fraction of this luminosity may 
come from external heating.

\begin{figure}
\begin{center}
\includegraphics[scale=1.0]{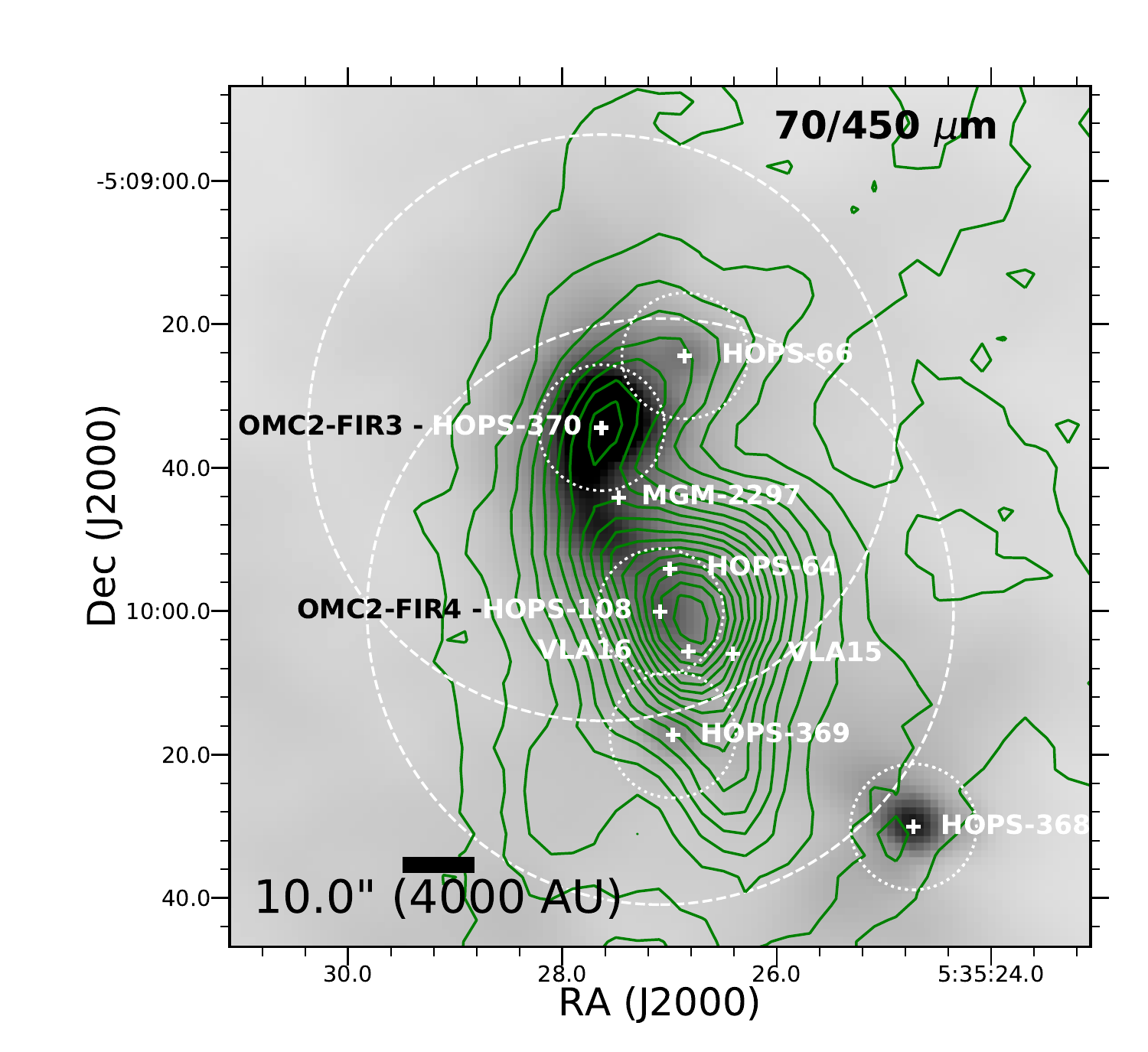}

\end{center}
\caption{Overview of the OMC2-FIR3/4 region. The grayscale image is the \textit{Herschel} 
70~\micron\ image and the green contours are the SCUBA 450~\micron\ emission \citep{johnstone1999}.
The large dashed circles mark the half-power point of the VLA primary beam at 9~mm, and the
smaller dotted circles mark the half-power point of the ALMA primary beam at 0.87~mm.
The 450~\micron\ contours start at 15$\sigma$ and increase on 10$\sigma$ intervals where 
$\sigma$=0.2 Jy~beam$^{-1}$. The positions shown are from \citet{furlan2014,furlan2016}, and
\citet{osorio2017}.
}
\label{overview}
\end{figure}

\citet{furlan2014} analyzed the SED of HOPS-108 using 
radiative transfer models, finding that the protostar
could have an internal luminosity as low as 37~L$_{\sun}$ or as high as 100~L$_{\sun}$.
This estimate is inconsistent with earlier luminosity claims of
700-1000~L$_{\sun}$ by \citet{lopez-sepulcre2013}, 
which was in part motivated by low-resolution ($\sim$6\arcsec) centimeter 
flux densities observed from the VLA \citep{reipurth1999} that 
were interpreted as an ultra-compact HII region and far-infrared flux densities from
the InfraRed Astronomy Satellite (IRAS).
\citet{osorio2017} spatially resolved the radio emission of FIR3 and FIR4.
With higher resolution VLA data ($\sim$0\farcs4) at 5 cm,
they found a jet-like feature with knots
that have moved away from FIR3 and toward FIR4
when compared with archival data taken $\sim$15 years prior.
This radio jet corresponds well with the jet mapped 
by \citet{gonzalez2016} in [OI] with \textit{Herschel}, where
the strongest [OI] emission was seen to originate near HOPS-108 and at the 
end of the radio jet, possibly
in a terminal shock. The lower luminosity of HOPS-108 from \citet{furlan2014} 
and the fact that the centimeter emission reflects a jet driven by FIR3 rather than a 
ultra-compact HII region, makes HOPS-108 inconsistent with being
a high-mass protostar. The low-resolution of \textit{Herschel} and 
uncertainty in the absolute positions relative to the much 
higher resolution VLA data, however, leave some ambiguity as to the nature
of HOPS-108 and the association of the [OI] shock.
We show an overview of the region in Figure \ref{overview}, with the previously known protostar
positions \citep{megeath2012, furlan2014, furlan2016} and the locations of
the compact radio continuum emission, likely tracing protostars, from \citet{osorio2017} overlaid.

In addition to the photometric, radio, and [OI] studies, 
FIR4 presents a diverse array of line emission from molecules that may be indicative
of chemical processes driven by a source of
locally generated energetic particles (i.e., cosmic rays) or photons that are 
catalysts for chemistry \citep{ceccarelli2014,gaches2018}. 
Most studies of this region, however, have been conducted at resolutions $\ge$3\arcsec, 
which are insufficient to resolve
the protostars completely from their environment \citep{favre2018}. 
The VLA 5~cm observations from \citet{osorio2017} do resolve many
protostars, but the presence of the radio jet makes positive 
identification of all sources difficult.

Building upon these previous studies, we have conducted VLA and ALMA 
observations at 9~mm and 0.87~mm, respectively, 
both with $<$0\farcs1 resolution, detecting and resolving 
the dust emission from the protostars within the FIR3 and FIR4 regions. Furthermore,
the molecular line emission contained within our ALMA bandpass enables 
us to further characterize the physical conditions of HOPS-108 
and HOPS-370 and the associated protostars in the region. 
This paper is structured as follows: the observations are 
presented in Section 2, our results are presented in Section 3, we 
discuss our results in Section 4,
and we present our conclusions in Section 5.

\section{Observations and Data Reduction}

The ALMA and VLA observations presented here are part of the
VLA/ALMA Nascent Disk and Multiplicity (VANDAM) Survey of the Orion molecular clouds.
Observations were conducted toward 328 protostars (148 for the VLA) in the Orion 
molecular clouds, all at $\sim$0\farcs1 resolution. The sample of 328 protostars is
derived from the HOPS sample \citep{furlan2016}, observing the bonafide protostars from 
Class 0 to Flat spectrum. The full survey results will be presented in an upcoming paper (Tobin et al. in prep.).

\subsection{ALMA Observations}

The ALMA observations of the HOPS-108/OMC2-FIR4 and HOPS-66
regions were conducted during three executions on
2016 September 4, 5, and 2017 July 19. The observations of
HOPS-370/OMC2-FIR3, HOPS-368, and HOPS-369 were conducted during
three executions, with
two executions on 2016 September 6 and the third on 2017 July 19. 
Between 34 and 42 antennas were operating during a given execution
and the on-source time per field was 0.3 minutes
during each execution, totaling $\sim$0.9 minutes per field. During the 
2016 observations, the baselines ranged from $\sim$16~m to 
2483~m, and the 2017 observations sampled baselines from $\sim$18~m to 3697~m. 
The largest angular scale recoverable by the observations is 
expected to be $\sim$1\farcs5. The precipitable water vapor was 0.43~mm and 0.42~mm during the 2016 September 6
and 2017 July 19 observations, respectively of HOPS-370/OMC2-FIR3, 
HOPS-368, and HOPS-369. Then for the 2016 September 4, 5, and 2017 July 19,
observations of HOPS-108/OMC2-FIR4 and HOPS-66 the 
precipitable water vapor was 0.73~mm, 0.53~mm, and  0.47~mm, respectively.
The ALMA observations are summarized in Table 1 and the phase centers along with 
the half-power points of the primary beam at 0.87~mm are shown in Figure \ref{overview}.

The correlator was configured with two basebands set to low spectral resolution continuum mode 
1.875 GHz bandwidth each, with 31.25~MHz ($\sim$27~\kms) channels. These continuum basebands were centered 
at 333 GHz and 344 GHz. The two remaining basebands were centered on \twco\ ($J=3\rightarrow2$) 
at 345.79599 GHz, having a total bandwidth of 937.5~MHz and 0.489~\kms\ channels, and
\thco\ ($J=3\rightarrow2$) at 330.58797~GHz, with a bandwidth of 234.375~MHz and
0.128~\kms\ channels. The line free-regions of the
basebands centered on \twco\ and \thco\ ($J=3\rightarrow2$) were then used for 
additional continuum bandwidth. The total aggregate continuum bandwidth was $\sim$4.75~GHz.

The calibrators used in the 2017 observations were
J0423-0120 (flux), J0510+1800 (bandpass), and J0541-0541 (complex gain).
During the 
first execution on 2016 September 6 (HOPS-368, HOPS-369, HOPS-370), the calibrators were 
J0510+1800 (bandpass and flux) and
J0541-0541 (complex gain), and during the second execution the calibrators were J0522-3627 (flux),
J0510+1800 (bandpass) and J0541-0541 (complex gain).
The calibrators used in the observations of HOPS-66 and HOPS-108 on 
2016 September 04 and 05 were J0510+1800 (bandpass and flux) and
J0541-0541 (complex gain).

The data were reduced manually by the Dutch Allegro ALMA Regional Center Node. 
The manual reduction was necessary to better correct
for variation of the quasar J0510+1800 that was used for absolute flux calibration in the
observations on 2016 September 04, 05, and 06. 
The flux calibration quasars are monitored regularly, but J0510+1800 had a flare and
had not been monitored at Band 7 (0.87~mm) between 2016 August 21 and 2016 September 19; 
however, monitoring had been conducted at Band 3 (3~mm) three times during this period. 
To extrapolate the Band 7 flux density of J0510+1800, the spectral index of the 
quasar from Band 3 to Band 7 was used, and the time variability of the spectral index 
was estimated from the contemporaneous Band 3 and 7 observations on 2016 August 21 and 2016 September 19.
The absolute flux calibration accuracy is expected to be 
$\sim$10\%, and comparisons of the observed flux densities for the 
science targets in the different executions are consistent 
with this level of accuracy.

 Following the standard calibration, three rounds of phase-only 
self-calibration were performed on continuum data to increase the signal-to-noise ratio (S/N). 
For each successive round, we used solution
intervals that spanned the entire scan length (first round), 12.08~s (second round), 
and 3.02~s which was the length of a single integration (third round). 
The self-calibration solutions were also applied to the \twco\ and \thco\ 
spectral line data. The continuum and spectral line data cubes were imaged
using the \textit{clean} task of the Common Astronomy Software Application (CASA). 
We used CASA 4.7.2 for all self-calibration and imaging. 

The continuum images were produced using the \textit{clean} task
in CASA 4.7.2 using Briggs weighting with a robust
parameter of 0.5, yielding a synthesized beam 
of 0\farcs11$\times$0\farcs10 (83~AU~$\times$~58~AU) full-width at 
half-maximum (FWHM). The continuum
image also only uses visibilities at baselines $>$25~k$\lambda$ (21.75~m) to 
mitigate striping resulting from large-scale emission
that is not properly recovered. The \twco\ and \thco\ spectral line data were 
imaged using Natural weighting for baselines
$>$50~k$\lambda$ (43.5~m) to mitigate striping, and an outer taper of 500~k$\lambda$ (435~m) was
applied to increase the sensitivity to extended 
structure. The resulting synthesized beams were 0\farcs25$\times$0\farcs24. Additional
spectral lines were imaged with an outer taper of 2000~k$\lambda$ (1740~m), resulting in a synthesized
beam of 0\farcs15$\times$0\farcs14. The inner uv-cuts 
applied to the data typically only removed one or two baselines from the imaging process.
The primary beam of the ALMA 0.87~mm observations was $\sim$17\arcsec\ in diameter, FWHM. 
However, we were able to detect sources beyond the FWHM and out to 11\farcs4 from the
field center. The resulting
RMS of the continuum, \twco, and \thco\ data are $\sim$0.31~mJy~beam$^{-1}$, 
1$\sim$7.7 mJy~beam$^{-1}$ (1~\kms\ channels), and $\sim$33.3 mJy~beam$^{-1}$ (0.44~\kms\ channels), 
respectively.

\subsection{VLA Observations}

The observations with the VLA were conducted in the A-configuration on 2016 October 26 (HOPS-370) and 
2016 December 29 (HOPS-108). During the
observation, 26 antennas were operating and the entire observation lasted 2.5 hours. The observations
used the Ka-band receivers and the correlator was used in its wide bandwidth mode (3-bit samplers) with
one 4~GHz baseband centered at 36.9~GHz (8.1~mm) and the second 4~GHz 
baseband centered at 29~GHz (1.05~cm). The absolute flux calibrator was 
3C48 (J0137+3309), the bandpass calibrator was 3C84 (J0319+4130),
and the complex gain calibrator was J0541-0541 in all observations. The observations were
conducted in fast-switching mode (2.6 minute calibrator-source-calibrator cycle times) to reduce phase decoherence 
in the high frequency observations and the total time on source was $\sim$64 minutes. We note that
the first observation was taken when the VLA was misapplying the tropospheric phase correction,
leading to position offsets when sources were at low elevation and/or far from their calibrator.
The HOPS-370 data were taken above elevations of 40\degr\ (the effect was worst below 35\degr) and
the calibrator distance was only $\sim$1\degr\ making the effects of this issue negligible
in the HOPS-370 dataset.
The VLA observations are summarized in Table 1 and the phase centers and
the half-power points of the primary beam at 9~mm are shown in Figure \ref{overview}.

The data were reduced using the scripted version of the VLA pipeline in CASA 4.4.0.
Phase-only self-calibration was conducted in two rounds
with solution intervals of 230~s (first round) and 90~s (second round), 
which corresponded to one solution for every two scans and one solution for each scan, 
respectively. The continuum was imaged using the \textit{clean} task in CASA 4.5.1 using
Natural weighting and multi-frequency synthesis with \textit{nterms=2} across both basebands. 
The final image has an RMS noise of 6.9 $\mu$Jy~beam$^{-1}$ and a synthesized beam of 
0\farcs08$\times$0\farcs07 (32~AU~$\times$~28~AU), FWHM. The primary beam of the 
VLA observations was $\sim$80\arcsec, FWHM; however, we were able to image a source 46\arcsec\
from the field center.

\section{Results}

\subsection{Protostellar Content}

The observations of the 0.87~mm and 9~mm continuum with ALMA and the VLA, respectively, detect
the compact dust emission originating from the protostars in the region with sufficiently
high dust mass (and temperature). The VLA 9~mm continuum can also have a contribution
from free-free emission. We detected the 10 known protostellar and compact radio 
continuum sources at both 0.87~mm and 9.1~mm at the 3$\sigma$ level (or above) and we detect
a new source at 0.87~mm and 9~mm for a total of 11 sources detected.
We fit Gaussians to these
sources using the \textit{imfit} task in CASA to measure their 
flux densities and positions. The detected sources have their properties
listed in Table 2 for ALMA and Table 3 for the VLA. Due to the number of 
source catalogs already published toward the region at different angular resolutions
and sensitivity, there are multiple identifiers available for many of the detected sources.
In light of this, we attempt to use the most common identifier possible for the sources
detected in the region. Only one source has not been previously cataloged at another wavelength
and we refer to this source as OMC2-FIR4-ALMA1 (hereafter ALMA1).

OMC2-FIR3 has four continuum sources associated with it located at the positions of HOPS-370, MGM-2297, and HOPS-66.
HOPS-66 contains two continuum sources and is a newly detected binary system separated by 2\farcs23 (892~AU);
these two sources are denoted HOPS-66-A and HOPS-66-B. HOPS-370 has a previous detection of an
infrared companion $\sim$3\arcsec\ south \citep{nielbock2003} that is not detected by ALMA nor the VLA.
This apparent companion is brighter at wavelengths
less than 12~\micron\, but HOPS-370 is dominant at longer wavelengths.
OMC2-FIR4 contains 6 continuum sources that are associated with HOPS-108, HOPS-64, VLA15, VLA16, 
HOPS-369, and ALMA1. Note that HOPS-369 is more closely associated with OMC2-FIR5, which corresponds
to the southern extension of dust emission from
OMC2-FIR4 (Figure \ref{overview}, but we still discuss it in relation to the other protostars in FIR4. 
The projected separations from HOPS-108 (measured at 0.87~mm) to the other sources associated with 
FIR4 are as follows: 6\farcs8 (2720 AU, VLA16), 6\farcs15 (2460 AU, HOPS-64), 11\farcs7 (4680 AU, VLA15),
10\farcs5 (4020 AU, ALMA1), and 17\farcs3 (6920 AU, HOPS-369).
HOPS-368 does not lie within OMC2-FIR4, but is just at the edge of the core.

\begin{figure}
\begin{center}
\includegraphics[scale=0.575]{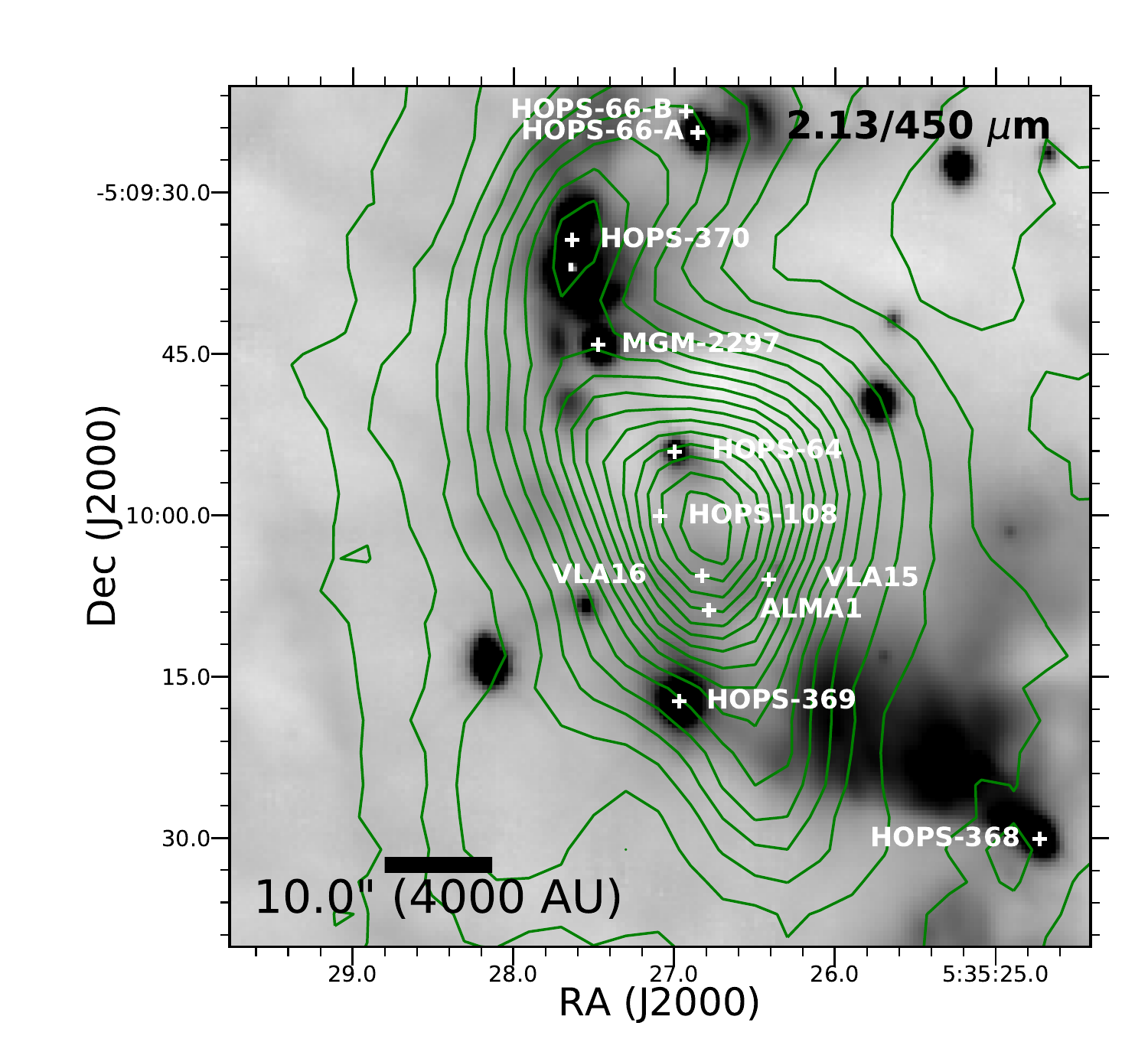}
\includegraphics[scale=0.575]{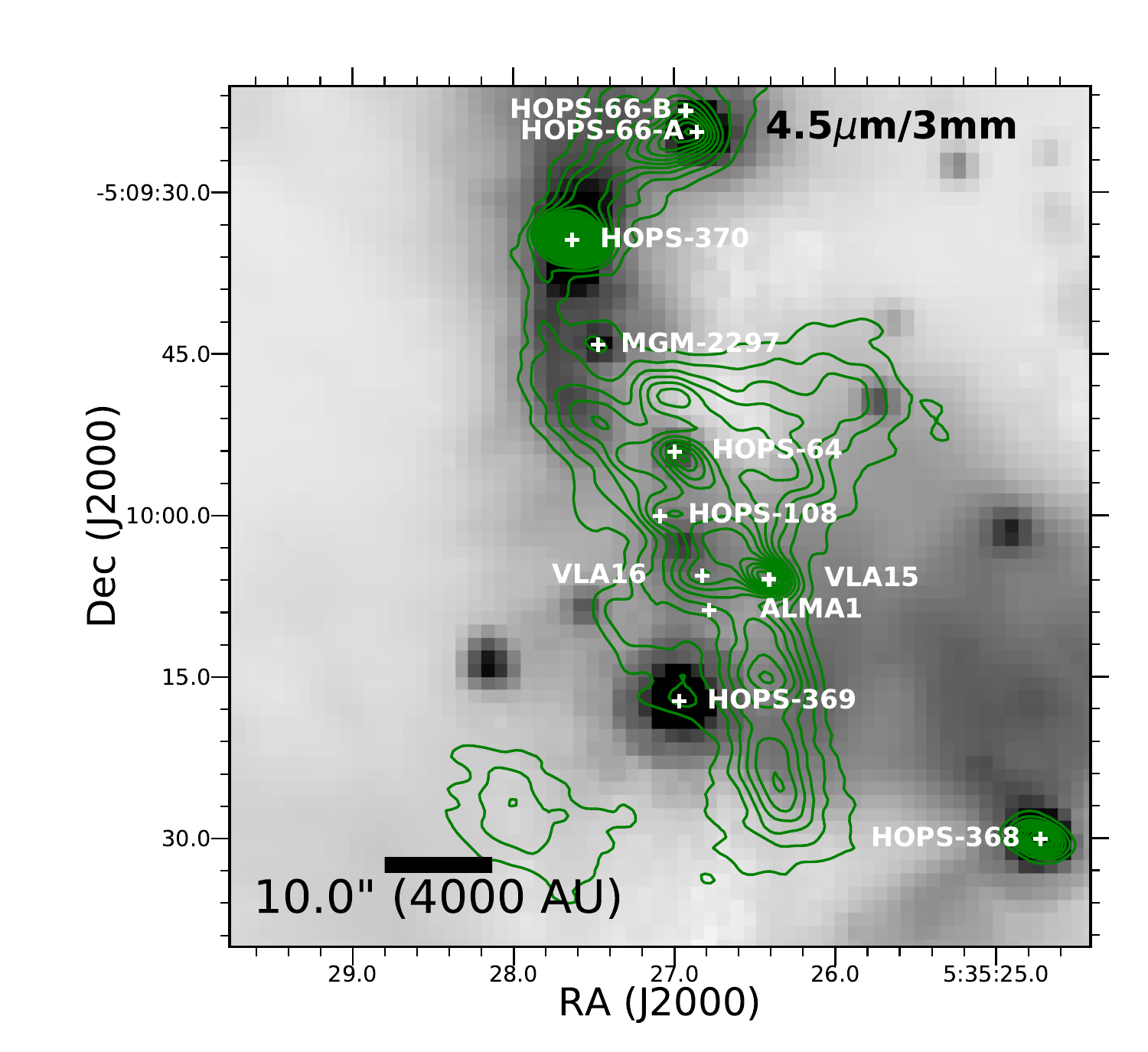}
\includegraphics[scale=0.575]{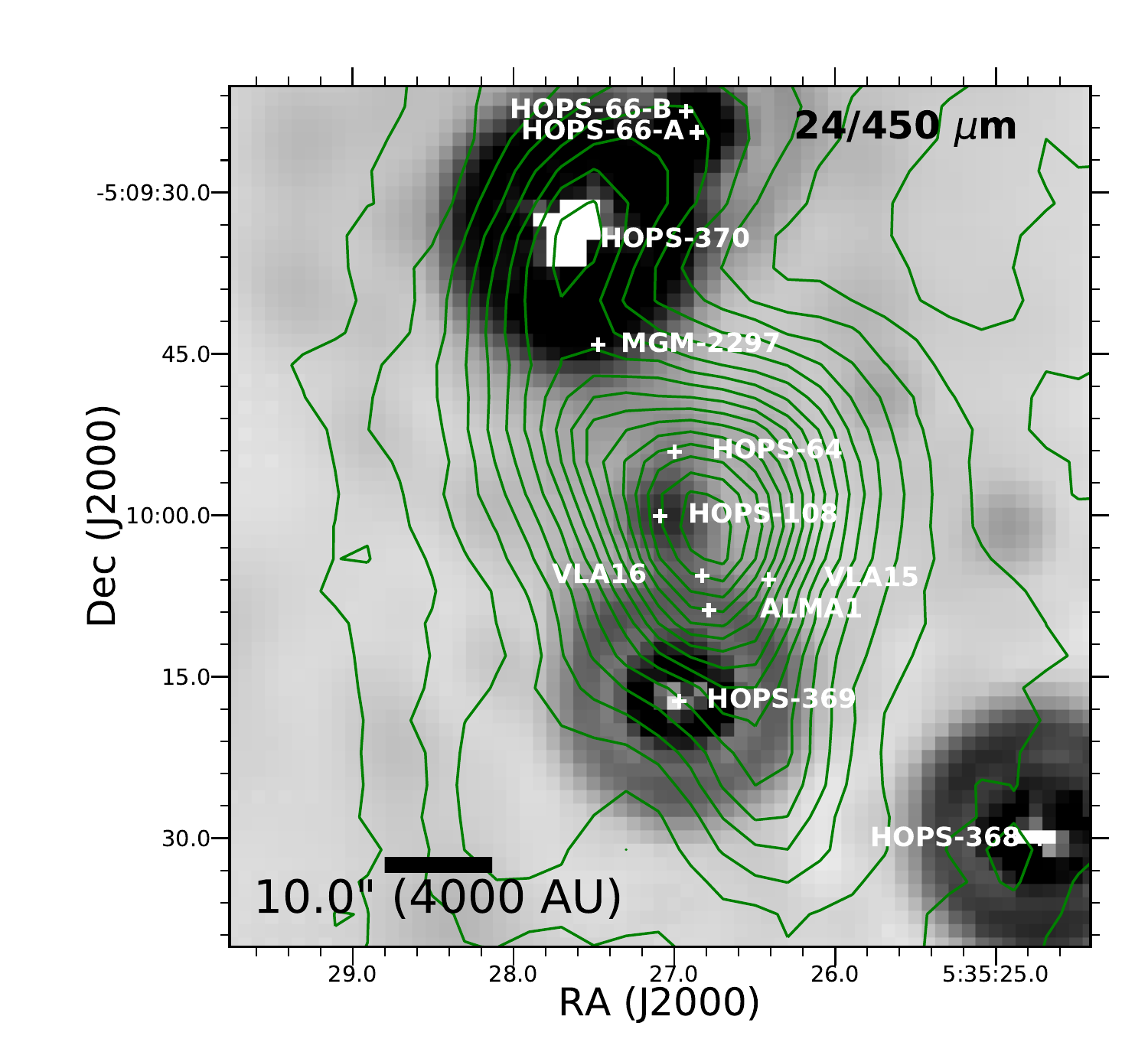}
\includegraphics[scale=0.575]{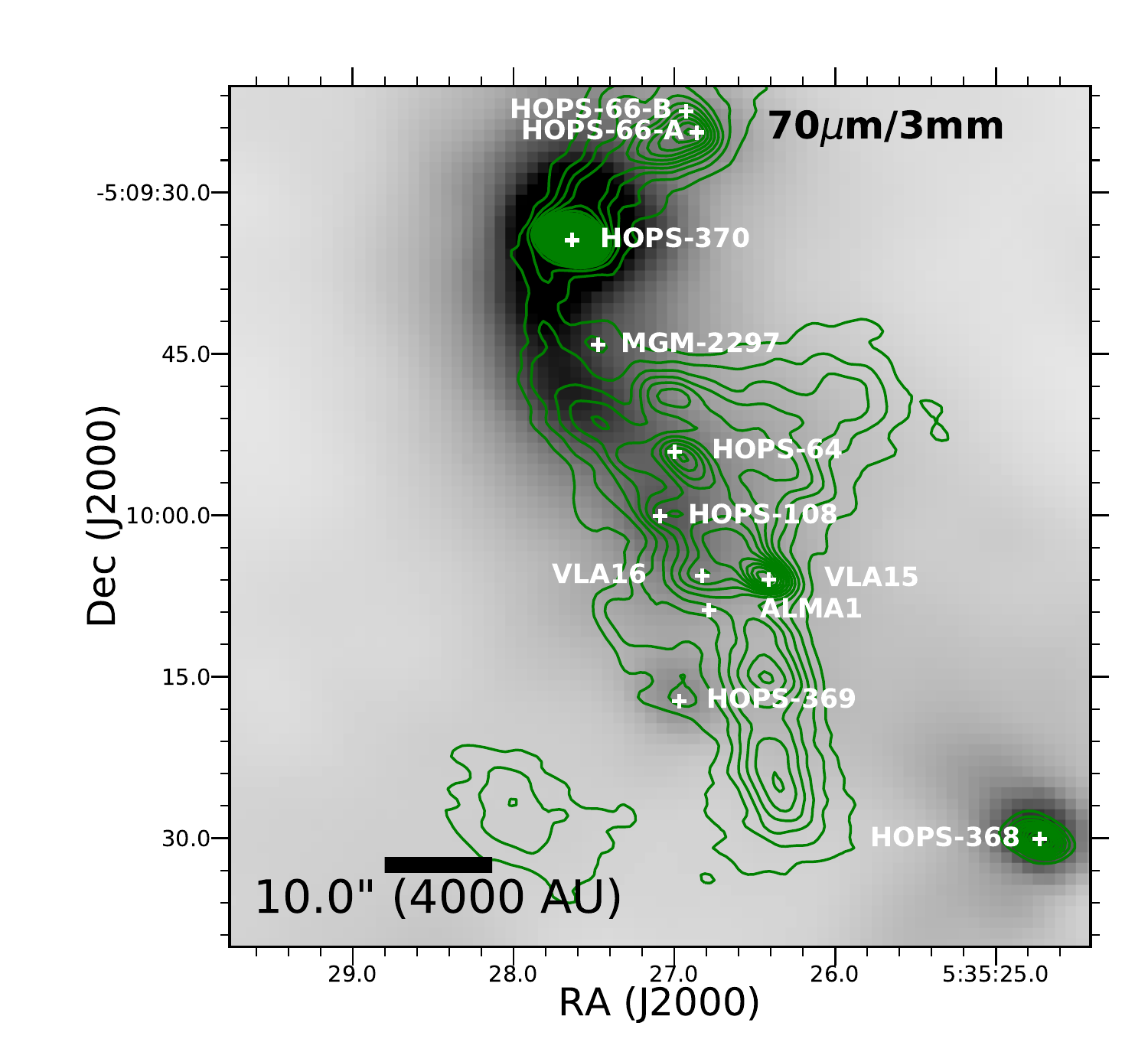}

\end{center}
\caption{The OMC2-FIR3/4 region is shown in grayscale at 2.13~\micron\ (top left), 
4.5~\micron\ (top right), 24~\micron\ (bottom left), and 70~\micron\ (bottom right).
The two left panels show contours of the SCUBA 450~\micron\ emission \citep{johnstone1999}
and the two right panels show contours of the ALMA+ACA 3~mm maps \citep{kainulainen2016}.
The source positions detected from the VLA and ALMA surveys are 
marked with white crosses in all panels. The 2.13~\micron\ and \textit{Spitzer} 4.5~\micron\ images show 
emission from young stars in the region arising from inner disk
emission, scattered light, and shocked molecular hydrogen emission. The
24~\micron\ and 70~\micron\ emission show where the bulk of the warm dust is
radiating due to the internal heating from the protostars, in addition to 
evidence for some external heating in the extended emission at 
70~\micron. The 450~\micron\ and 3~mm primarily show cold dust emission,
the prominent peak at the center of the image is classically known as OMC2-FIR4.
Some of the most deeply embedded
protostars detected with the VLA and ALMA (VLA16 and VLA15) do not have distinct 4.5~\micron\
emission located at their position. The stars that are 
detected with at 2.13 and 4.5~\micron\ and not detected by ALMA and the VLA are likely more evolved YSOs and
not embedded protostars.
Several of the most centrally-located sources in FIR4 (HOPS-108, VLA16, HOPS-64, and VLA15) 
have local peaks in 3~mm emission in the ALMA maps.
}
\label{overview-detail}
\end{figure}

We compare our identification 
of the continuum emission associated with the protostellar sources
to the multi-wavelength imaging that has been conducted toward the region. Figure \ref{overview-detail} shows
the continuum positions overlaid on several images: ground-based 2.13~\micron, 
\textit{Spitzer} 4.5~\micron, \textit{Spitzer} 24~\micron, and \textit{Herschel} 70~\micron, 
along with SCUBA 450~\micron\ and ALMA 3~mm contours 
overlaid \citep{megeath2012,furlan2014,johnstone1999,kainulainen2016}. 
The two shorter wavelengths show a combination of emission from the embedded protostars and
pre-main sequence stars with disks; it is clear, however, that HOPS-108, VLA15, VLA16,
and ALMA1 exhibit very little emission in these bands. 
On the other hand, HOPS-370 and HOPS-66
have prominent emission at 2.13 and 4.5~\micron, and the
Class II source (MGM-2297) south of HOPS-370 \citep{megeath2012} is also apparent at
these wavelengths. MGM-2297 may be located in 
the foreground and not be directly associated with OMC2-FIR3.

The non-detections of HOPS-108, VLA16, VLA15, and ALMA1 at 2.13~\micron\ and 4.5~\micron,
is expected for protostars too deeply embedded for their 
emission to be detected at short wavelengths. 
The 2.13 and 4.5~\micron\ images can trace the presence of shock-excited
H$_2$ emission, scattered light in outflow cavities, and the continuum emission 
from the warm inner disks surrounding pre-main sequence stars. 
There is some compact infrared emission adjacent to them (within $\sim$1\arcsec) that could be 
associated with scattered light, but this possibility is difficult to substantiate
with the $\sim$1\arcsec\ seeing of the 2.13~\micron\
image and the 1\farcs2 angular resolution of the \textit{Spitzer} IRAC images. 
HOPS-64 and HOPS-369 both have the most directly associated, 
point-like 4.5~\micron\ emission of all detected sources within FIR4, and HOPS-64 has some evidence
of a scattered light cone extending southwest in addition to a well-resolved conical
scattered light feature in \textit{Hubble Space Telescope} data \citep{kounkel2016}.
There is also a spot of emission at 4.5~\micron\ located southwest of HOPS-108, 
between it and VLA16, which may be associated with a shock 
from the HOPS-370 jet,
identified as VLA12S in \citet{osorio2017}.

It is important to compare the 24 and 70~\micron\ maps in Figure \ref{overview-detail}
with the ALMA and VLA detections because the peaks in those maps will generally
signify the internally-generated
luminosity from protostars and their accretion disks. 
HOPS-370 is the brightest source in the field
at both 24~\micron\ and 70~\micron, and the
70~\micron\ emission extends southwest of the source.

Toward the peak of FIR4, HOPS-108 is the 
brightest infrared source at wavelengths between 24~\micron\ to 70~\micron\ 
\citep{furlan2014,adams2012}. At longer wavelengths,
the emission that can be directly associated with HOPS-108 is blended with the extended
emission associated with the surrounding FIR4 core \citep{furlan2014}.
HOPS-369 is brighter than HOPS-108 at 24~\micron, but it is located $\sim$17\arcsec\ 
from the peak 450~\micron\ emission, near the FIR5 region \citep{mezger1990}.
HOPS-369 is also fainter than HOPS-108 at 70~\micron\ and longer wavelengths, and it does not appear
as deeply embedded in its core, especially given its detection at near-infrared wavelengths
(Figure \ref{overview-detail}). HOPS-368 is the second brightest
source in the field at 24~\micron\ and 70~\micron\, but it is located outside the FIR4
core to the southwest by $\sim$46\arcsec. HOPS-64 is detected but blended with HOPS 108.
Then, VLA16, VLA15,
and ALMA1 are not well-detected at 24~\micron\ or 70~\micron, possibly
due to blending with nearby sources at the $\sim$7\arcsec\ resolution of the data at these wavelengths.

The 450~\micron\ intensity is highest near HOPS-108, VLA16, VLA15, and ALMA1
indicating significant column densities of cold dust and large gas masses \citep{furlan2014}.
The emission extends north and has a peak associated with HOPS-370 and further
extends toward HOPS-66.
Furthermore, the higher-resolution 3~mm map from ALMA \citep{kainulainen2016}
shows local peaks of emission associated with HOPS-108, HOPS-64, VLA16, and VLA15. To the north
there are also 3~mm peaks associated with HOPS-370, HOPS-66, and MGM-2297.
\citet{kainulainen2016} also detect two other 
potential substructures at 3~mm south of VLA15 associated with the FIR5 region, 
but they lack ALMA/VLA detections at high-resolution. 
HOPS-369 has a weaker peak associated with its position relative to the others
and ALMA1 does not have a 3~mm peak
associated with its position. Thus, HOPS-108, HOPS-64, VLA16, and VLA15
are the most likely sources to be young and embedded within FIR4. Of these protostars,
HOPS-64 appears to be the least embedded, with detections 
even at optical wavelengths \citep{rodriguez2009}. \citet{furlan2016}
classified HOPS-64 as a 
Class I protostar because its SED longward of 24~\micron\ is blended with the 
surrounding sources; the lack of a detection longward of 24~\micron\ by 
\citet{adams2012} demonstrates that its SED is not steeply rising with increasing wavelength and
may not be embedded within an envelope. The peak at 3~mm at the location
of HOPS-64 and its detection at optical and near-infrared wavelengths could mean that
it is physically associated with the OMC2-FIR4 but near the edge in the foreground.
Taken together, the correspondence of HOPS-108 with the brightest 
24~\micron\ and 70~\micron\ detections within FIR4, 
its proximity to the 450~\micron\ peak, and its lack of direct detection 
shortward of 8~\micron\ make it most likely to be the 
most luminous protostar within OMC2-FIR4.

\subsection{ALMA and VLA Continuum Images}

\begin{figure}
\begin{center}
\includegraphics[scale=0.35]{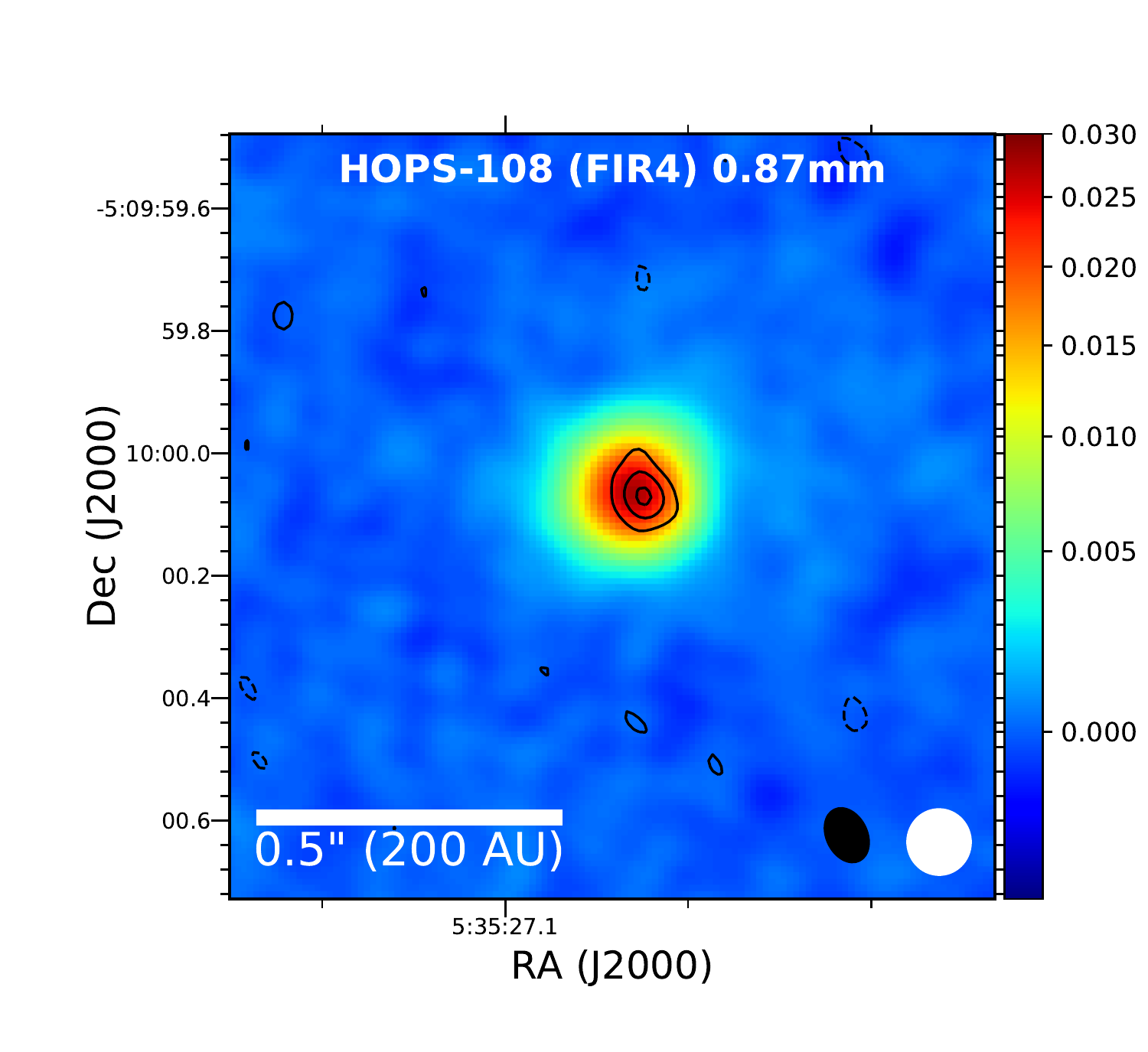}
\includegraphics[scale=0.35]{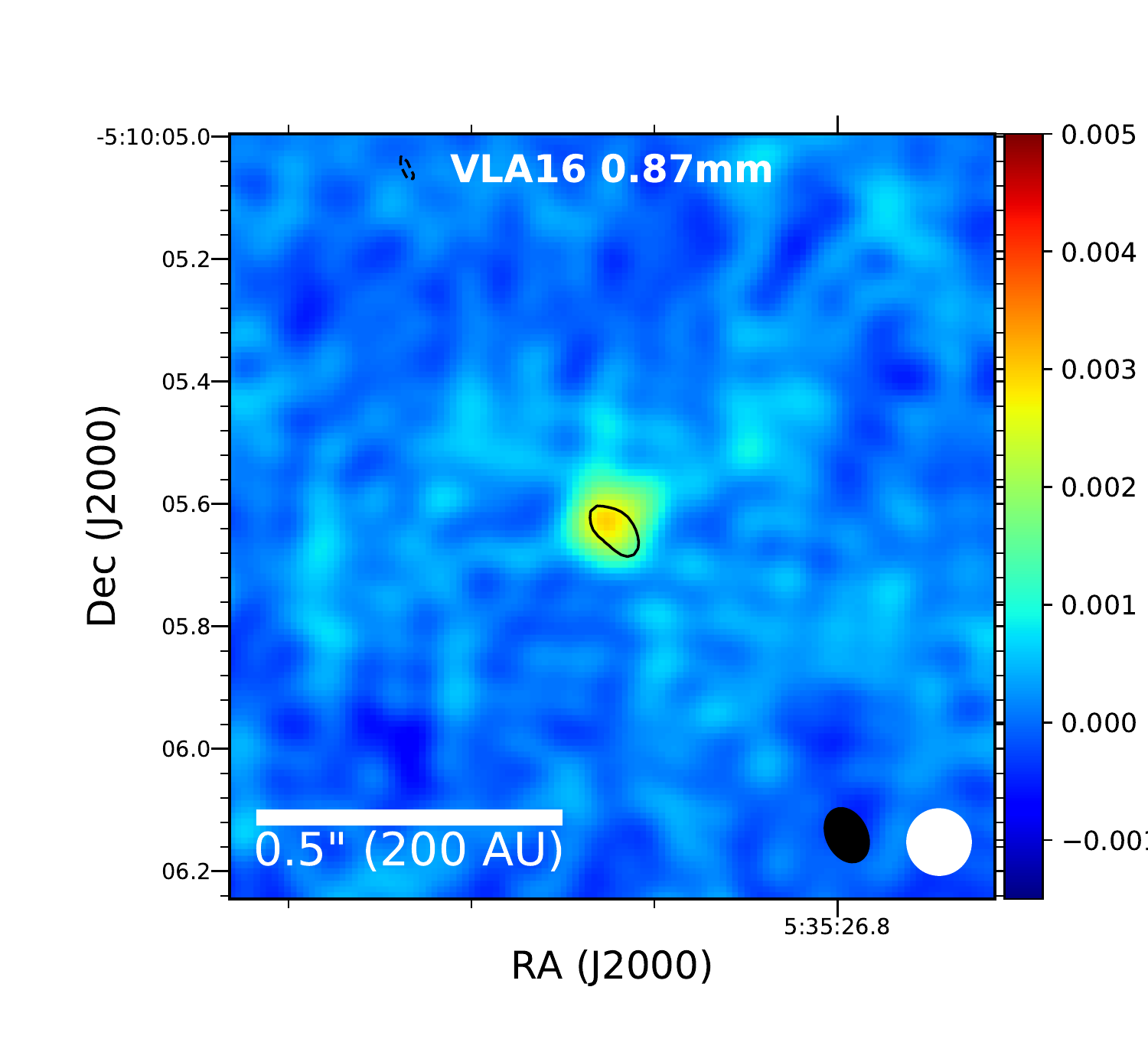}
\includegraphics[scale=0.35]{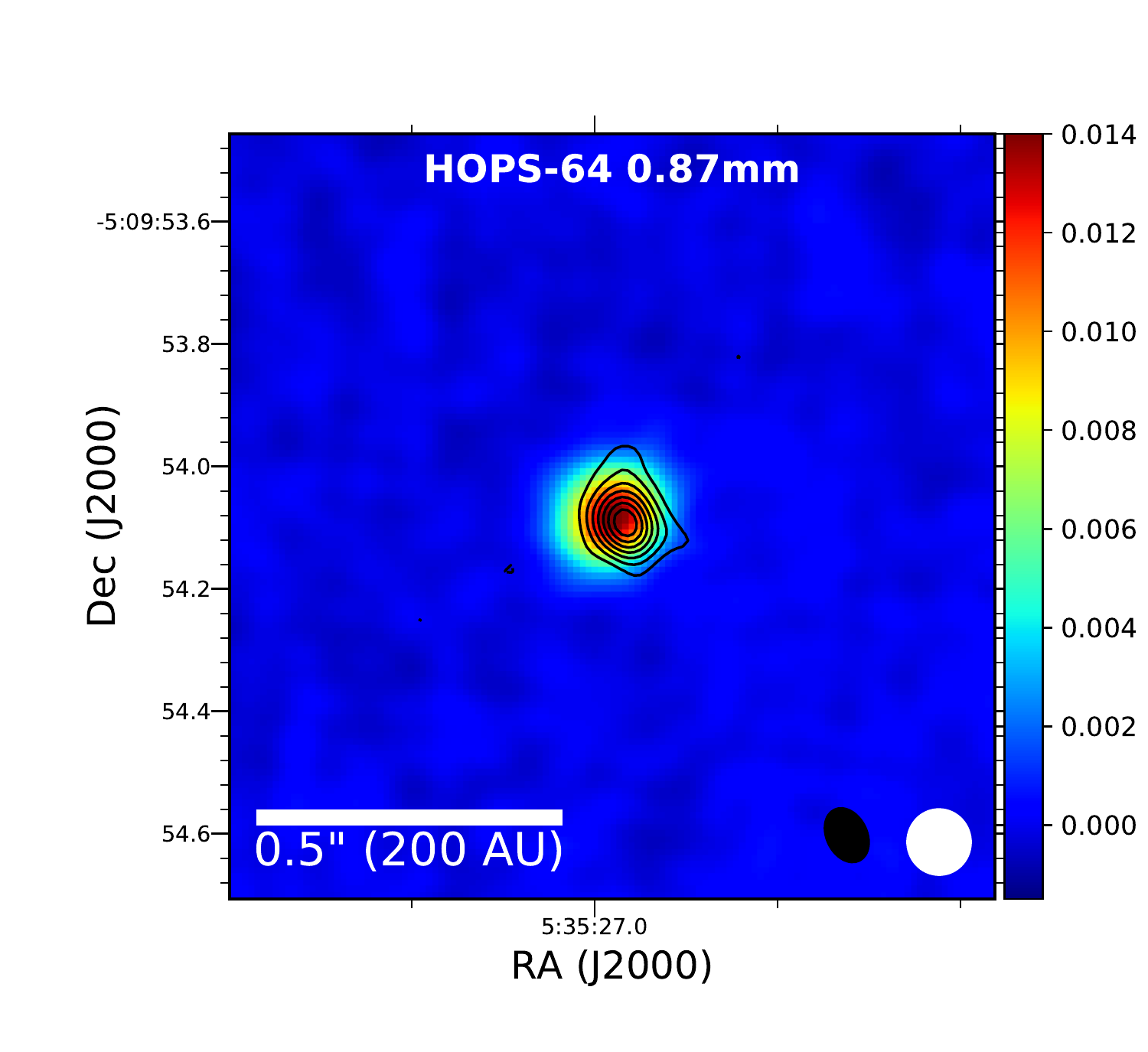}
\includegraphics[scale=0.35]{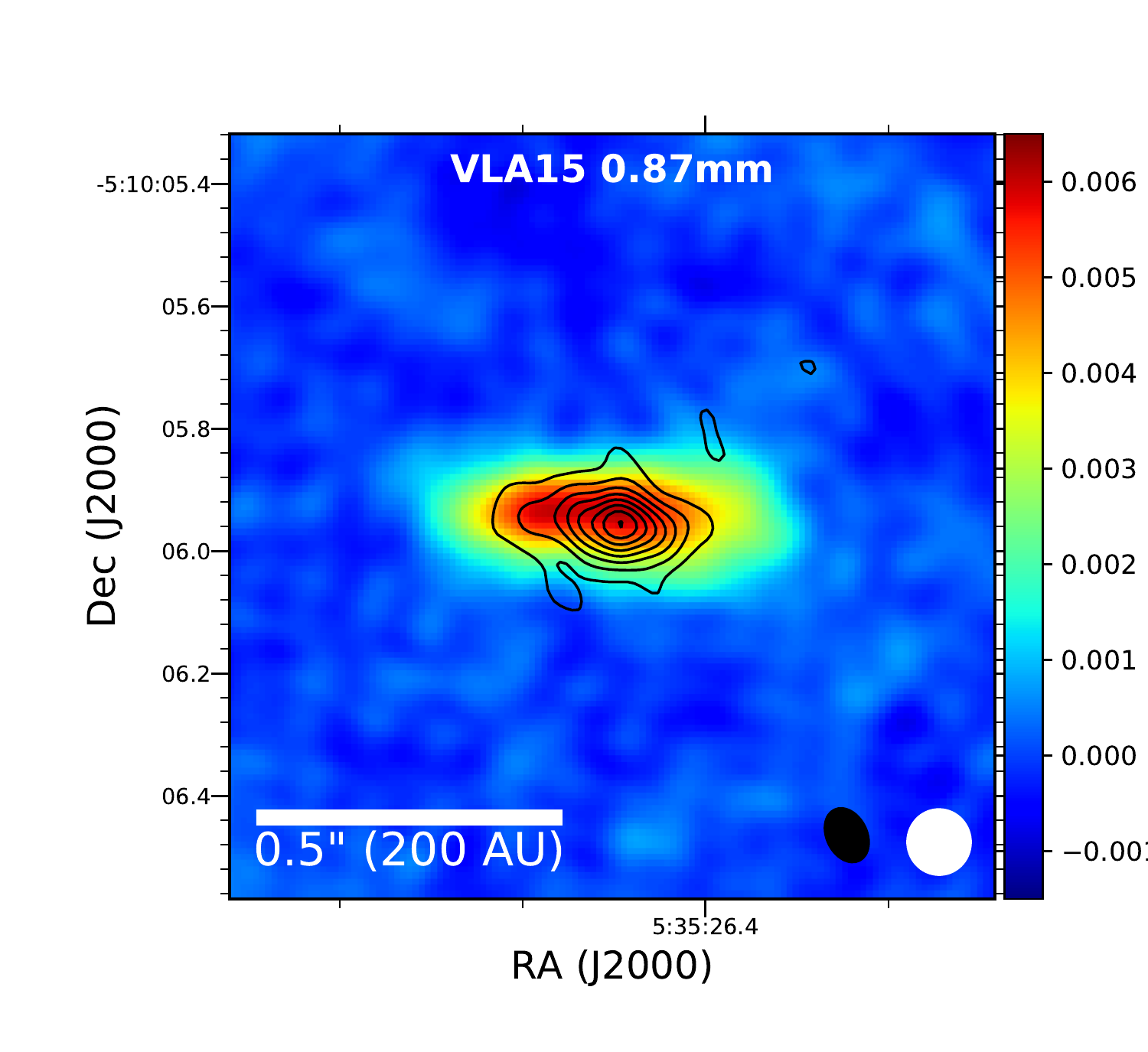}
\includegraphics[scale=0.35]{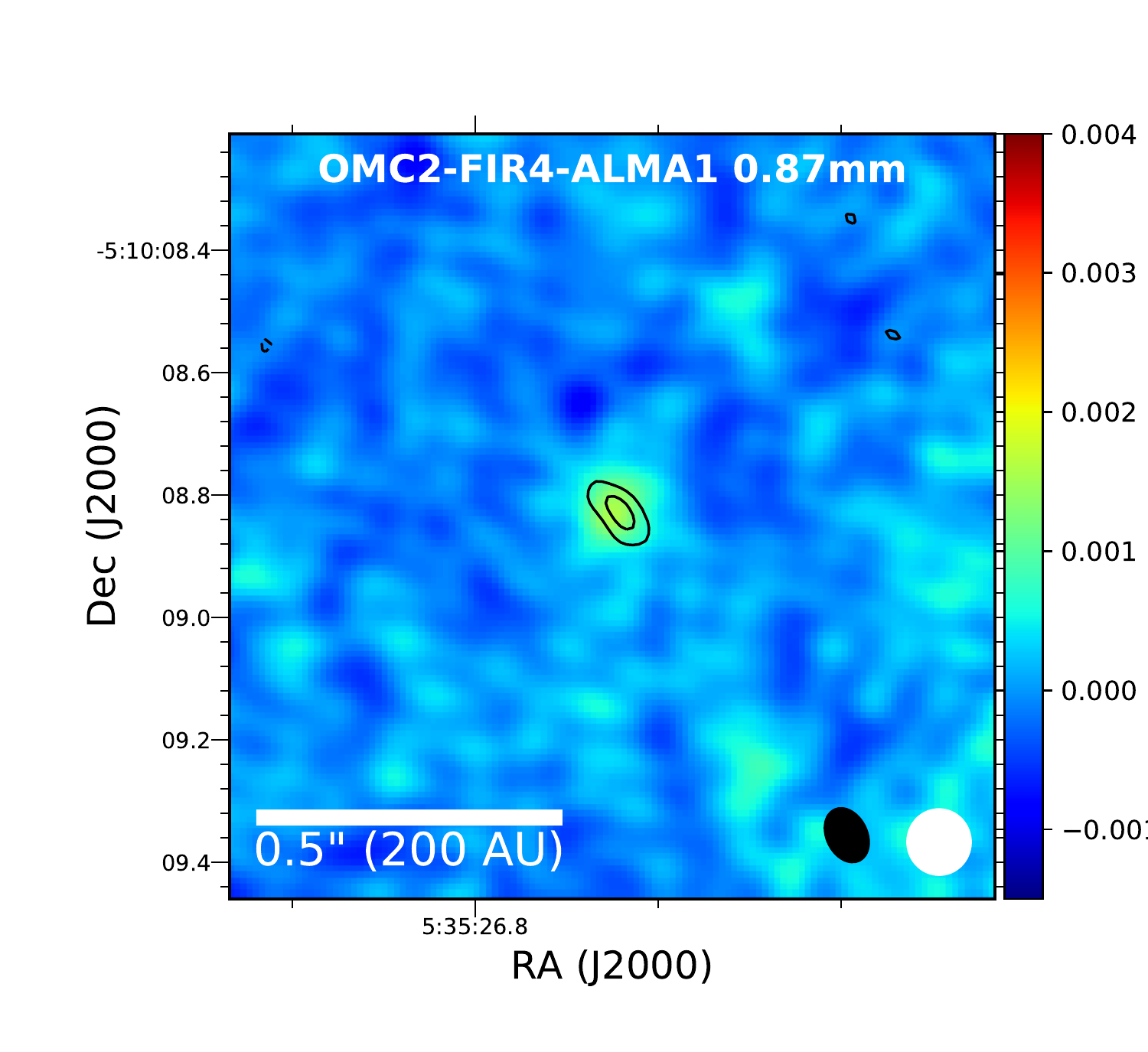}
\includegraphics[scale=0.35]{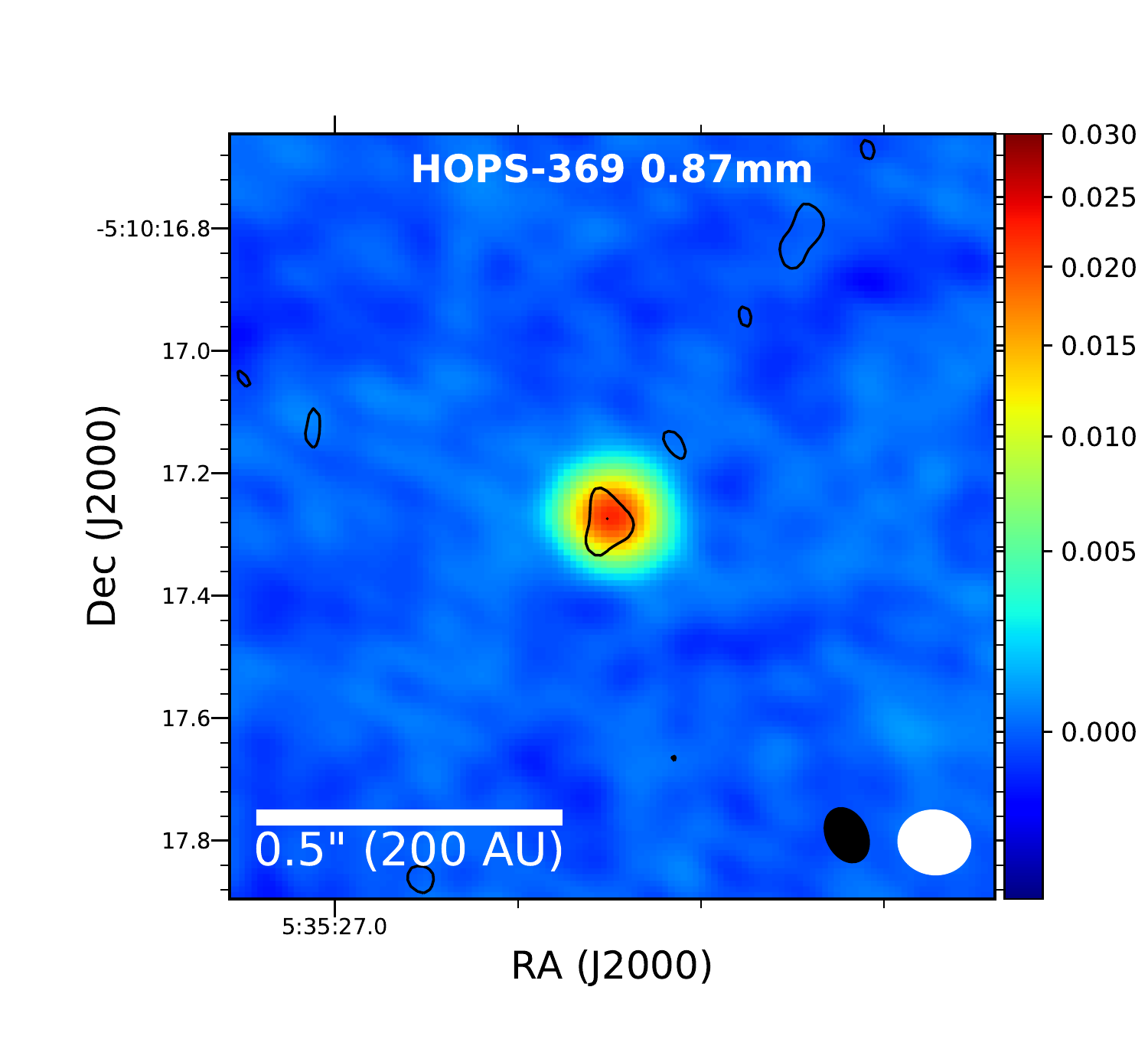}
\includegraphics[scale=0.35]{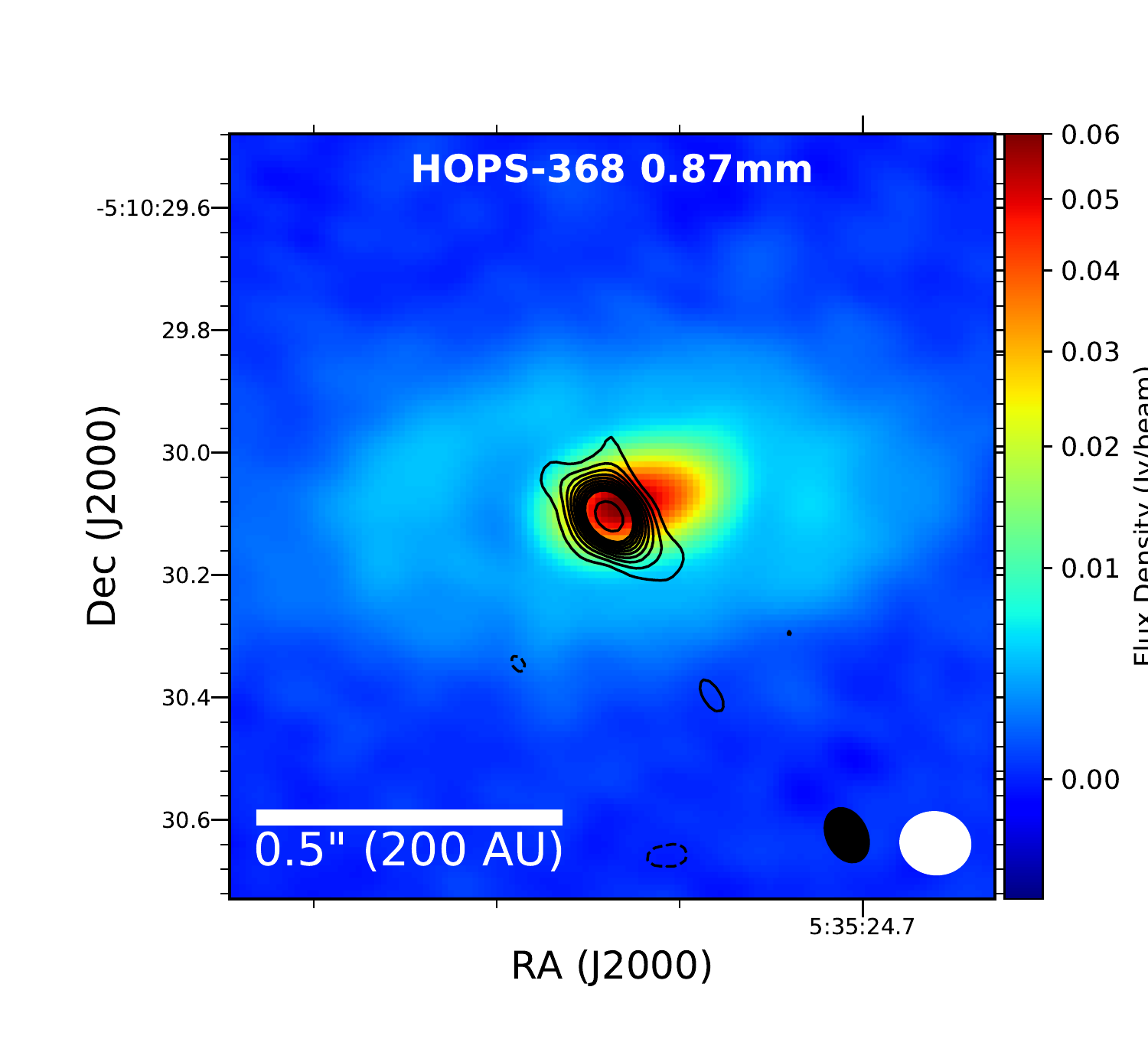}
\includegraphics[scale=0.35]{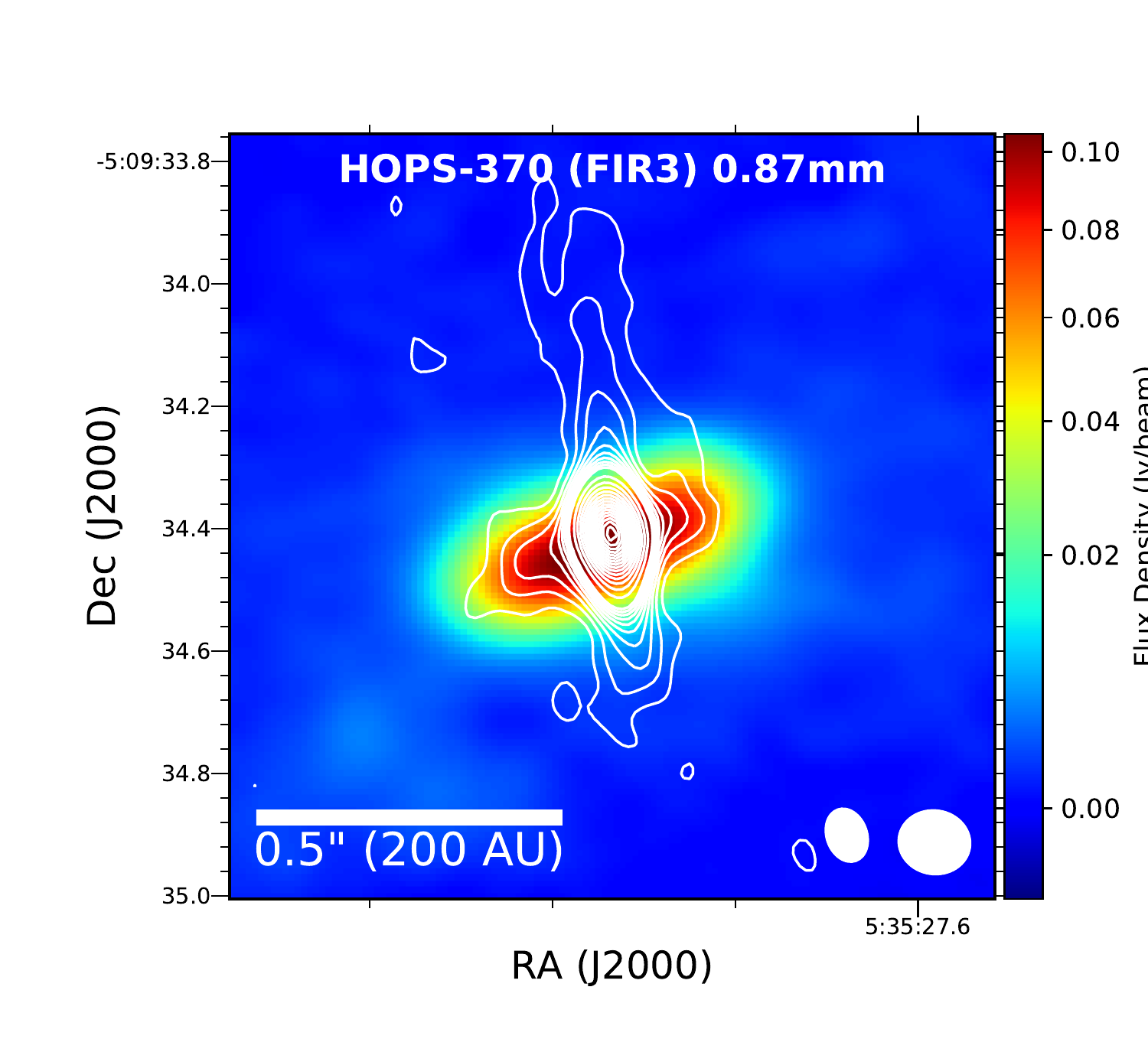}
\includegraphics[scale=0.35]{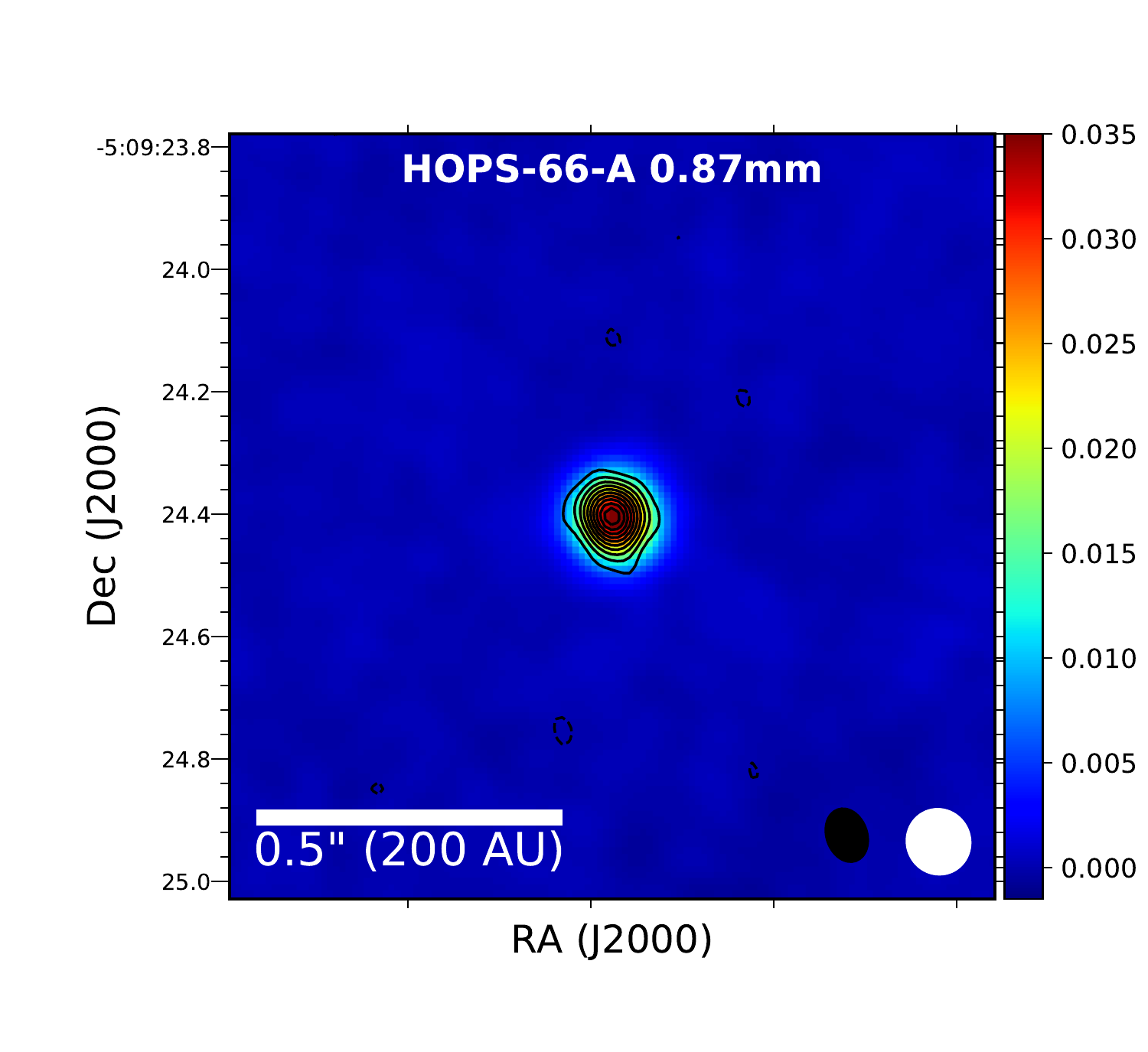}
\includegraphics[scale=0.35]{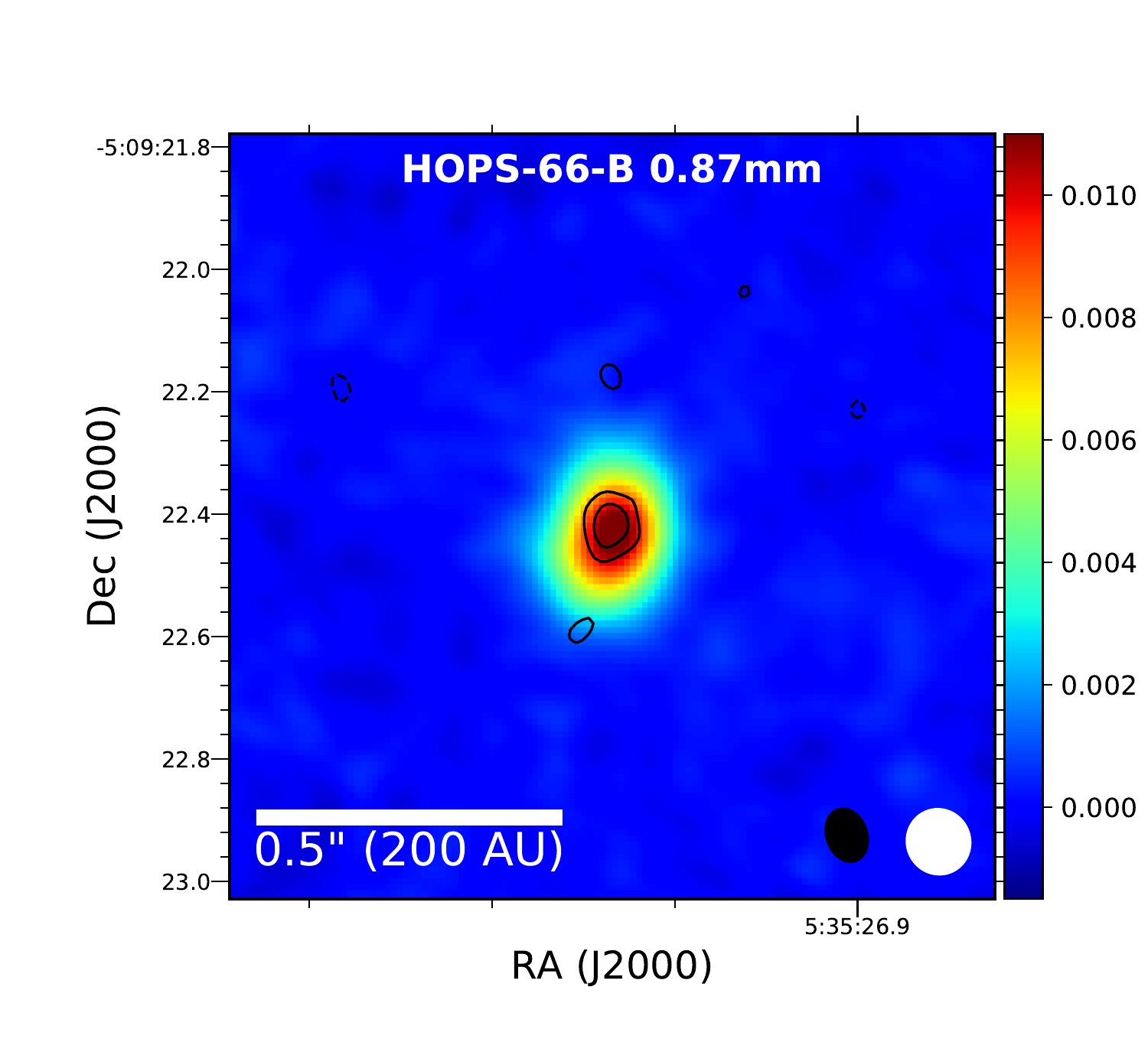}
\includegraphics[scale=0.35]{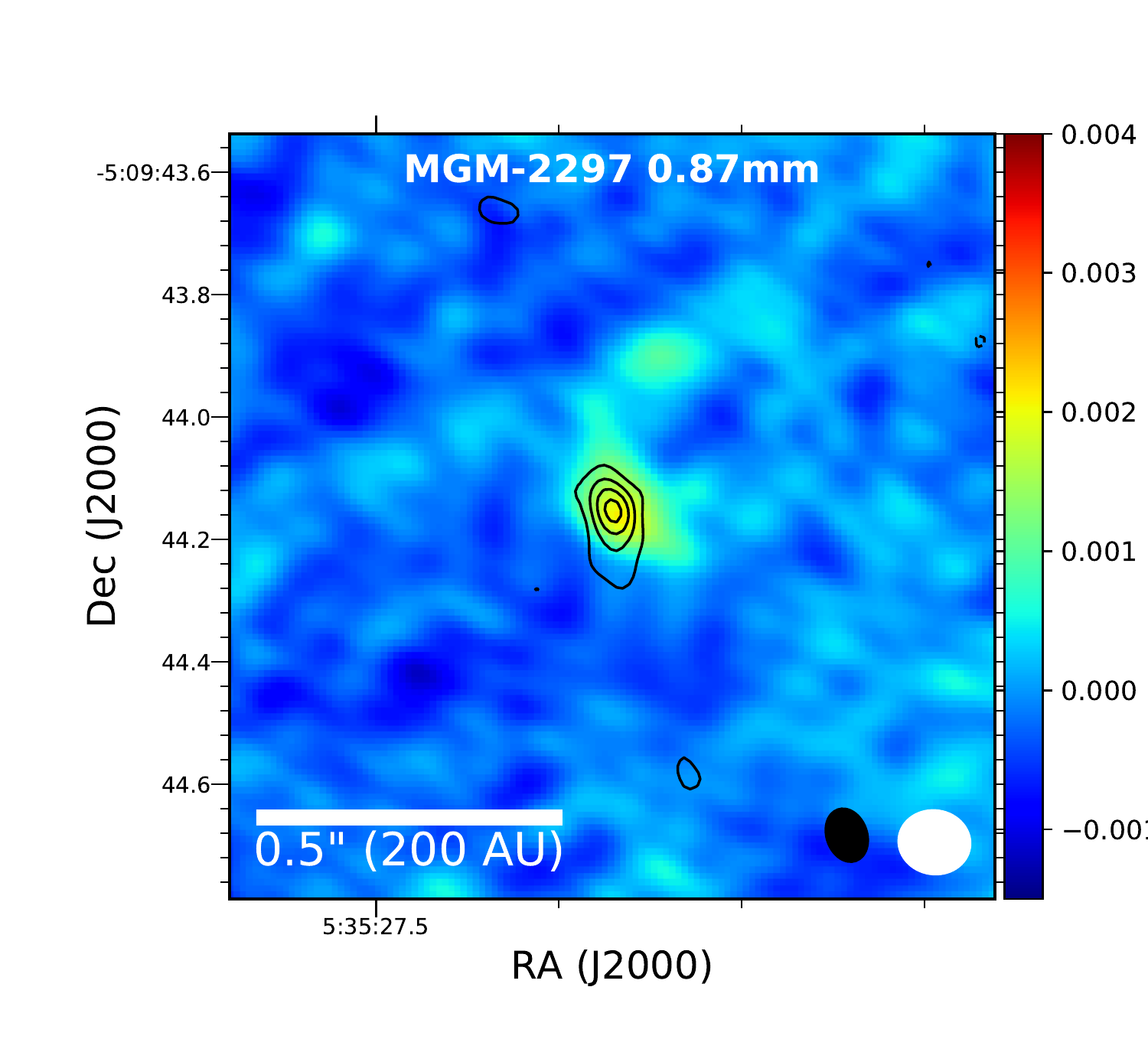}
\end{center}
\caption{Continuum images of the ALMA and VLA-detected protostars in the
vicinity of OMC2-FIR4 and OMC2-FIR3. The ALMA 0.87~mm images are shown in color while the 
VLA 9~mm images are represented as black contours (white for HOPS-370 to enhance visibility). 
HOPS-108 is marginally resolved, but does not appear disk-like, though this could be
due to a low inclination. HOPS-64 is also marginally resolved
and appears to show an elongation in the SE to NW direction. VLA15 is very well-resolved
both at 0.87~mm and 9~mm possibly tracing an edge-on disk; it is the only source in FIR4 that is well-resolved at 9~mm. HOPS-370 is also well-resolved at 0.87~mm and 9~mm with contributions
from both its disk and jet. The 9~mm contours start at and increase by $\pm$3$\sigma$ until 30$\sigma$ where the
contours begin to increase on 15$\sigma$ intervals $\sigma$ for each sources is listed in Table 3. Note
that the images are not primary beam corrected. The beam size 
of the ALMA images is 0\farcs11$\times$0\farcs10 (44~AU~$\times$~40~AU) (white ellipse) and
the VLA beam size is 0\farcs08$\times$0\farcs07 (32~AU~$\times$~28~AU) (black ellipse; white for HOPS-370).
}
\label{continuum}
\end{figure}

We show the ALMA (0.87~mm) and VLA (9~mm) continuum images toward the sources 
within the OMC2-FIR4 and OMC2-FIR3 regions in Figure \ref{continuum}. All the sources are
detected at both wavelengths indicating robust detections. This is important
given the very high resolution of these observations. 
The continuum emission at 0.87~mm on these scales is expected to trace 
mostly emission from the disks surrounding the protostars, but some
emission could result from the inner envelope.

The continuum emission of HOPS-108 has flux densities of $\sim$30~mJy at 0.87~mm
and $\sim$100~$\mu$Jy at 9~mm. 
HOPS-108 appears marginally-resolved at 0.87~mm, but no elongation
or substructure is apparent, and the 9~mm detection is a point source.
Furthermore, HOPS-108 has lower flux densities than several of the
surrounding protostars in the region at these wavelengths (Tables 2 and 3). 
The 0.87~mm emission could be tracing a disk at a low inclination
(close to face-on), which could explain its
near circularly symmetric appearance. VLA15 was
identified at 5~cm by \citet{osorio2017} and exhibits the morphology of an edge-on 
disk at both 0.87~mm and 9~mm, but the asymmetry at 9~mm could also indicate that
this protostar is a close binary.
HOPS-64 has detections in both continuum bands and appears marginally resolved and
elongated as expected for a disk, 
and VLA16 is point-line and faint at both wavelengths.

We stated earlier that HOPS-66 was a binary system separated by 2\farcs23. 
HOPS-66-A appears point-like at both
0.87~mm and 9~mm, while HOPS-66-B appears resolved at 0.87~mm, but point-like at 9~mm.
HOPS-370 is well-resolved at both 9~mm and 0.87~mm, and has a companion at shorter
wavelengths that is not detected at 0.87~mm or 9.1~mm. At 0.87~mm it is clearly disk-like
in appearance, while at 9~mm it has a cross-like morphology. The emission in the east-west direction 
is coincident with the resolved 0.87~mm emission, while the north-south emission is orthogonal
to the major axis and corresponds to the jet direction observed at 5~cm \citet{osorio2017}. Thus, at
9~mm we are detecting both dust emission from its disk and free-free emission
from the jet. 

It is clear that some 9~mm detections appear offset from the 0.87~mm sources. All the FIR4 associated
sources were observed within the same field as HOPS-108 at 9~mm, while all the sources associated with FIR3 (HOPS-370, HOPS-66, and MGM-2297)
were observed within the same 9~mm field as HOPS-370). Some sources show a marginal offset (HOPS-108, HOPS-64, HOPS-66-B), while HOPS-368 shows a large
offset from the center of the bright $\sim$0\farcs5 extended feature found toward it. The dust emission from HOPS-368 
at 0.87~mm is brighter toward the 9~mm position, possibly indicating that the extended feature might reflect two 
blended sources at 0.87~mm, with only one being detected at 9~mm. The slight offset toward HOPS-66-B also appears real 
given that the correspondence of HOPS-66-A is very close. The offsets toward HOPS-108 and HOPS-64, however, may not be real. HOPS-108, HOPS-64, VLA16,
and VLA15 were all observed within the same ALMA field and both HOPS-108 and HOPS-64 are offset in the same direction, and the low S/N of VLA16 at 9~mm
and the extended nature of VLA15 are compatible with a systematic offset. Given that a systematic offset appears most likely, this could
be the result of a systematic phase offset in the case of the ALMA observations which might have resulted from the phase transfer from the
calibrator to the sources or from self-calibration. This offset is 
$\sim$0\farcs03 and does not substantially affect our analysis.

We also investigated how much flux was recovered in our observations relative to the
APEX 0.87~mm observations presented in \citet{furlan2014}. The flux density
measured by the APEX observations was 12.3~Jy in a 19\arcsec\ aperture centered on HOPS-108. 
We summed the flux densities of all the FIR4-associated sources listed in Table 2, finding
a total flux density of 0.257~Jy. Thus, we are only recovering $\sim$2\% of the overall
flux density from this region in our observations.

\subsection{Dust Continuum Mass and Radius Estimates}

We used the integrated flux densities measured with elliptical Gaussian fits 
to analytically calculate the mass of each continuum source within the FIR3 and FIR4 region.
We make the simplifying assumption that the 
dust emission is isothermal and optically thin, 
enabling us to use the equation
\begin{equation}
\label{eq:dustm}
M_{dust} = \frac{D^2 F_{\nu} }{ \kappa_{\nu}B_{\nu}(T_{dust}) }.
\end{equation}
In this equation, D is the distance ($\sim$400~pc), 
F$_{\nu}$ is the observed flux density, B$_{\nu}$ is
the Planck function, T$_{dust}$ is the dust temperature, and $\kappa_{\nu}$ is the
dust opacity at the observed wavelength (0.87~mm for ALMA and 9~mm for the VLA.) 
We adopt $\kappa_{0.87mm}$~=~1.84~cm$^2$~g$^{-1}$ 
from \citet{ossenkopf1994} column 5 (thin ice mantles, 10$^6$ cm$^{-3}$ density), 
and we extrapolate the opacity to 9~mm using the 1.3~mm opacity 
(0.89~cm$^2$~g$^{-1}$) from \citet{ossenkopf1994} and adopting a dust 
opacity spectral index ($\beta$) of 1. Note that our adopted dust opacity at 9~mm
is not from a continuous dust model, but yields masses in agreement with shorter wavelength
studies \citep[e.g.,][]{tychoniec2018,andersen2019}. Otherwise dust masses 
from the 9~mm data are unphysically large. We multiply the calculated dust mass by
100, assuming a dust to gas mass ratio of 1:100 \citep{bohlin1978}, to obtain the gas
mass. The average dust temperature we adopt for a protostellar system is given by
\begin{equation}
T_{dust} = T_{0}\left(\frac{L}{1~L_{\odot}}\right)^{0.25}
\end{equation}
where T$_{0}$ = 43~K, derived from a radiative transfer model grid of disks
embedded within an envelope that 
is described in Tobin et al. (submitted). The average dust temperature of 
43~K is reasonable for a $\sim$1~\lsun\ protostar at a
radius of $\sim$50~AU \citep{whitney2003a,tobin2013}. The luminosity is the \lbol\ for each 
protostellar system measured from the SED \citep{furlan2016}.
If a system does not have a measured \lbol\ (e.g., VLA16 and VLA15), then 1~\lsun\ is assumed.

The masses derived from the continuum sources are listed in Table 4, as well as the
radii derived from the Gaussian fits. The continuum emission from the protostars
is likely to be partially optically thick, thus the masses are likely lower limits, especially
at 0.87~mm. The half-width at half-maximum (HWHM) of the continuum emission
multiplied by the distance to Orion ($\sim$400~pc) is used as an estimate of
the source radius.
We note that there is often disagreement between the continuum 
masses measured at 0.87~mm and 9~mm. This can be due to both the uncertainty in 
scaling the dust mass opacity to 9~mm, but also there is likely free-free emission contributing
to the 9~mm flux density and thus inflating the mass estimates \citep[e.g.,][]{tychoniec2018}.
The spectral indices determined from 8.1~mm to 10.1~mm using the full bandwidth of the
VLA observations, also shown in Table 4,
are evidence for free-free emission with spectral indices
less than 2 found for several sources. A spectral index less than 2 is shallower than
optically thick dust emission thereby requiring an additional emission mechanism.

\subsection{Methanol Emission Toward HOPS-108 and HOPS-370}

We detected strong emission from three methanol transitions toward
HOPS-108 and HOPS-370 within the spectral window containing \twco. 
Methanol (CH$_3$OH) is a complex organic molecule (COM), referring to molecules
containing carbon and a total of 6 or more atoms \citet{herbst2009} that are
typically formed on the surfaces of icy dust grains \citep[e.g.,][]{chuang2016}.

We examined the kinematics of the lines using integrated intensity maps of the blue and
red-shifted emission. The blue- and red-shifted contours of three methanol
transitions are shown in Figure \ref{methanol}.
The lowest excitation methanol line ($J=5_4\rightarrow6_3$) exhibits an 
east-west velocity gradient in both HOPS-108 and HOPS-370.
The higher excitation
methanol lines toward HOPS-108 ($J=16_1\rightarrow15_2$ and $J=18_3\rightarrow17_4$) 
have velocity gradients from southeast to northwest. 
The shift in the position angle is $\sim$135\degr, demonstrating that the different transitions
may be arising from different physical environments in HOPS-108. 
However, toward HOPS-370 the higher-excitation lines trace an east-west velocity gradient,
appearing to trace a rotation pattern across the disk detected in dust continuum.
The methanol emission toward HOPS-370 appears reduced at the regions of brightest
continuum emission and the brightest methanol emission is above and below the continuum disk 
on the northeast and southwest sides, not unlike HH212-MMS \citep{lee2018}.

Additional molecular line emission was detected toward HOPS-108 and HOPS-370, but not toward the
other sources in the field. The molecular line emission toward HOPS-108 and HOPS-370 is
analyzed and discussed in more detail in Appendix A.

\subsection{Outflows in $^{12}$CO}

\begin{figure}
\begin{center}

\includegraphics[scale=0.375]{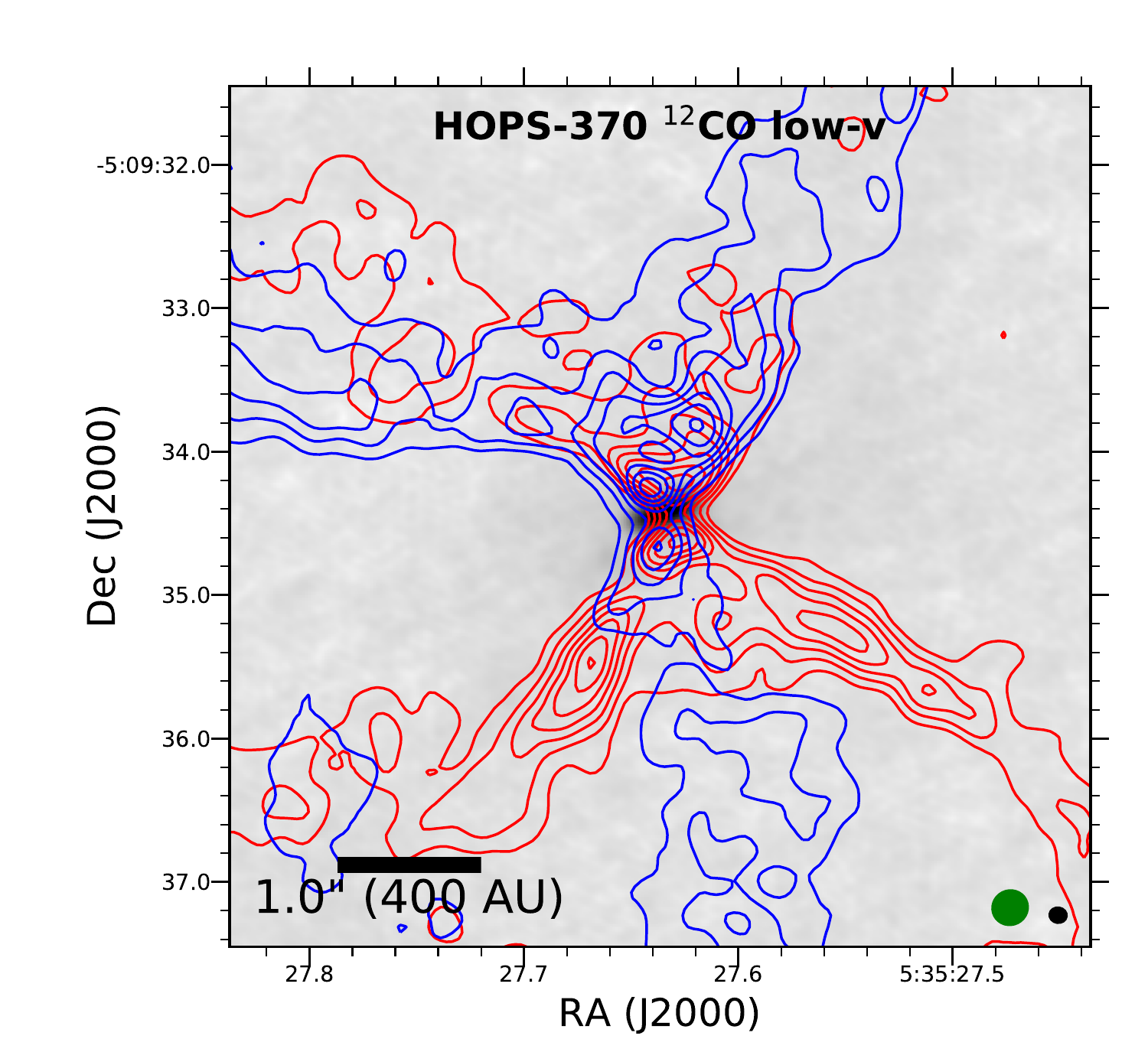}
\includegraphics[scale=0.375]{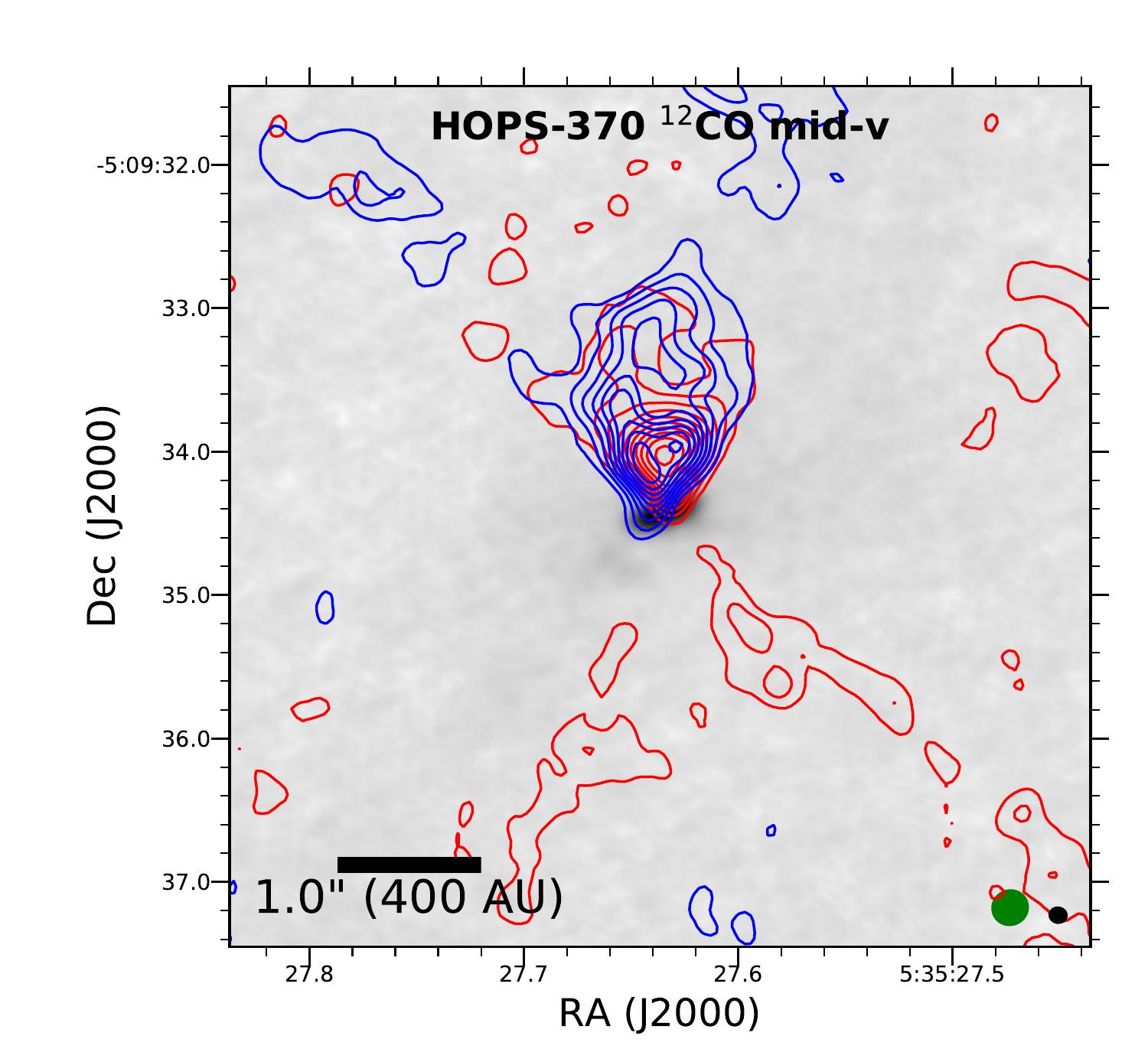}
\includegraphics[scale=0.375]{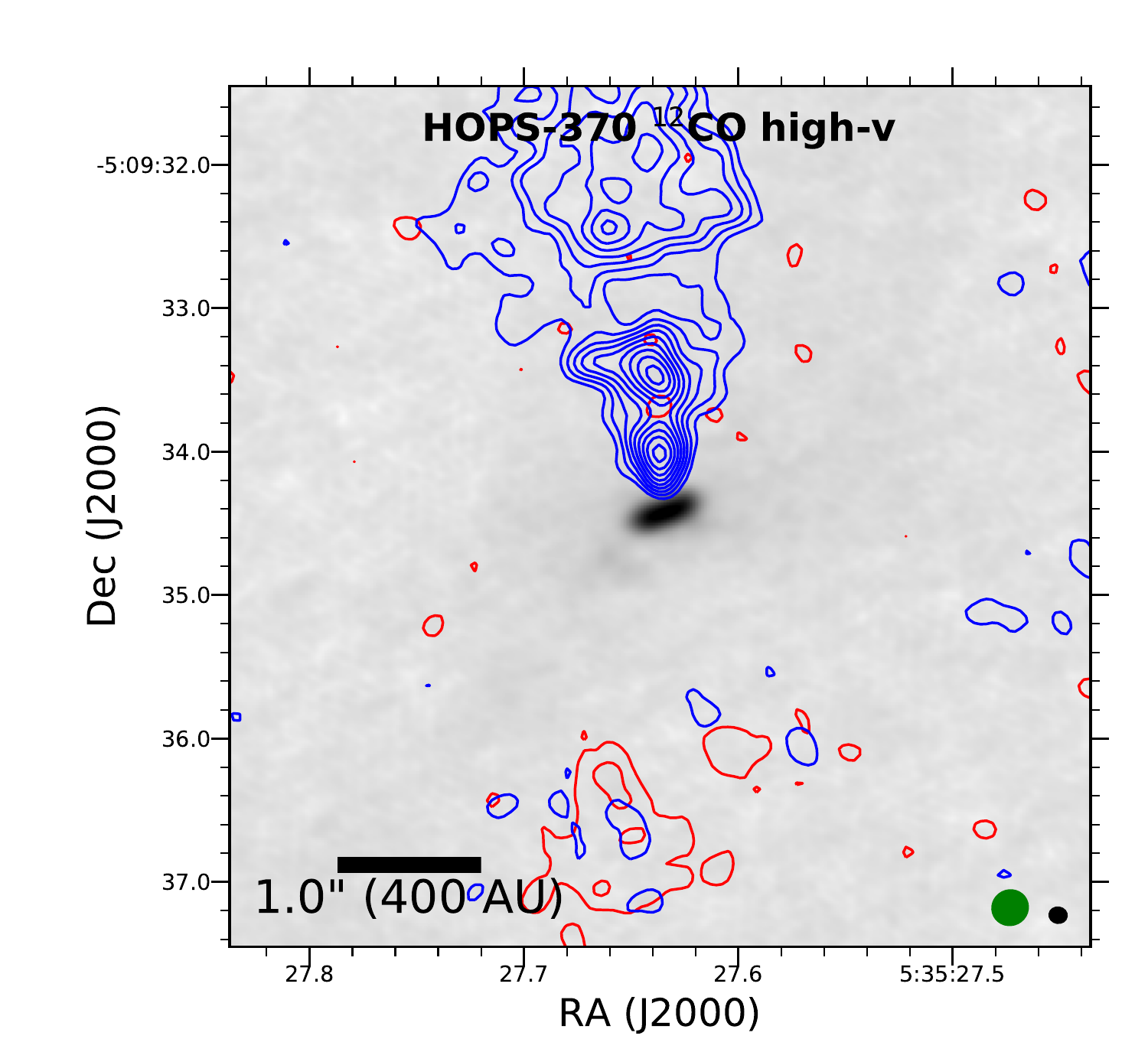}
\end{center}
\caption{
ALMA \twco\ blue- and red-shifted integrated intensity maps toward HOPS-370
overlaid ALMA 0.87~mm continuum (grayscale). The three panels (from left to right) correspond to
low-velocity (-26 to -3~\kms and 3 to 12~\kms), medium-velocity (-39 to -27~\kms\ and 13 to 45~\kms), and 
high-velocity (-56 to -40~\kms and 46 to 66~\kms). The velocity ranges are with 
respect to the system velocity of $\sim$11.2~\kms. The low and medium velocity panels show evidence 
for spatial offset between the blue and red-shifted emission at the base of the outflow.
The contour levels in each panel start at 5$\sigma$ and increase on 3$\sigma$ intervals.
In the low-velocity, mid-velocity, and high-velocity panels, 
$\sigma_{low}$=0.12 (0.09) Jy~beam$^{-1}$, $\sigma_{mid}$=0.053 (0.076) Jy~beam$^{-1}$,
and $\sigma_{high}$=0.056 (0.063) Jy~beam$^{-1}$, respectively, with the red-shifted
level being given in parentheses. The $^{12}$CO beam is 
0\farcs25$\times$0\farcs24. 
}
\label{H370-outflow}
\end{figure}

HOPS-370 exhibits a clear high-velocity outflow in \twco\ ($J=3\rightarrow2$) 
shown in Figure \ref{H370-outflow} that is in agreement with the larger-scale
CO outflow detected by \citet{shimajiri2008}. The blue-shifted outflow is oriented
in the northeast direction, while the redshifted outflow is in the southeast direction.
There is spatial overlap within the blue- and red-shifted lobes in the low and mid velocity
ranges due to the source being located near-edge-on. Also in the mid-velocity panel,
the origin of the blue- and red-shifted outflows appears to be offset on either side of the 
disk in continuum emission.

We examined the \twco\ data toward HOPS-108 to
see if an outflow is detectable from it. 
We show integrated
intensity maps of the red- and blue-shifted \twco\ emission toward HOPS-108 in Figure \ref{12CO-H108}.
Similar to the HOPS-370 images in Figure \ref{H370-outflow} we break the \twco\ emission into different velocity
ranges and overlay them the VLA 5~cm maps from \citet{osorio2017}. 
The extended 5~cm emission northeast and southwest (VLA 12C and VLA 12S),
are knots from the HOPS-370 jet emitting synchrotron emission, while HOPS-108 at
the center is emitting thermal free-free emission \citep{osorio2017}. The \twco\ 
emission has significant complexity; the low-velocity ($\pm$3-10~\kms)
 emission does not appear very organized, but there is an arc-like
feature $\sim$4\arcsec\ southwest of HOPS-108 that is coincident with emission detected at 5~cm.
Furthermore, in the low-velocity map there is a hint of blue- and red-shifted emission
extending $\sim$1\farcs5 on either side of the continuum source that could trace an outflow
at a position angle of $\sim$45\degr, but this feature is highly uncertain and perhaps
spurious. We examined the \thco\ emission, but the emission was not strong enough to 
detect a clear outflow at low-velocities.

The medium velocity ($\pm$10-20~\kms) emission remains complex, the blue-shifted 
emission is dominated by a linear feature northeast
of HOPS-108 that does not appear to trace back to HOPS-108. 
The red-shifted emission in this velocity range
has a morphology that resembles an elliptical ring or loop, possibly centered on 
and surrounding
HOPS-108. Northeast of the protostar, extended 5~cm emission appears within the loop-shaped feature 
traced by the red-shifted \twco.
There is still blue- and red-shifted emission coincident
with the bright 5~cm emission to the southwest of HOPS-108, but the blue-shifted emission there
is fainter.

Lastly, at the highest velocities (-20 to -30~\kms\ and 15 to 25~\kms) there is
 no corresponding blue-shifted emission near HOPS-108, but 
red-shifted emission is still apparent. The loop seen at medium velocities is now
smaller and appears pinched toward
HOPS-108 along the minor axis of the loop. Also,
the higher-velocity red-shifted \twco\ emission seems to anti-correlate 
with that the spatial distribution of the 5~cm emission within this region.

A clear outflow driven by HOPS-108 
cannot be positively identified in the ALMA observations, though 
there could be a hint of one at
low velocities. It is possible that the red-shifted \twco\ emission
observed is tracing an outflow from HOPS-108, but the morphology of
the emission only changes northeast of the protostar and not southwest. It is possible that
the protostar is oriented face-on, a possibility indicated by the marginally-resolved
and circularly symmetric continuum emission. In this case, the morphology of the outflow would
appear more complex at this resolution if it has a wide opening angle. 
It is difficult, however, to reconcile the appearance of the loops 
surrounding HOPS-108 with a typical bi-polar outflow.
Also, the \twco\ emission in the vicinity of HOPS-108 could be
complex due to the outflow from HOPS-370 (FIR3) crossing this region
\citep{shimajiri2008,gonzalez2016}. Other searches for outflows in Orion from 
$^{13}$CO emission \citep{williams2003}, $^{12}$CO \citep{shimajiri2008, hull2014, kong2018},
and the near-infrared \citep{davis2009,stanke2006} are also not conclusive for HOPS-108.
The highly embedded nature of HOPS-108 and the density of nearby sources reduces the utility of
near-infrared outflow indicators and the previous molecular line observations that could have
traced the outflow; both had low angular resolution (even in the near-infrared)
 and were confused with the outflow from HOPS-370.

We also examined the \twco\ emission toward VLA16, HOPS-64, VLA15, and 
ALMA1, and did not find evidence for outflows from any of these sources. The strong CO emission
from the molecular cloud and the spatial filtering, however, make
these non-detections far from
conclusive and observations with higher S/N and imaging fidelity are required to properly
establish the presence or lack of CO outflows from these sources. Finally, the
apparent edge-on nature of VLA15 will make its outflow difficult to disentangle from
the molecular cloud because any outflow is not expected to have a large velocity
separation from the cloud.

\section{Discussion}

Most previous studies of OMC2-FIR4 and OMC2-FIR3 have been limited to modest spatial resolution.
The highest resolution millimeter continuum maps from ALMA and NOEMA had $\sim$3\arcsec\ 
resolution or worse \citep{lopez-sepulcre2013,kainulainen2016,favre2018}. This 
limitation has resulted in significant ambiguity of the actual content
and location of discrete protostellar sources within FIR4. The cm-wave maps from the
VLA were useful in identifying likely young stellar objects \citep{osorio2017}, 
but the presence of the extended jet from HOPS-370 (FIR3) in emission at 5~cm 
makes it difficult to positively infer a protostellar nature from the 5~cm detections in
the region. The observations presented here with $<$0\farcs1 (40~AU) resolution from 
both VLA at 9~mm and ALMA at 0.87~mm enable us to more
conclusively identify the protostellar content from their compact dust emission at these
wavelengths. Hence, these data shed new light on the star formation activity that is taking
place within these massive cores.

\subsection{A Young Stellar Group in OMC2-FIR4}

The source known as FIR4 has long been known to not simply be
a discrete protostar, but possibly a collection of several sources \citep{shimajiri2008}. The
designation of FIR4 refers to the $\sim$30\arcsec\ region (12000 AU) 
centered on the large, massive core ($\sim$30~\msun) identified at 
1.3~mm by \citet{mezger1990} and followed-up by \citet{chini1997}. 
Further analysis by \citet{furlan2014} fit a modified blackbody to the emission at 
wavelengths longer than 160~\micron, finding a temperature of 22~K, a mass of 27~\msun, 
and a luminosity of 137~\lsun. The observed total integrated luminosity 
of FIR4 is $\sim$420~\lsun\ \citep{mezger1990}, 
but much of this luminosity originates at wavelengths longer than 70~\micron\ and includes
contributions from multiple protostars
and likely external heating.

Several studies have suggested that FIR4 is
a proto-cluster. \citet{shimajiri2008} resolved FIR4 into 11 cores at $\lambda$=3~mm, 
but compared to the ALMA $\lambda$=3~mm maps from \citet{kainulainen2016} with superior sensitivity and our detected source positions, 
some of the fragmentation within FIR4 detected by
\citet{shimajiri2008} is in fact
spurious due to interferometric imaging artifacts. 
The maps from \citet{kainulainen2016} identify about 6 fragments 
within FIR4 at $\sim$3\arcsec\ (1200 AU) resolution, while \citet{lopez-sepulcre2013} 
identify 2 main fragments at $\sim$5\arcsec\ (2000 AU)
resolution.  These observations, however, were optimized for examining 
fragmentation of the FIR4 core on larger scales,
while the compact dust emission that we are 
detecting on $\sim$0\farcs1 (40 AU) scales 
is likely to be directly associated with forming protostars within FIR4. 

HOPS-108, VLA16, HOPS-64, VLA15, and ALMA1
all appear to be associated with the FIR4 core at least in projection. The projected
separations of these sources with respect to HOPS-108 are given in Section 3, and
they have an average projected separation of 10\farcs4 (4160 AU). 
HOPS-369 is on the outskirts of FIR4, separated by 17\farcs3 from HOPS-108.
HOPS-369 is classified as a more-evolved Flat Spectrum
protostar, and HOPS-64 might also be a Flat Spectrum protostar, despite its
classification as a Class I, due to blending at longer wavelengths. 
Furthermore, the near to mid-infrared characteristics and detections at short wavelengths 
toward HOPS-64 and HOPS-369
point to them being more evolved and possibly located toward the
edge of the FIR4 core, in the foreground. HOPS-108, VLA16, VLA15, and ALMA1, however, 
all appear as compact continuum sources and do not have obvious direct detections
at wavelengths shorter than 8~\micron. Their lack of short wavelength detections are
indicative of their youth and likely
physical association with the FIR4 core and embedded within it. 
We note, however, that we cannot rule-out 
some sources being located in the foreground or background for FIR4. 
For example, ALMA1 lacks a peak in the ALMA 3~mm map at 3\arcsec\ resolution meaning 
that it does not have a significant amount of dust emission concentrated at its position.

While previous studies of HOPS-108 indicate that it is likely the most luminous
source within FIR4, VLA16 and VLA15 also lie close to the center of the core. 
Numerical studies have shown that even monolithic collapse of a massive core
could lead to the formation of a young stellar group, but those fragments 
are generally formed via disk fragmentation \citep{krumholz2009,rosen2016}.
There is, however, no protostar at the exact center of the FIR4 core
that would likely have formed via monolithic collapse, 
and the several widely-separated protostars within FIR4 could point to 
competitive accretion within the core if these 
protostars are physically associated with FIR4
and actively accreting material \citep[e.g.,][]{zinnecker1982,bonnell2001,hsu2010}. 
Indeed, VLA16 and VLA15 could continue to accrete mass and 
evolve into intermediate-mass stars. However, the only observational data on
VLA16 and VLA15 are their millimeter and centimeter flux densities and there are no
current constraints on their luminosities or kinematics.

There has also been debate about what is driving the complex chemistry that is observed
on larger scales within FIR4 \citep{lopez-sepulcre2013, ceccarelli2014,favre2018}. 
In \citet{lopez-sepulcre2013}, they denote their fragments as main and west, in addition to 
south which appears in molecular lines. The region denoted `main' is most closely associated 
with our detection of the 
compact continuum toward HOPS-108. Also, there is bright methanol and DCN emission 
from the location of main/HOPS-108 indicative of a heating source evaporating methanol, 
especially since an offset between DCN and DCO$^+$ could indicate that 
DCN is forming via high-temperature chemistry with 
CH$_2$D$^+$ \citep[e.g.,][]{parise2009,oberg2012}.

\citet{lopez-sepulcre2013} suggested that FIR4 might contain
 an embedded B star with $\sim$1000~\lsun, based on 
their observations of complex organics and the 
marginally-resolved source detected at 3.6~cm wavelengths 
with $\sim$6\arcsec\ (2400 AU) resolution \citep{reipurth1999}.
Thus, they interpreted the detection of 3.6~cm emission as an ultracompact HII region.
With higher-resolution cm data, \citet{osorio2017} showed
that the centimeter-wave emission contains contributions from both HOPS-108 and 
knots in the HOPS-370 outflow.  
These knots show both proper motion away from HOPS-370 and non-thermal 
spectral indices. Hence, the data in \citet{osorio2017} showed conclusively that the emission 
is not from an ultracompact HII region (see Section 4.2). An ultracompact
HII region would have a spectral index reflecting thermal free-free emission, and
the non-thermal spectral index and proper motions observed by are
inconsistent with that interpretation.

\subsection{Outflow Interaction with HOPS-370 (FIR3)?}

\begin{figure}
\begin{center}

\includegraphics[scale=0.375]{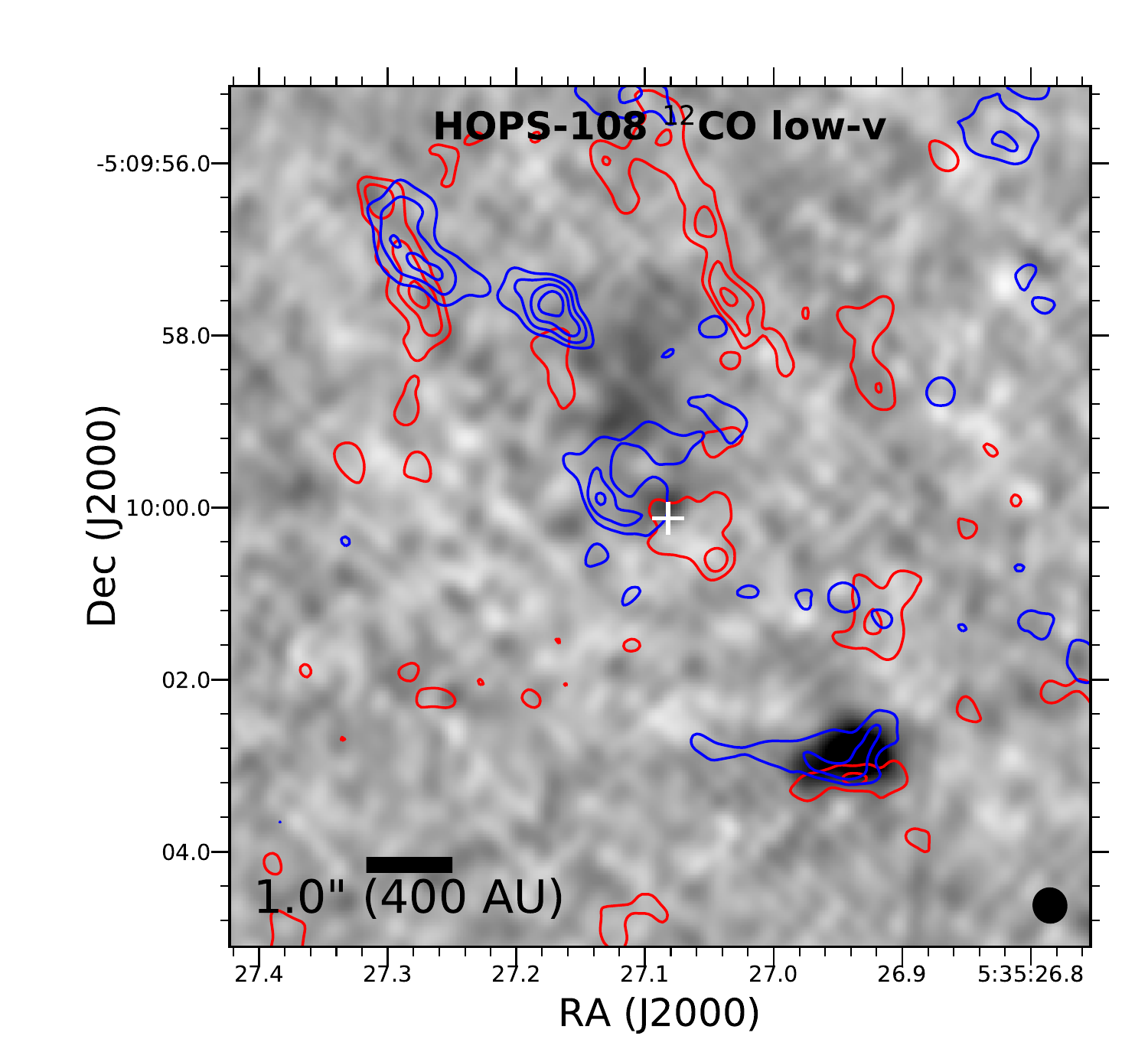}
\includegraphics[scale=0.375]{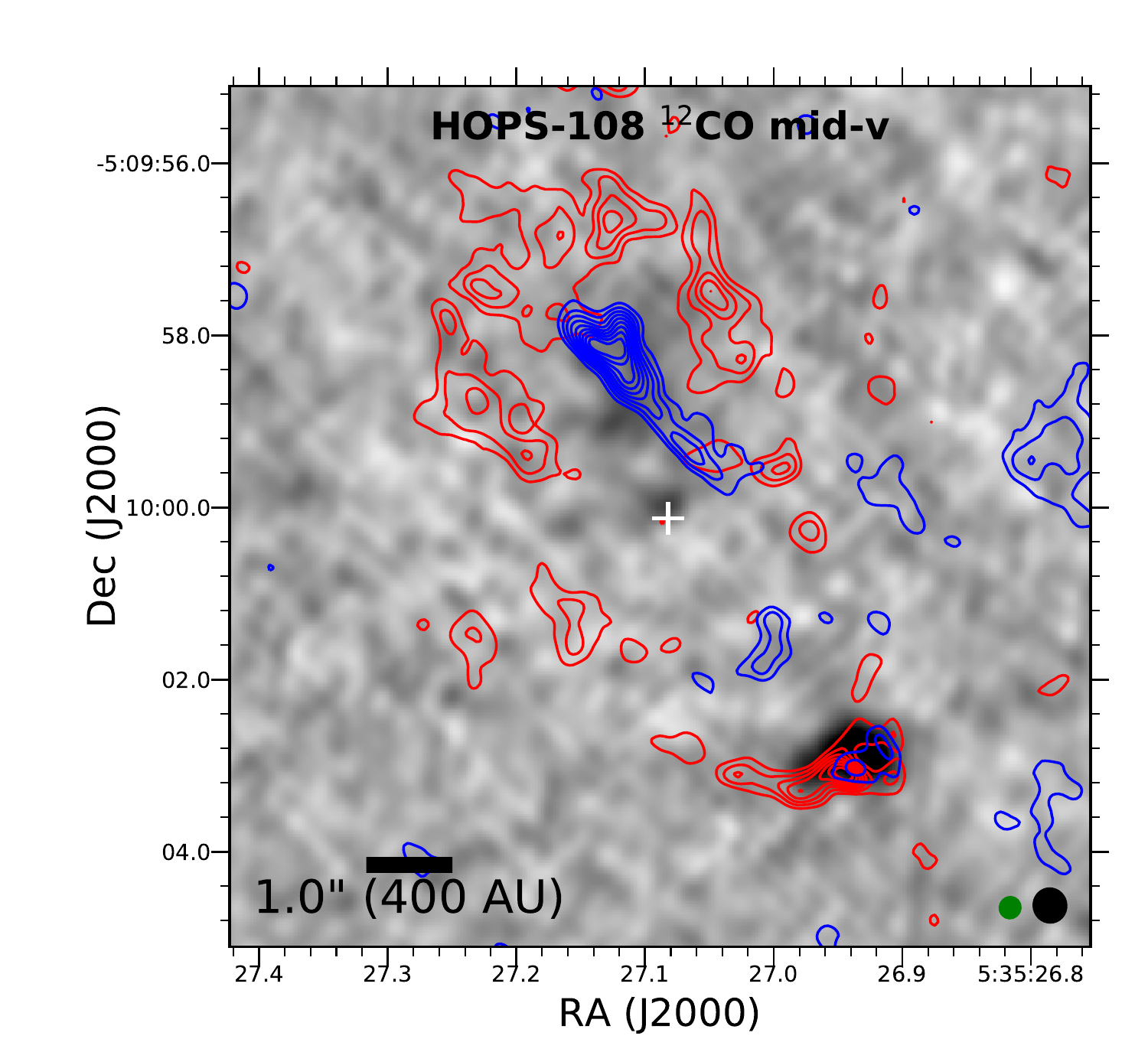}
\includegraphics[scale=0.375]{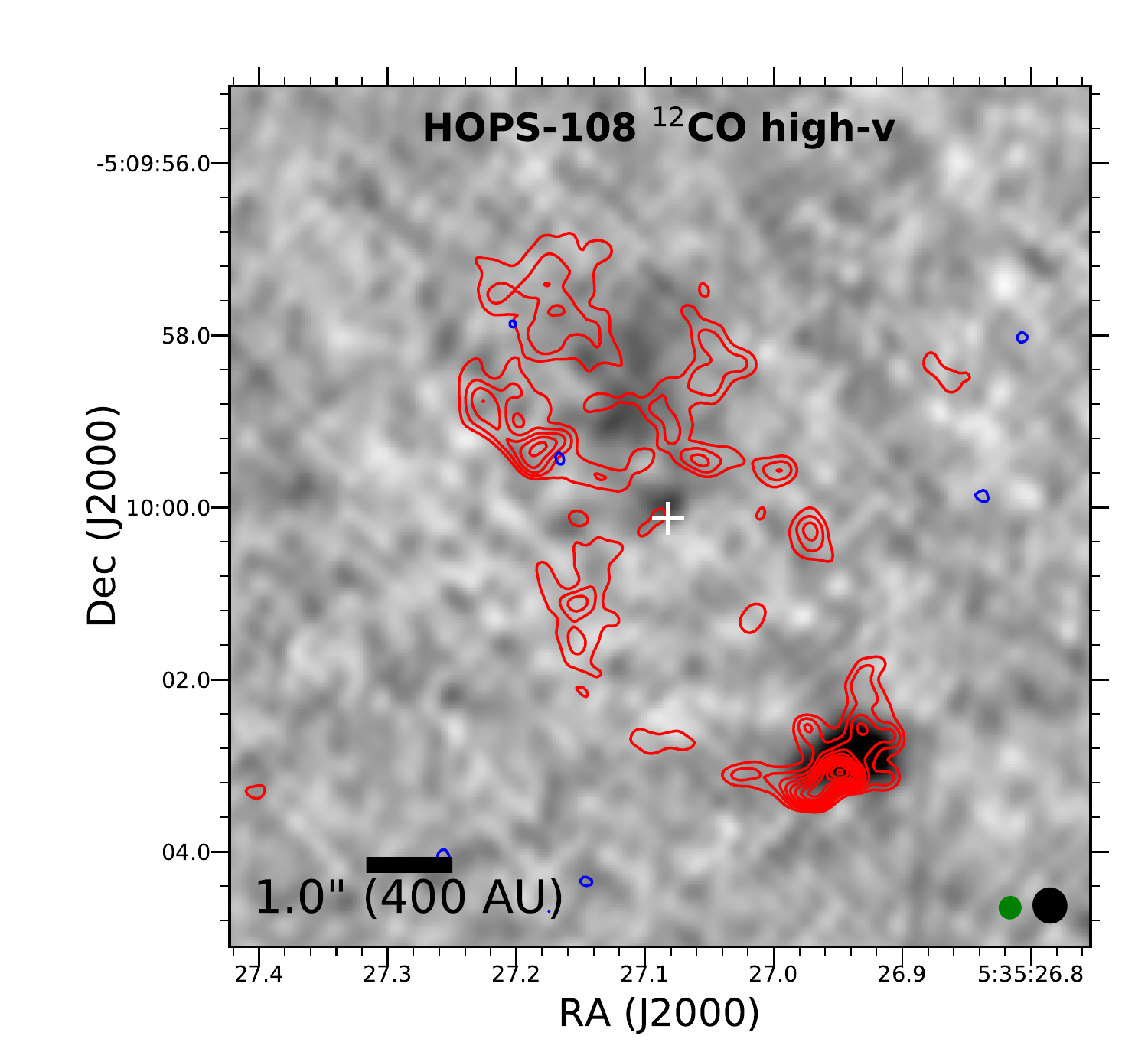}
\end{center}
\caption{ALMA \twco\ blue- and red-shifted integrated intensity maps
overlaid on VLA 5~cm emission (grayscale). The position of HOPS-108 from the ALMA continuum
data is marked with a white cross. The three panels (from left to right) correspond to
low-velocity ($\pm$3 to 10~\kms), medium-velocity ($\pm$10 to 20~\kms), and 
high-velocity (-20 to -30~\kms and 15 to 25~\kms). The velocity ranges are with 
respect to the system velocity of $\sim$12.6~\kms.
The brightest and southern-most 5~cm emission feature ($\sim$4\arcsec\ south west of HOPS-108)
coincides well with the southern-most blue- and red-shifted clump of \twco\ emission. 
At medium and high-velocities the red-shifted \twco\ traces an elliptical 
feature with a position angle
from northeast to southwest. Northeast of HOPS-108, diffuse 5~cm emission ($\sim$2\arcsec\
northeast of HOPS-108) fills in some of the structure that is lower intensity 
in the \twco\ emission.
The contour levels in each panel start at 5$\sigma$ and increase on 3$\sigma$ intervals.
In the low-velocity, mid-velocity, and high-velocity panels, 
$\sigma_{low}$=0.19 (0.19) Jy~beam$^{-1}$, $\sigma_{mid}$=0.092 (0.12) Jy~beam$^{-1}$,
and $\sigma_{high}$=0.051 (0.089) Jy~beam$^{-1}$, respectively, with the red-shifted
level being given in parentheses.
The beams at 5~cm and $^{12}$CO are shown in the lower right as black and
green ellipses, respectively. The 5~cm beam is 0\farcs39$\times$0\farcs37 and the $^{12}$CO beam is 
0\farcs25$\times$0\farcs24. 
}
\label{12CO-H108}
\end{figure}

It is known that the powerful jet from HOPS-370 is directed toward the east
side of the FIR4 core (Figures \ref{H370-outflow} and \ref{12CO-H108}, and 
HOPS-108 in particular appears coincident with this jet (at least in projection). 
\citet{shimajiri2008} first suggested that the outflow from 
FIR3 was directly impinging on FIR4 and possibly triggering star formation there.
\citet{gonzalez2016} presented \textit{Herschel} [OI] maps that show the brightest
emission is located near HOPS-108, but there is also a clear jet seen in [OI] emission
extending from HOPS-370 to HOPS-108. \citet{favre2018} noted, however, that there 
was not definitive evidence for interaction in the gas temperatures of c-C$_3$H$_2$, but 
the resolution of these observations was relatively low, $\sim$9\arcsec$\times$6\arcsec\ 
(3600 AU $\times$ 2400 AU) and $\sim$5\arcsec$\times$3\arcsec, and the upper-level 
excitation of the highest energy transition observed was just 16~K.

The VLA 5~cm maps presented by \citet{osorio2017} show that the jet from HOPS-370 
has strong shocks that are producing centimeter-wave radio
emission with a spectral slope indicative of synchrotron emission. 
One knot has passed the position of HOPS-108 already (VLA12S) and is located $\sim$4\arcsec\
southwest. The other is located $\sim$2\arcsec\ northeast from HOPS-108 (VLA12C; see Figure \ref{12CO-H108}). 
The VLA12S is clearly interacting with molecular gas, given that we observe 
both blue and red-shifted \twco\ emission coincident with it, possibly reflecting a
terminal shock. Toward the knot located northeast of HOPS-108 (VLA12C), 
the diffuse 5~cm emission seems to be surrounded by $^{12}$CO emission.
Also, the knots show proper motion from northeast to 
southwest, and the observed $^{12}$CO morphology is consistent with the jet moving through
and interacting with this medium. 

\citet{ceccarelli2014} have suggested that there is 
a source of high-energy particles within FIR4 that is helping to drive the observed
chemistry, albeit under the assumption that all molecules within the beam are co-spatial. 
The shocks driven by the jet from HOPS-370 are
emitting synchrotron emission. \citet{padovani2016} and \citet{gaches2018} 
suggested that such jet shocks would be a natural source for 
high-energy particles, without the requirement 
for a particularly massive protostar within HOPS-108. However, the shocks in the 
jet may not be strong enough to drive the chemical abundance ratios found
by \citet{ceccarelli2014}, but accretion shocks $>$10$^{-6}$~\msun\ yr$^{-1}$ 
could \citep{gaches2018}. It is also important to note that the
molecular column densities in \citet{ceccarelli2014} were derived from low-angular
resolution \textit{Herschel} HIFI observations that include the entire core. Thus, it
is not clear if the molecules used to infer the need for high cosmic ray ionization 
are spatially coincident and physically associated. Furthermore, \citet{gaches2019} 
argued that the ratio of HCO$^{+}$ to N$_2$H$^+$ may not 
accurately reflect the cosmic ray ionization rate, which was the basis of the arguments
by \citet{ceccarelli2014}.

Despite the indications of interaction between the HOPS-370 jet and the molecular gas in 
FIR4, the observed interaction is not necessarily impacting HOPS-108. 
Indeed, the interaction could be happening in front of or behind HOPS-108 itself. 
If we consider that the HOPS-370 jet has a full opening angle of 2\degr, 
equivalent to the jet subtending 400 AU at the physical distance to HOPS-108 of 11000~AU
(the approximate size of the shocks from the HOPS-370 jet are $\sim$400 AU \citep{osorio2017}.)
Then the ratio of this angle to 180\degr\ corresponds
to the random probability of the HOPS-370 jet crossing HOPS-108 in 
projection. The probability
of the HOPS-370 jet overlapping HOPS-108 in projection by chance is only 0.011. 
A similar calculation
is possible for a direct interaction in three dimensions. Since we know that the
jet already crosses HOPS-108 we can reduce the dimensionality to two and only consider
the jet width and the depth of the cloud. If we assume that the cloud has
a depth equivalent to its projected size (22000 AU), then the probability
of a direct interaction is $\sim$0.02. Thus, it is possible, but
perhaps not likely that the HOPS-370 jet is directly impacting 
HOPS-108. \citet{osorio2017} suggested
that perhaps the jet impact triggered the formation of HOPS-108, 
similar to the scenario proposed by \citet{shimajiri2008}. A direct
impact by a jet or outflow generally tends to disperse material rather than collect
it \citep{arce2006,offner2014,tafalla2017}, but an oblique impact could lead to further 
gas compression. Given that the probability of the jet directly impacting HOPS-108 is low,
an oblique impact near HOPS-108 could be feasible.

\subsection{Origin and Implications of the Compact Methanol Emission}

HOPS-108 in FIR4 and HOPS-370 within FIR3
are the only sources
with compact, high-excitation methanol emission that we detect,
as well as compact emission in other molecules
and COMs, see Section 3.4 and Appendix A. This result
does not mean that other sources in FIR4/FIR3 do not also emit methanol, but they are below our 
sensitivity limit on $<$200~AU spatial scales. 
The methanol emission that we detect is very compact, centered
on the continuum sources of HOPS-108 and HOPS-370 (Figures \ref{methanol}). 
The methanol lines that
we detect have E$_{up}$~$\ge$~115~K
and rotation temperatures of 140~K and 129~K
for HOPS-108 and HOPs-370, respectively (see Appendix A, Table 5).
Therefore, warm conditions are required to excite these transitions. 
This points to a source of moderate to high luminosity both to 
evaporate methanol out to several tens of AU from the protostar and to 
excite these particular transitions, unless it is heated by interaction
with its own outflow (as opposed to its luminosity from the protostar and
accretion) \citep[e.g.,][]{lee2018}. The methanol emission, however,
is known to extend out to large radii in lower-excitation lines, 
encompassing much of the core \citep{lopez-sepulcre2013}. Thus, we are very 
much detecting the `tip of the iceberg' in our high-resolution observations. 
Detections
of high excitation methanol emission centered on HOPS-108
and HOPS-370, while not elsewhere in the core indicate that 
presence of gas-phase methanol is the direct result of the internal 
heating from the protostars. The more extended
methanol emission could be due to the ambient heating in the cluster environment.
Indeed, there are several cases of extended methanol emission from pre-stellar cores in the absence
of a direct internal heating source \citep{jimenezserra2016, bacmann2012}.

The presence of these high-excitation methanol lines, in addition to 
the other transitions detailed in Appendix A, are all typical tracers of hot molecular
cores \citep[e.g.,][]{schilke1997, hatchell1998}, usually associated 
with high-mass protostars. These hot cores, however, typically have
luminosities of $\sim$10$^3$-10$^4$~\lsun;  HOPS-108 has a luminosity that is
constrained from its SED to be $\leq$100~\lsun, 
and HOPS-370 has a luminosity of 360~\lsun.

The presence of compact COM emission, coupled with 
the relatively low-luminosities of HOPS-108 and HOPS-370 (compared to hot cores)
are consistent with hot corinos, lower-luminosity
protostars that have rich molecular spectra, similar to hot cores. Some
examples of hot corinos are NGC 1333 IRAS2A, NGC 1333 IRAS 4A2, L483, HH212 MMS and IRAS 16293-2422 \citep{taquet2015, 
ceccarelli2004,jacobsen2018,drozdovskaya2016,lee2018}.

HOPS-370 is clearly the most luminous protostar within FIR3 and has
strong continuum emission in the submillimeter and centimeter.
On the other hand, HOPS-108 has
compact and not particularly strong continuum emission at high-resolution
(and submillimeter/centimeter wavelengths), but HOPS-108 appears to harbor the most luminous
protostar within FIR4. It is
consistent with having a luminosity that is at least high enough to
evaporate methanol off dust grains in its immediate vicinity and excite the observed
high-excitation transitions. The radius of the methanol emitting region around
HOPS-108 from the HWHM of methanol integrated intensity maps is $\sim$50 AU (0\farcs125).
Assuming that methanol has an evaporation temperature of 120~K \citep{collings2004}, the 
luminosity required to heat dust to this temperature at a radius of 50~AU is 
$\sim$86~\lsun, calculated assuming thermal equilibrium.
This is consistent with the range of luminosities favored by \citet{furlan2014}.
Ice mixtures, however, can increase the
evaporation temperature to $\sim$160~K, which would then require a luminosity
of $\sim$270~\lsun; higher than the most likely luminosity range defined by \citet{furlan2014}. 
However, the luminosity of protostars is known to be variable
\citep{fischer2019, safron2015,hartmann1996}, and outbursts from low-mass stars
have been shown to release complex organics out to relatively large radii \citep{vanthoff2018}.
The release of molecules from the ice happens nearly instantaneously \citep{collings2004}.
Then, when the outburst fades, the molecules can take
100-10000 yr to freeze-out again (depending on the density), 
leaving an imprint of outburst in the chemical richness 
of submillimeter and millimeter spectra \citep{jorgensen2015,visser2015,frimann2016}.
Hence, the inconsistency in the luminosities inferred from the SED and the 
evaporation temperature may be explained by such luminosity variations. 

\subsection{The Luminosity and Ultimate Mass of HOPS-108}

Based on the analysis from Furlan et al. (2014), our high-resolution
continuum maps, and the compact methanol emission, it is clear that
HOPS-108 is
the most luminous protostar within FIR4. Several studies of the
near-to-far-infrared observations \citep{adams2012, furlan2014} 
used SED modeling to determine that the internal luminosity of HOPS-108
was between 37~\lsun\ and 100~\lsun. Much of the ambiguity in the luminosity
results from the emission being blended in the mid-to-far-infrared and the heating from multiple
protostars illuminating the clump at wavelengths longer than 70~\micron, in addition to
the unknown inclination of the source.
The constraints on the luminosity of the protostar both from the SED and the
extent of the compact methanol emission, taken with the lack of a clear and powerful
outflow, suggest that HOPS-108 is not currently
a high-mass protostar but more likely a low to intermediate-mass protostar. Indeed, much of
the luminosity from HOPS-108 could result from accretion luminosity, and the observed
radius of the COMs could reflect past luminosity bursts of the protostar 
and possibly not the current luminosity. However, assuming that the luminosity
necessary to liberate the COMs out to the observed radii was the luminosity during
a burst, we can calculate the estimated accretion rate necessary using 
protostellar structure models \citep{hartmann1997,palla1993}.

Accretion luminosity from gas in free-fall onto the protostellar surface can be estimated
from the equation L$_{acc}$~$\simeq$~GM$_{ps}$$\dot{M}$/R$_{ps}$, where G is the gravitation
constant, M$_{ps}$ is the protostellar mass,
$\dot{M}$ is the accretion rate on to the protostar,
and R$_{ps}$ is the protostellar radius.
We first adopt the case of a 1~\msun\ protostar. \citet{hartmann1997} find that
the radius of the protostar at a given mass depends on its accretion rate. 
A 1~\msun\ protostar that has been accreting at 2.0$\times$10$^{-6}$~\msun~yr$^{-1}$ 
would have a radius of $\sim$2.1~R$_{\odot}$ and a protostar accreting at 
$\sim$1$\times$10$^{-5}$~\msun~yr$^{-1}$ would have a radius of $\sim$4.5~R$_{\odot}$. 
For these protostellar radii, the 
luminosity from the protostellar photosphere is expected to be $\sim$3~\lsun\ 
and $\sim$10~\lsun, respectively. With the above stellar radii and a 
mass of 1~\msun, accretion rates of $\sim$1.8$\times$10$^{-5}$~\msun~yr$^{-1}$ 
to $\sim$3.7$\times$10$^{-5}$~\msun~yr$^{-1}$ are necessary to reach 
a total luminosity of 270~\lsun. 

If the protostar mass is currently 2~\msun, the stellar radius is
expected to be $\sim$4.5~R$_{\odot}$ \citep{palla1993}
and the luminosity from the protostellar photosphere would be $\sim$10~\lsun. 
These stellar parameters also require $\sim$1.9$\times$10$^{-5}$~\msun~yr$^{-1}$ 
to reach a total luminosity of 270~\lsun. 
These inferred mass accretion rates
could also explain the luminosity of HOPS-370 if scaled upward by a factor of 1.33.

It is difficult to estimate the mass of a protostar from its current luminosity, meaning
that both HOPS-108 and HOPS-370 could be low-to-intermediate-mass protostars,
with their current luminosities
is set by their accretion rates. Once a protostar is much more massive than $\sim$3~\msun, its
luminosity becomes dominated by the stellar photosphere rather than accretion \citep{palla1993,
offner2011}. The inferred accretion rates are sufficiently high to produce significant
cosmic-ray ionization, as predicted by \citet{padovani2016} and \citet{gaches2018}, even if
the current luminosity of HOPS-108
is as low as 37~\lsun. Thus, it remains possible that
the protostellar accretion, even though not from a high-mass star, could be driving the
chemistry through local production of energetic particles or photons 
as suggested by \citet{ceccarelli2014}.

The FIR4 core has $\sim$30~\msun\ surrounding HOPS-108, while the 
FIR3 core is substantially less massive at $\sim$17~\msun\ \citep{nutter2007}.
Therefore, both HOPS-108 and HOPS-370
could potentially grow into at least an intermediate-mass star 
given its apparent central location, at least in projection.
With the inferred accretion rate of HOPS-108
required to generate a total luminosity of 272~\lsun\, 
it would take $\sim$1 Myr to accrete all this mass \citep[assuming
a 33\% star formation efficiency;][]{offner2017,offner2014,machida2013}. 
A timescale of 1~Myr is
very long relative to the estimated length of the
protostellar phase \citep{dunham2014}. The current low accretion rate (long accretion time) and its
lack of strong mid-infrared emission could indicate that HOPS-108 is in an `IR-quiet' phase
of high-mass star formation \citep[e.g.,][]{motte2017}, a short-lived phase 
prior to becoming extremely luminous with a high accretion rate.
Furthermore, the other embedded protostars in the region (VLA15 and VLA16) could also
gain enough mass via accretion to become intermediate-mass stars.

\subsection{Remaining Questions}
There remain several inconsistencies between our results and other observations
of HOPS-108. For example, no clear outflow has been detected from HOPS-108 itself in
$^{12}$CO molecular line emission, and the free-free continuum source
associated with HOPS-108 is very weak in comparison with HOPS-370. There is a known correlation between
\lbol\ and free-free continuum emission \citep{anglada1995,shirley2007,tychoniec2018,anglada2018}.
If HOPS-108 is consistent with the correlation derived by \citet{tychoniec2018}, then
a 100~\lsun\ source is expected to have a 4.1~cm flux density of $\sim$1.5~mJy
at the distance to Orion. HOPS-108 is $\sim$30$\times$ fainter at
4.1~cm than expected from its measured bolometric luminosity \citep{osorio2017}. Thus, 
if outflow activity is correlated with accretion, the lack of such activity from HOPS-108 
may be at odds with the high accretion rate needed to explain 
its high-luminosity. There
is significant scatter in the correlation between \lbol\ and free-free continuum
and the low 4.1~cm flux density does not rule-out HOPS-108 having a luminosity of $\sim$100~\lsun.

The lack of an obvious outflow from HOPS-108
has implications for the interpretation of its far-infrared CO emission.
HOPS-108 is among the strongest far-infrared CO emitters, significantly above the relationship
found by \citet{manoj2016}. Thus, HOPS-108 may not actually be responsible
for generating the CO emission and instead it is dominated 
by the terminal shock from the nearby HOPS-370 outflow as suggested by
\citet{gonzalez2017}. This scenario would make HOPS-108 much more consistent with the 
\lbol\ vs. L$_{\rm CO}$ relationship derived for the majority of protostars \citep{manoj2016}.

There are also alternative explanations for the rich molecular line spectrum
observed toward HOPS-108. Shock-heating could explain their presence toward 
HOPS-108 and enable it to have a low-luminosity. For example, HH212 MMS is
found to be exhibiting COM emission from the surface of its disk, presumably
from mechanical heating by the outflow \citep{lee2018}. Furthermore, the kinematics
of the higher-excitation methanol transitions have different velocity gradient
directions with respect to the lowest excitation transition. Thus, 
we cannot rule-out that some COM emission could result from shock 
heating by a nascent outflow that is
not obvious in \twco. It is very unlikely that the COM emission results from the
HOPS-370 jet directly impacting HOPS-108 on a 100 AU scale where 
the COMs are detected. If that were the case, we would expect the
COM emission to be more extended and 
associated with the outflow
knows observed (VLA12S and VLA12C). Instead,
the observed emission is 
concentrated on the compact continuum of
HOPS-108. Although it is difficult to rule out all mechanical or shock-heating, 
the COM emission generated as a result of thermal evaporation from the luminosity
of HOPS-108 is the simplest explanation.

\section{Conclusions}

We have used ALMA and the VLA, in conjunction with previous near to far-infrared,
single-dish submillimeter data, and interferometric mapping at millimeter
wavelengths to identify and characterize the protostellar 
content of OMC2-FIR3 and FIR4. Furthermore, 
serendipitous detections of compact methanol emission
toward HOPS-108 and HOPS-370
enable us to better characterize the nature of the protostellar sources.
Our main results are as follows.

\begin{itemize}

\item We detect six distinct continuum sources at 0.87~mm and 9~mm 
that are spatially coincident with the OMC2-FIR4 core: HOPS-108, 
 VLA16, HOPS-64, VLA15, ALMA1, 
and HOPS-369. HOPS-108 is the most centrally
located object in OMC2-FIR4 and is deeply embedded. HOPS-108 is marginally
resolved at 0.87~mm, but it does not show significant structure 
at the observed angular resolution. HOPS-108 has faint
9~mm emission, fainter than expected for a protostar with a luminosity of potentially 100~\lsun.
HOPS-64 is also coincident with the FIR4 core, 
but is more evolved and likely viewed in projection
in the foreground given its detectability at optical/near-IR wavelengths. 
VLA15 also appears to have an edge-on disk, given
its continuum morphology at 0.87~mm and 9~mm.

\item We detect for continuum sources associated with
OMC2-FIR3. HOPS-370 is at the position of FIR3 and accounts for the bulk of the luminosity
from the region, and we also detect a binary system, HOPS-66-A and HOPS-66-B, separated
by 2\farcs23 ($\sim$892~AU). HOPS-370 is also an apparent binary with $\sim$3\arcsec separation, 
but its companion is only detected at wavelengths shorter than 24~\micron.
A more-evolved source MGM-2297 is also detected at both wavelengths further south from HOPS-370.

\item We detect compact methanol emission from three transitions
toward HOPS-108 and HOPS-370, in addition to
emission from other molecules.
This indicates that HOPS-108 and HOPS-370 could be hot corinos.
The molecular line emission originates 
from $\sim$100~AU scales 
coincident with the HOPS-108 and HOPS-370
continuum sources. This is consistent with 
the protostars
generating at least 
enough luminosity to desorb a significant
amount of methanol out to $\sim$50~AU radii.
The only efficient route to forming methanol, however, is within ices, so 
the observed methanol emission
must result from ice evaporation. We argue that thermal evaporation due to
the luminosity of HOPS-108 is the simplest explanation for the methanol emission, but we
cannot rule-out shock heating from a nascent outflow from HOPS-108.
The methanol emission
in HOPS-370 has a clear velocity gradient along the major axis of the disk, likely tracing rotation.

\item We detect spatially and kinematically complex \twco\ emission in the vicinity 
of HOPS-108 and do not positively detect an outflow from HOPS-108.
We do, however, tentatively detect a candidate outflow at low-velocities that 
is in a similar direction to the two higher excitation methanol emission lines.
The \twco\ emission also appears
to trace the interaction of the outflow/jet from nearby HOPS-370 (OMC2-FIR3) within the region
surrounding HOPS-108. VLA 5~cm emission is coincident with structures observed
in \twco\ and the proper motion of the northern 5~cm feature is inconsistent with
it coming from HOPS-108. 
\end{itemize}

We conclude
that HOPS-108 is the most luminous protostar within OMC2-FIR4. It is likely a low to
intermediate-mass protostar but could potentially grow into a high-mass 
star with continued accretion. Higher resolution and sensitivity mapping from the far-infrared
to millimeter wavelengths in both continuum and molecular lines will shed further
light on the nature of the protostellar sources within OMC2-FIR4 and their 
relationship to the OMC2-FIR4 core.
\newline

We thank the anonymous referee for useful feedback that improved the quality
of the manuscript and acknowledge useful discussions of this work with F. van der Tak, P. Schilke,
and H. Beuther. We are grateful for the support from L. Maud at the Dutch Allegro ALMA Regional Center Node
for his efforts in reducing the data.
JJT is acknowledges support from the Homer L. Dodge Endowed Chair,
 grant 639.041.439 from the Netherlands Organisation for Scientific Research (NWO),
and from the National Science Foundation AST-1814762.
ZYL is supported in part by NASA 80NSSC18K1095 and NSF AST-1716259.
SO acknowledges support from NSF AAG grant AST-1510021. 
GA, MO, and AKD-R acknowledge financial support from the State Agency for Research of the
Spanish MCIU through the
AYA2017-84390-C2-1-R grant (co-funded by FEDER) and through the ``Center of Excellence
Severo Ochoa'' award for the Instituto de Astrof\'{\i}sica de Andaluc\'{\i}a
(SEV-2017-0709). 
AS greatfully acknowledges funding through Fondecyt regular (project code
1180350) and Chilean Centro de Excelencia en Astrof\'isica y
Tecnolog\'ias Afines (CATA) BASAL grant AFB-170002.
Astrochemistry in Leiden is supported by the Netherlands Research School for Astronomy (NOVA), 
by a Royal Netherlands Academy of Arts and Sciences (KNAW) professor prize, and by the European Union A-ERC grant 291141 CHEMPLAN. 
This paper makes use of the following ALMA data: ADS/JAO.ALMA\#2015.1.00041.S.
ALMA is a partnership of ESO (representing its member states), NSF (USA) and 
NINS (Japan), together with NRC (Canada), NSC and ASIAA (Taiwan), and 
KASI (Republic of Korea), in cooperation with the Republic of Chile. 
The Joint ALMA Observatory is operated by ESO, AUI/NRAO and NAOJ.
The National Radio Astronomy 
Observatory is a facility of the National Science Foundation 
operated under cooperative agreement by Associated Universities, Inc.
This research made use of APLpy, an open-source plotting package for Python 
hosted at http://aplpy.github.com. This research made use of Astropy, 
a community-developed core Python package for 
Astronomy (Astropy Collaboration, 2013) http://www.astropy.org.

 \facility{ALMA, VLA, Spitzer, Herschel, Mayall}
\software{Astropy \citep[http://www.astropy.org; ][]{astropy2013,astropy2018}, APLpy \citep[http://aplpy.github.com; ][]{aplpy}, scipy \citep[http://www.scipy.org; ][]{scipy}, CASA \citep[http://casa.nrao.edu; ][]{mcmullin2007}}

\appendix

\section{Molecular Line Emission}
\begin{figure}
\begin{center}
\includegraphics[scale=0.45,trim=0.0cm 4cm 0.0cm 0.0cm,clip=true]{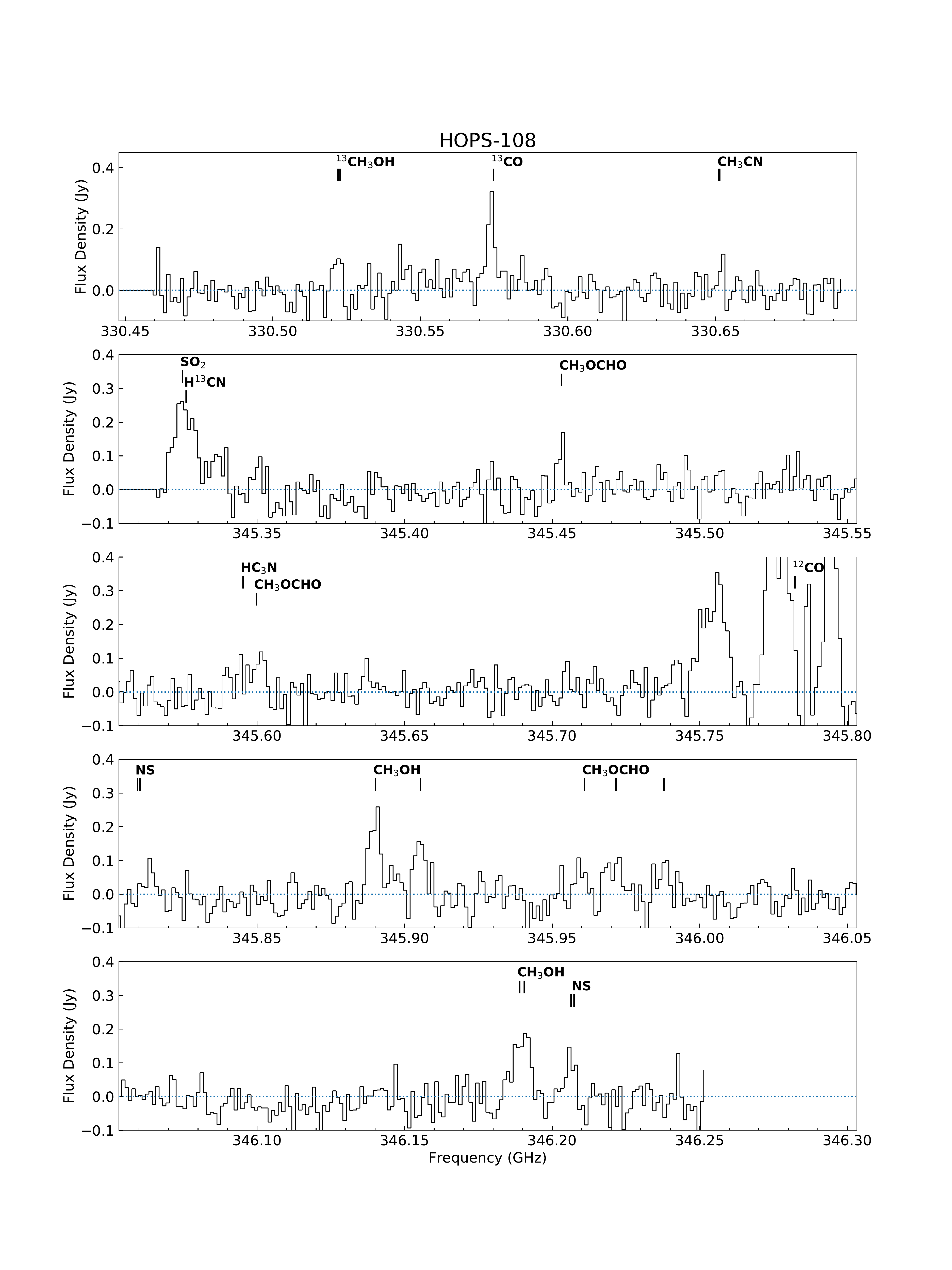}

\end{center}
\caption{Spectra of HOPS-108 centered at 330.575 GHz (top) and 345.8~GHz 
(lower 4 panels), showing the presence of various molecules
in emission toward the compact continuum source. The spectra were extracted 
from a 0\farcs5 diameter circle centered on the continuum source. The major 
identified features are labeled, and the horizontal dashed
line marks the zero flux level in the spectra. Note that the structure around
the $^{12}$CO ($J=3\rightarrow2$) transition (345.735 GHz to 345.80 GHz) 
is complex due to spatial filtering and
not the result of additional molecular features.
}
\label{spectrum-H108}
\end{figure}

\begin{figure}
\begin{center}
\includegraphics[scale=0.45,trim=0.0cm 4cm 0.0cm 0.0cm,clip=true]{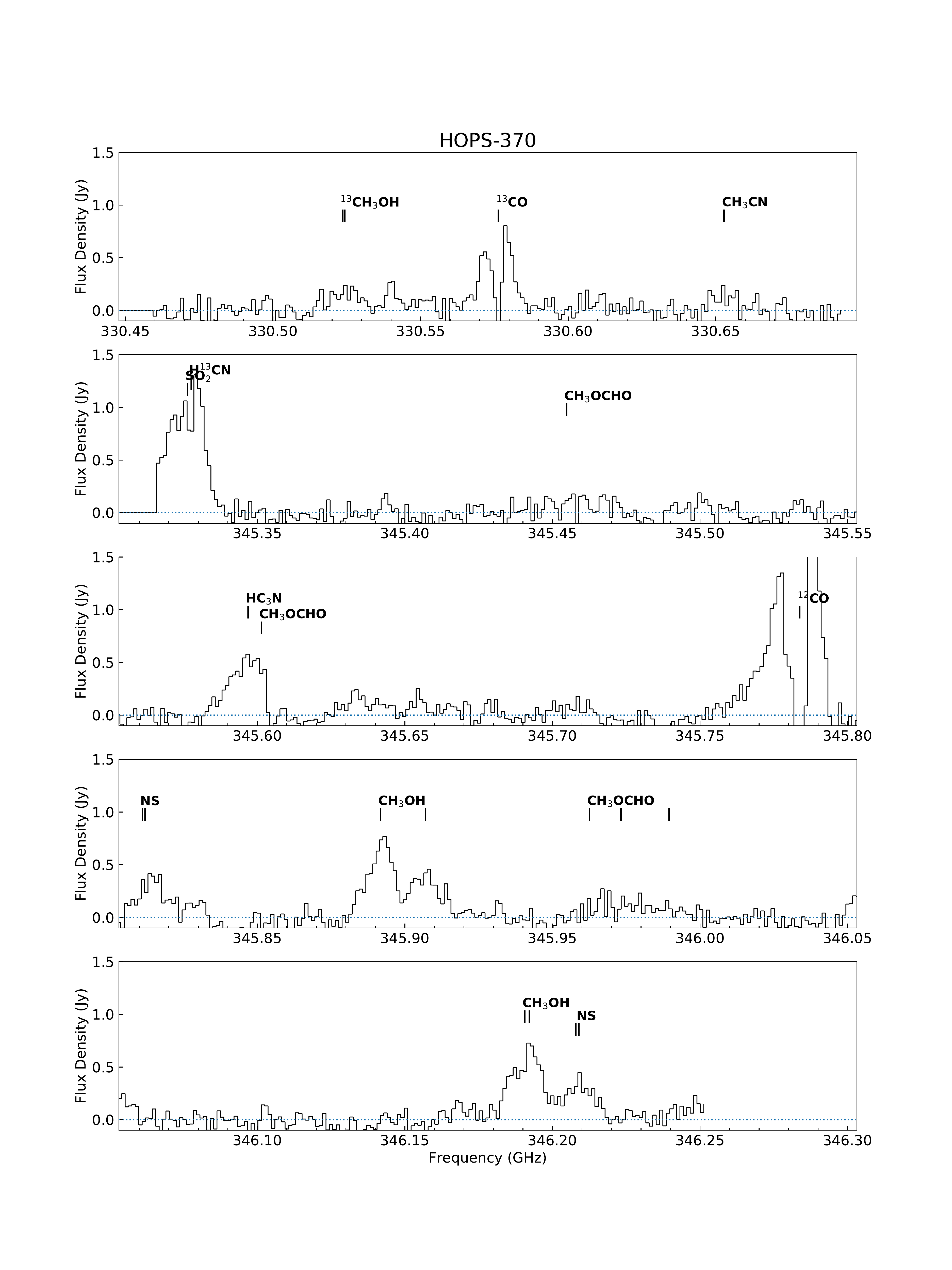}

\end{center}
\caption{Same as Figure \ref{spectrum-H108}, but for toward HOPS-370}.
\label{spectrum-H370}
\end{figure}

We detected numerous molecular lines associated with HOPS-108 and HOPS-370 in our ALMA 
data (Table 5). We observed
two bands with high spectral resolution centered on the \twco\ and \thco\ ($J=3\rightarrow2$)
transitions. In addition to these two targeted lines, a number of other
molecular lines are present within the \twco\ and \thco\ bands (Figure \ref{spectrum-H108} and \ref{spectrum-H370}).
Several of these lines originate from 
complex organic molecules \citep[COMs, molecules containing carbon and a 
total of 6 or more atoms][]{herbst2009}.

We examined the spectra toward all continuum sources in the FIR4 region, and HOPS-108 is the only
one to exhibit emission from COMs and molecules other than \twco. The spectrum of 
HOPS-108, centered on \twco\ and \thco, is shown in Figure \ref{spectrum-H108}.
We detect several molecules that are typically detected in hot cores or hot corinos, such
as methanol (CH$_3$OH), methyl formate (CH$_3$OCHO), and NS (nitrogen sulfide) 
\citep[e.g.,][]{schilke1997,hatchell1998}, as well as a strong H$^{13}$CN/SO$_2$ line toward
HOPS-108. There are also tentative detections
of HC$_3$N blended with another methyl formate line, as well as possible
detections of emission from $^{13}$CH$_3$OH, and CH$_3$CN. Details of the detected
molecular transitions are provided in Table 5. The velocity center and linewidth
of each line is fit with a Gaussian function using the \textit{curve\_fit} function
of \textit{scipy}. The system velocity of HOPS-108 is found to be 12.6~\kms, which is 
red-shifted by about $\sim$1-2~\kms\ with respect to molecules observed on
larger scales by \citet{lopez-sepulcre2013}. The average line width from Table 5 (using unblended
lines and a single value from the blended lines) is $\sim$1.5~\kms.

We also examined the spectra of protostars in the FIR3 region and toward HOPS-370 we detect
methanol, NS, H$^{13}$CN/SO$_2$, and HC$_3$N blended with methyl formate and we show the spectrum
in figure \ref{spectrum-H370} and list the line properties in Table 5.
The detected lines toward HOPS-370 have higher flux densities and 
larger linewidths compared to HOPS-108; these features are evident from a comparison of
the spectra Figures \ref{spectrum-H108} and \ref{spectrum-H370}. The system velocity of HOPS-370 is $\sim$11.2~\kms, with an
average linewidth of $\sim$4.8~\kms. 
The larger linewidth may explain the lack of a clear detections for 
the three methyl formate lines between 345.95 and 346.0 GHz. We also tentatively detect an
additional NS feature at 345.81 GHz, and this feature may also be present in HOPS-108. This feature
is contaminated by the high-velocity wings of the HOPS-370 outflow, so the line flux density
is uncertain. Also, there is a methyl formate transition at a similar frequency that could
also potentially contaminate the NS emission.

The non-detections of emission from molecules other than \twco\ toward other 
sources could in part be due 
to primary beam attenuation. HOPS-64 is situated at the $\sim$72\% power point in our
data, making it unlikely 
that the non-detection toward HOPS-64 is due to the 
primary beam attenuation if the emission were comparable in strength to HOPS-108.
VLA15, however, is at the $\sim$32\% power point making detections much 
more difficult. 
The typical peak line flux densities are 0.25 toward HOPS-108 and dividing this
by a factor of 3 would result in a peak line flux density of ~0.08 which would be difficult
to distinguish from noise. Hence, we would not expect to detect emission from VLA15
even if it was at the same level as HOPS-108.

\begin{figure}
\begin{center}
\includegraphics[scale=0.45]{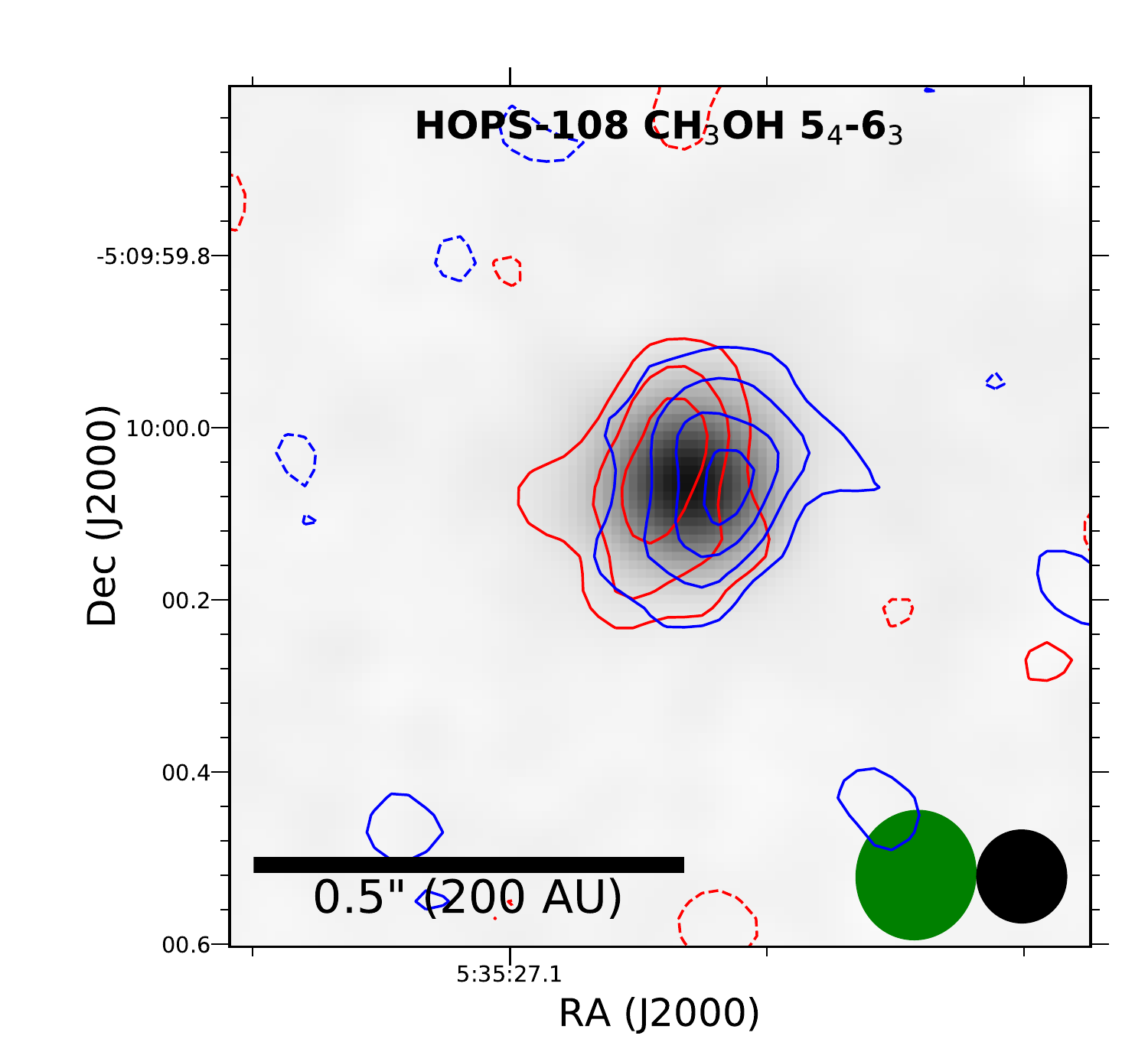}
\includegraphics[scale=0.45]{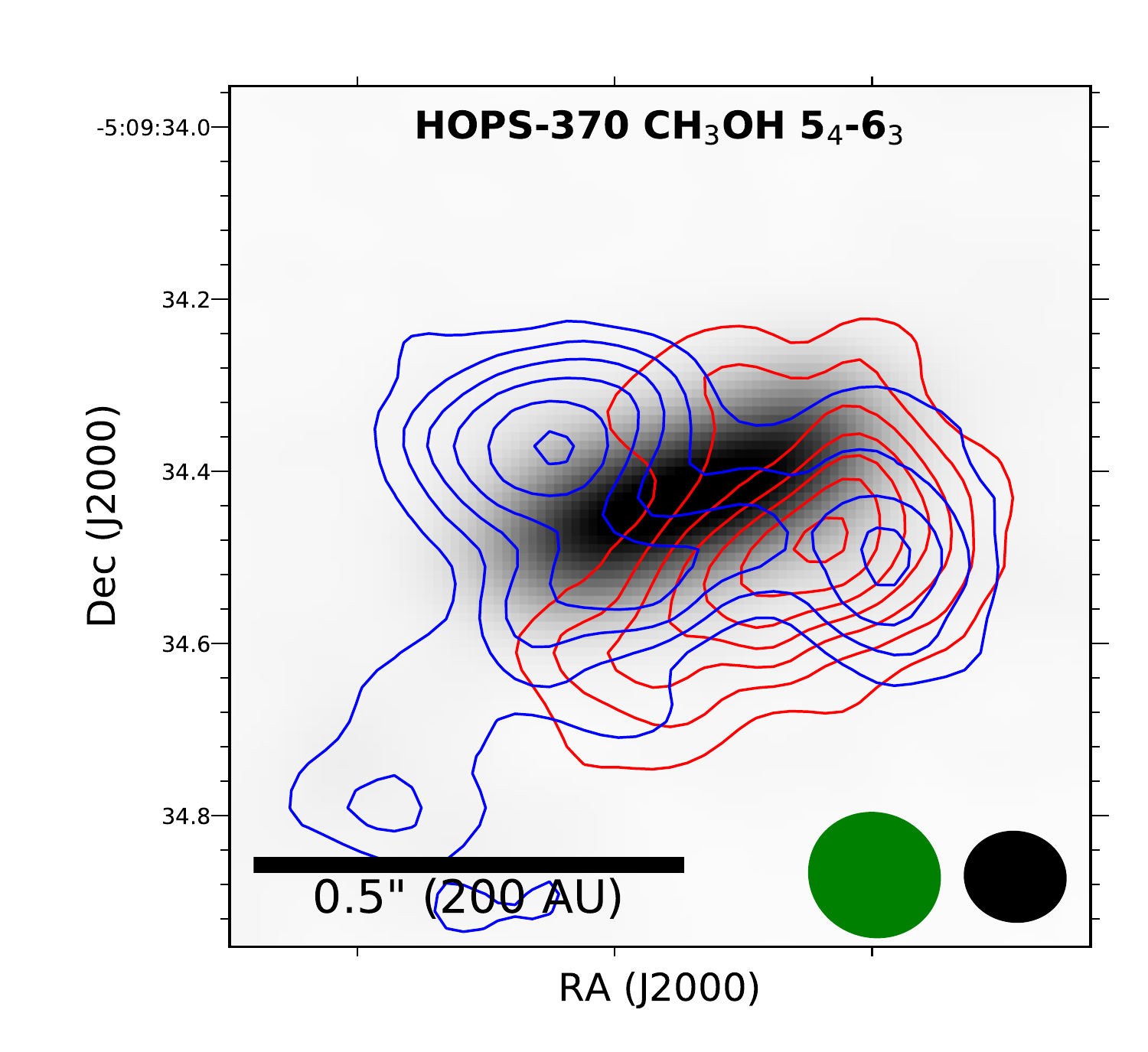}
\includegraphics[scale=0.45]{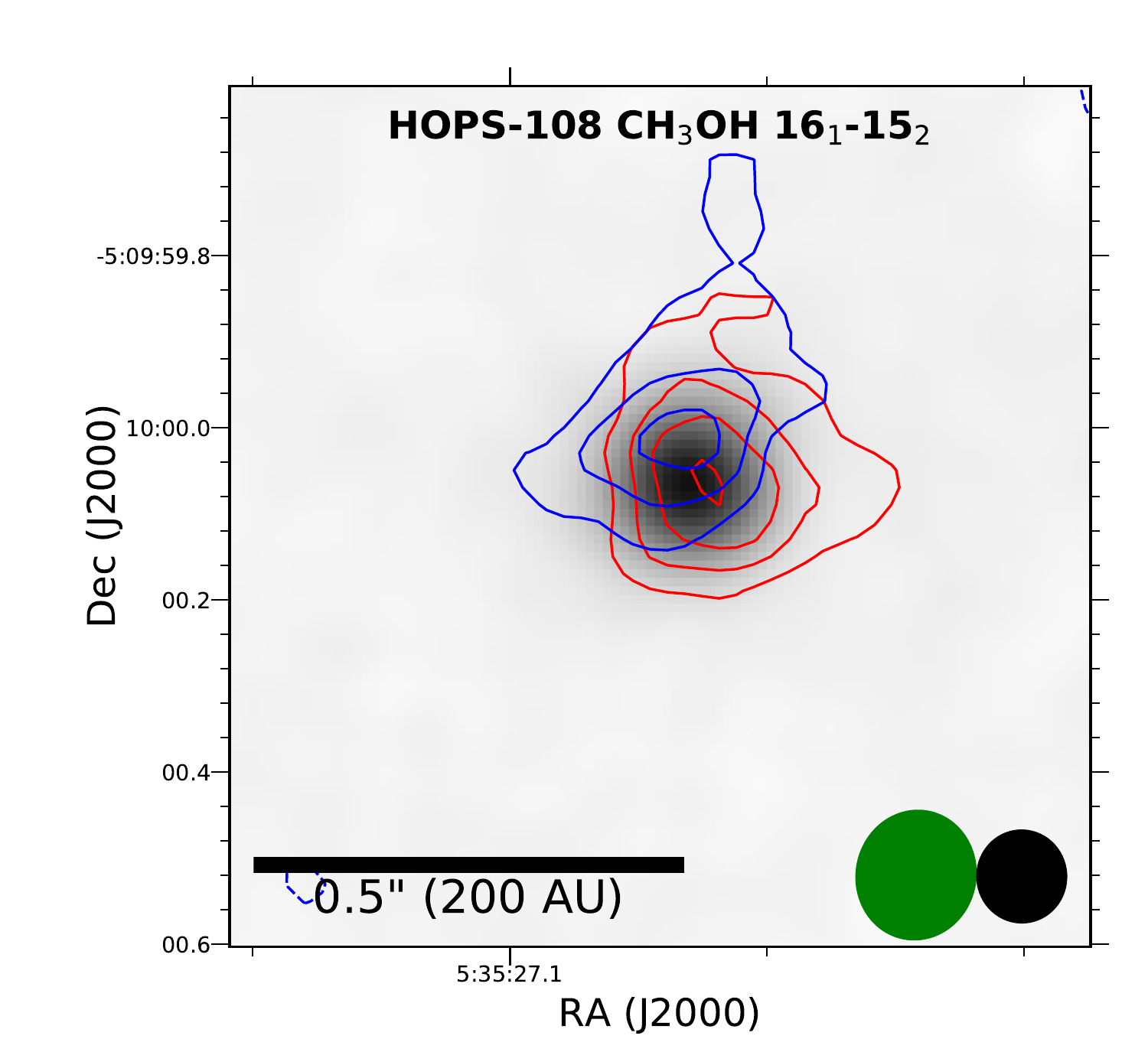}
\includegraphics[scale=0.45]{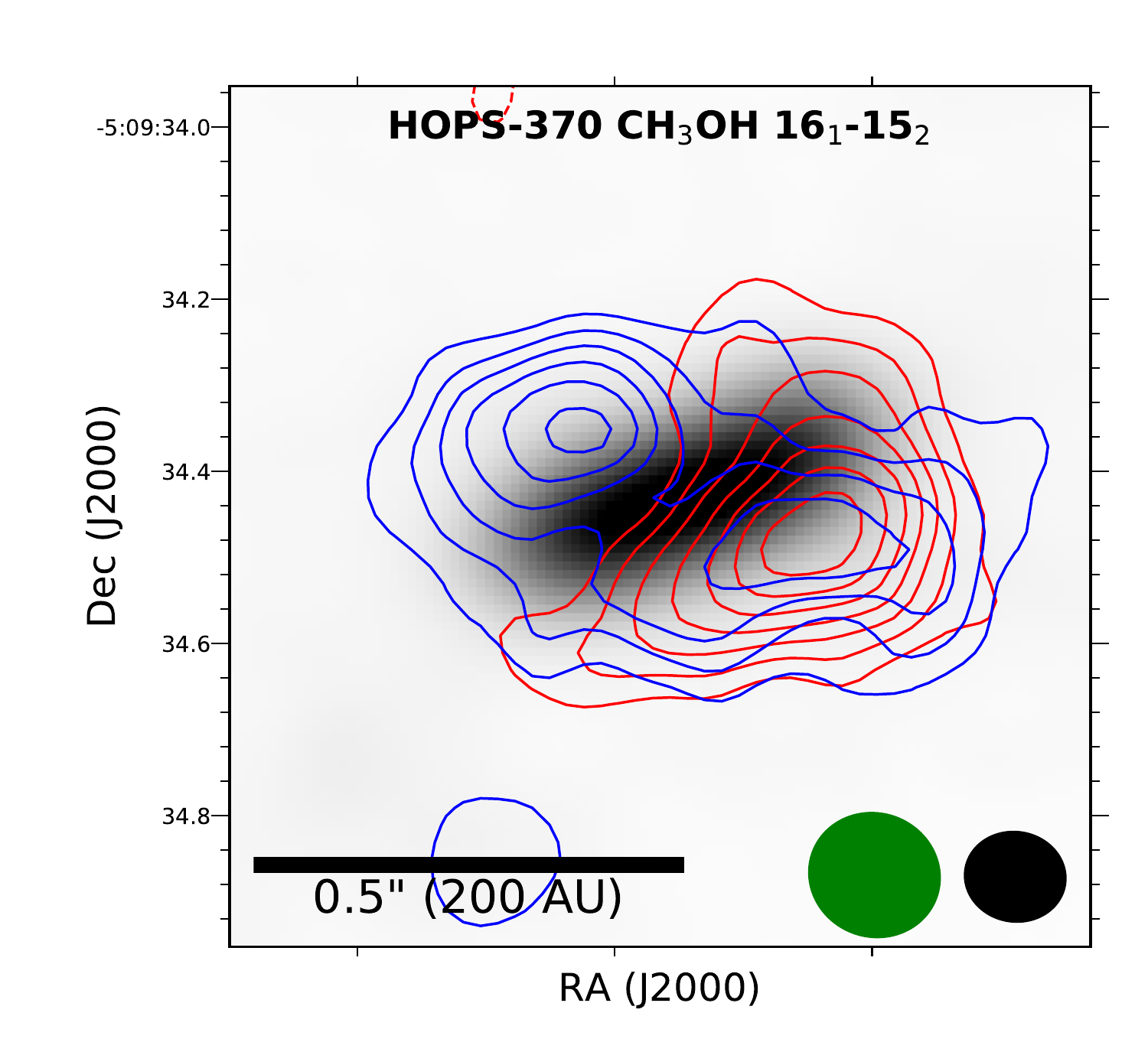}
\includegraphics[scale=0.45]{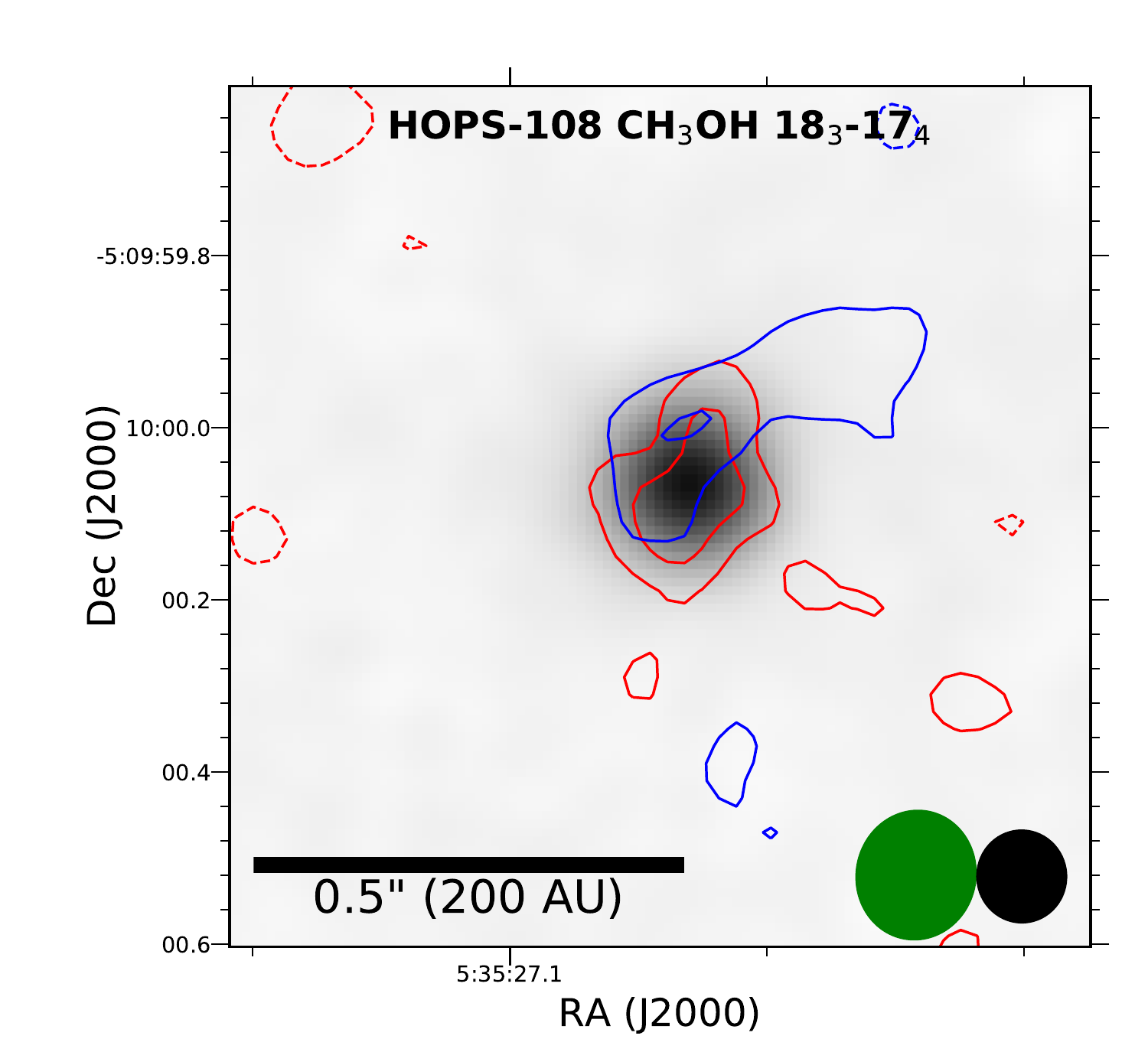}
\includegraphics[scale=0.45]{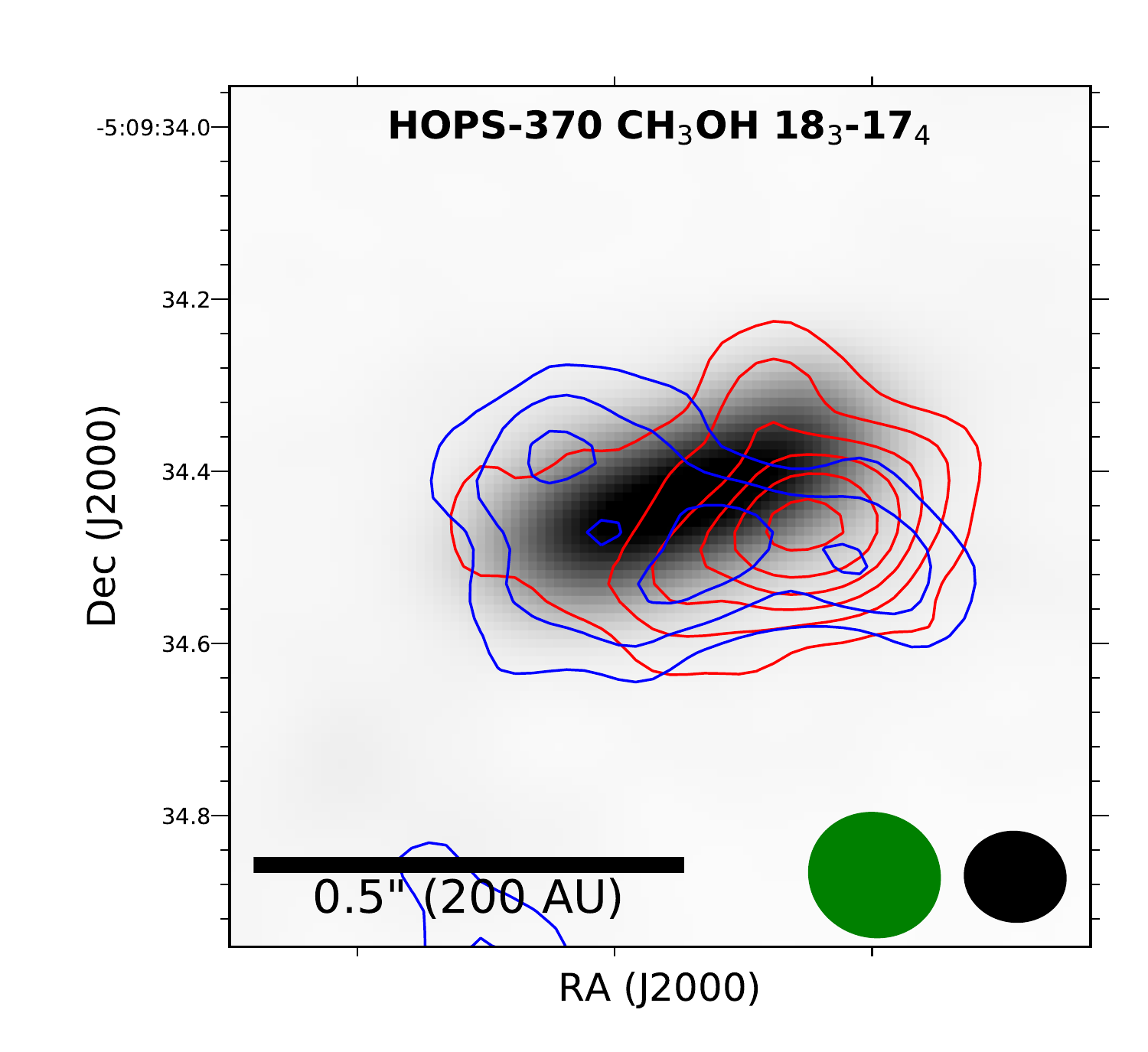}
\end{center}
\caption{Integrated intensity maps of 
methanol emission toward HOPS-108 
(left) and HOPS-370 (right)
overlaid on the 0.87~mm continuum (grayscale).
The CH$_3$OH ($J=5_4\rightarrow6_3, J=16_1\rightarrow15_2, J=18_3\rightarrow17_4$)
are shown in the top, middle, and bottom panels, respectively.
The integrated intensity maps 
of CH$_3$OH are separated into blue- and red-shifted velocities and
plotted with blue and red contours, respectively.
The contours start at 3$\sigma$ and increase on 2$\sigma$ intervals.
 See Appendix A for more details
of the particular molecular transitions shown.
These molecular lines are indicative of
a compact, warm object associated with the continuum. The velocity gradient
of CH$_3$OH changes from the lowest energy transition to two higher transitions for HOPS-108,
perhaps suggesting the presence of rotation and outflow motion 
HOPS-370 in contrast is consistent with rotational motion across all the transitions.
The beams of the
continuum and molecular line data are shown in the lower right as black
and green ellipses, respectively. The continuum beam is 0\farcs11$\times$0\farcs10 
and the molecular line beams are $\sim$0\farcs15$\times$0\farcs14.
}
\label{methanol}
\end{figure}

\begin{figure}
\begin{center}
\includegraphics[scale=0.45]{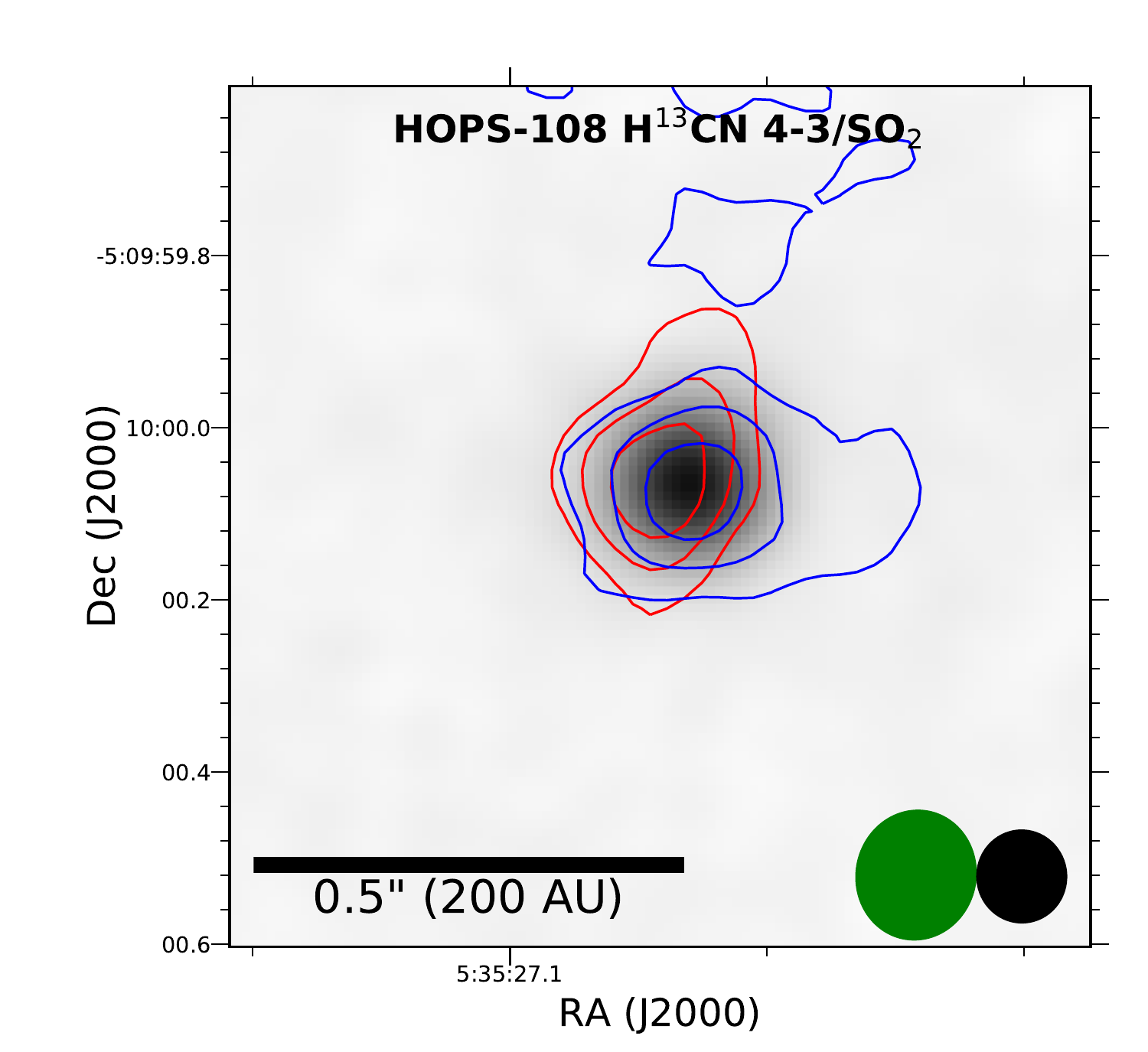}
\includegraphics[scale=0.45]{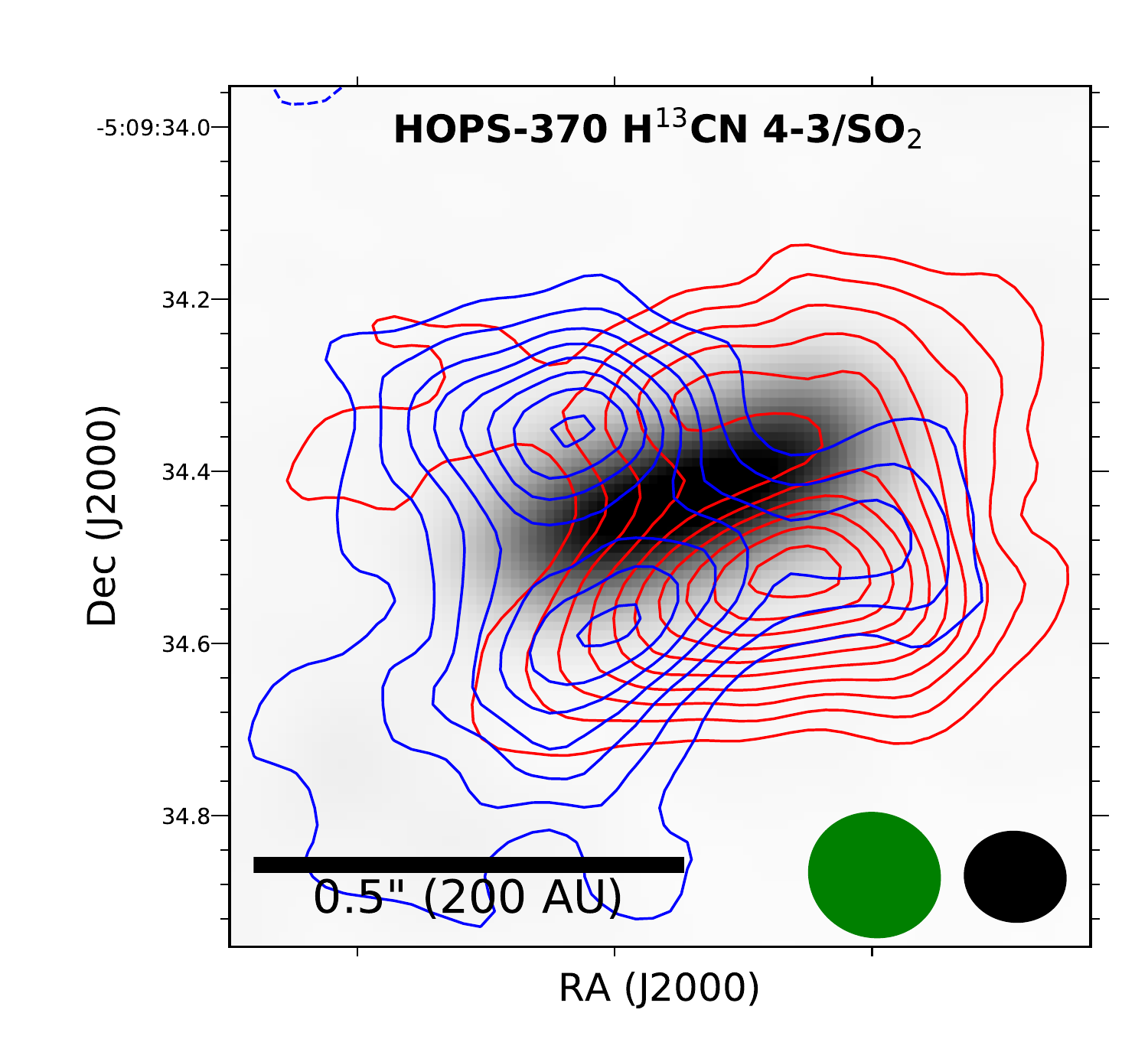}
\includegraphics[scale=0.45]{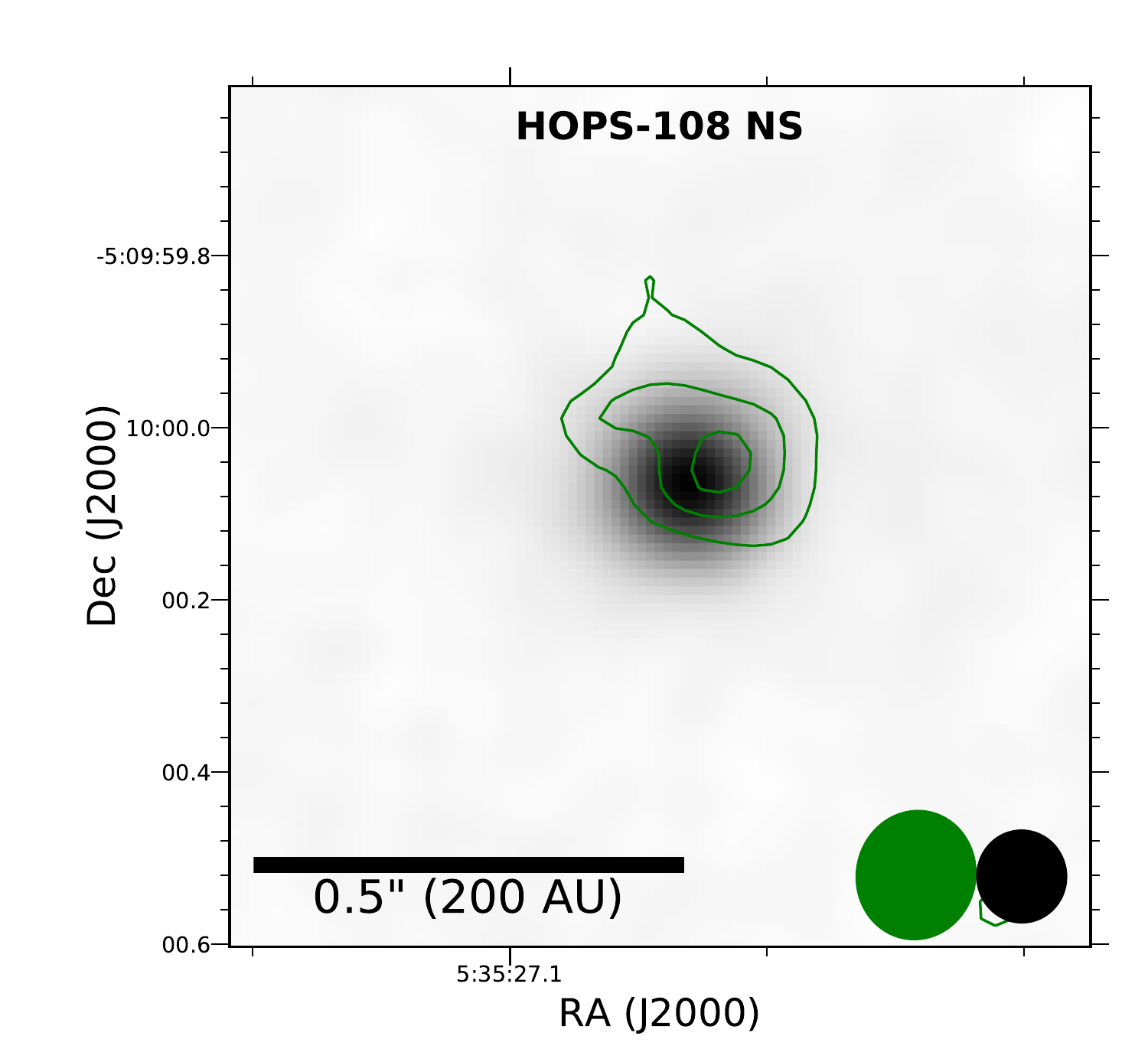}
\includegraphics[scale=0.45]{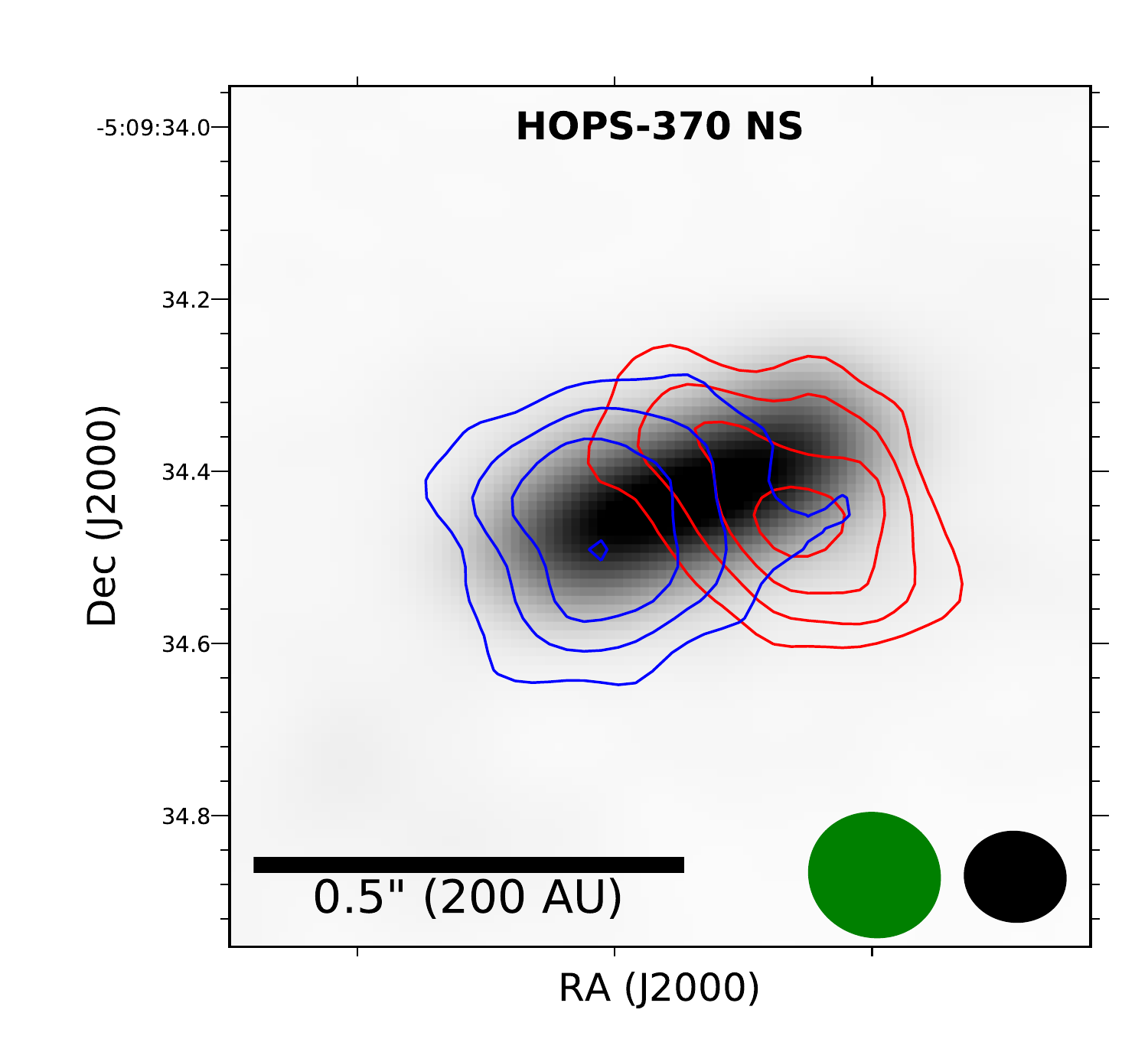}
\includegraphics[scale=0.45]{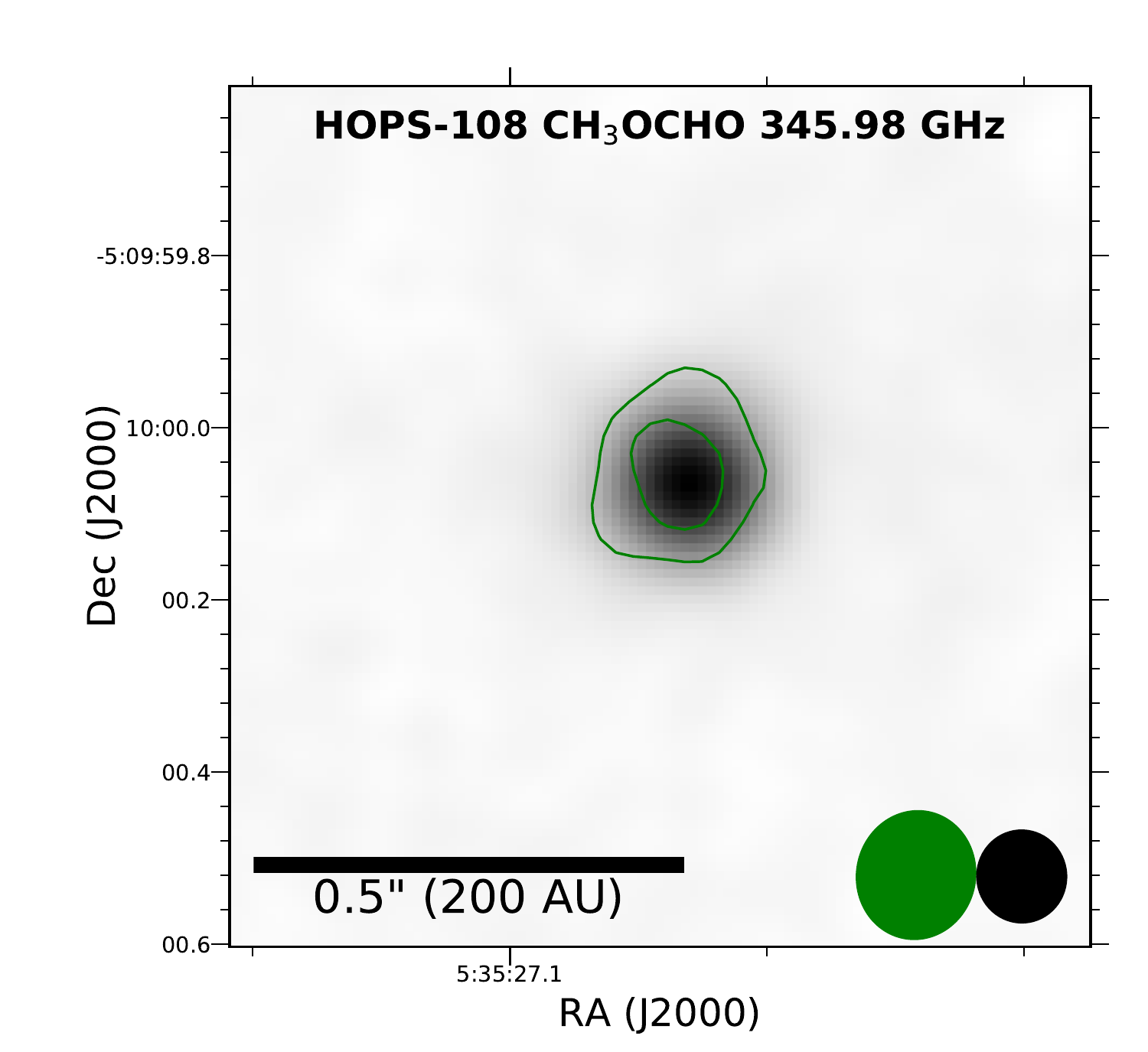}
\includegraphics[scale=0.45]{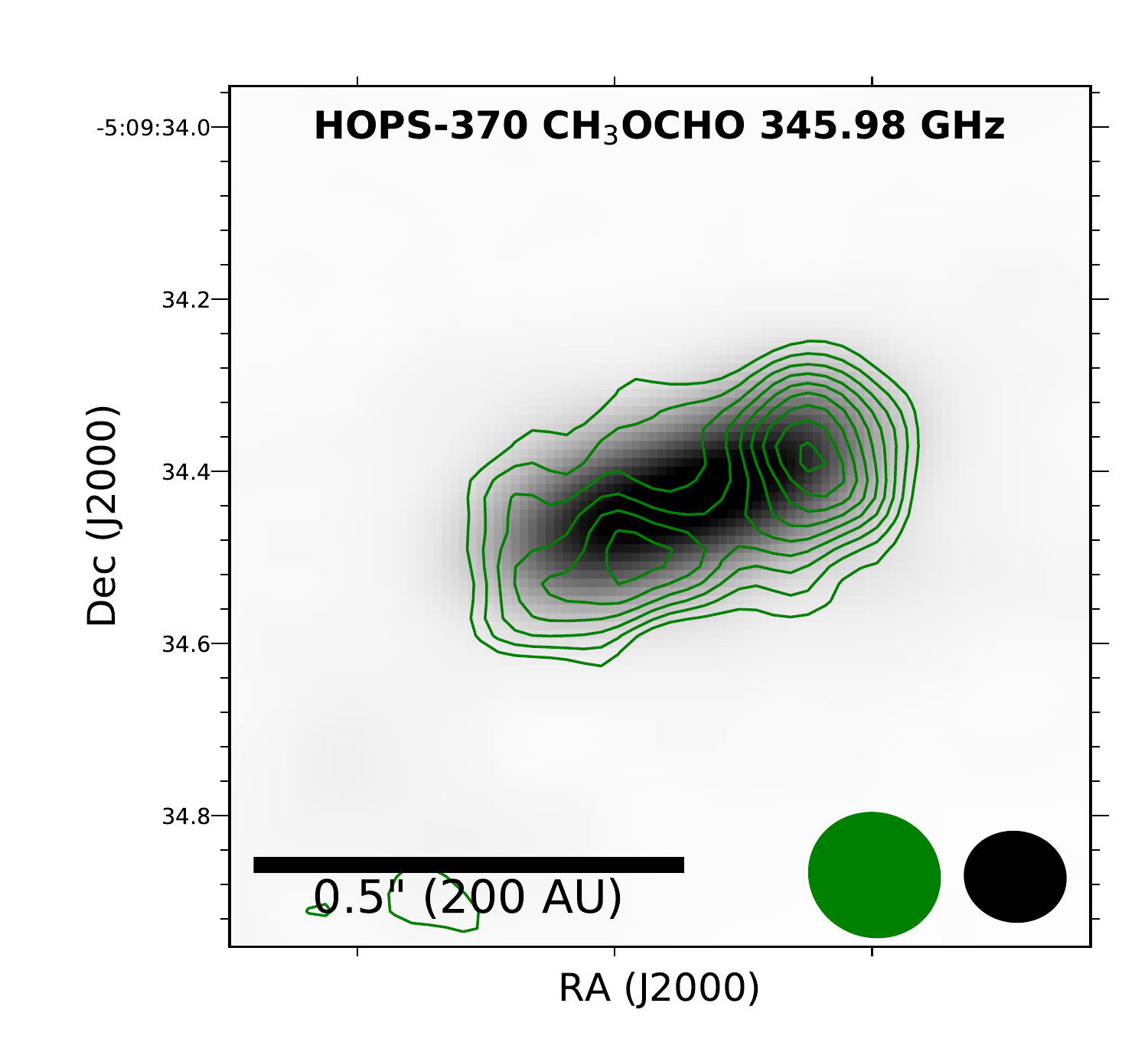}
\end{center}
\caption{Integrated intensity maps toward HOPS-108
(left) and HOPS-370 (right)
overlaid on the 0.87~mm continuum (grayscale) of the
molecular lines H$^{13}$CN ($J=4\rightarrow3$) blended with SO$_2$ ($13_{2,12}\rightarrow12_{1,11}$),
NS ($J=15/2\rightarrow13/2$), 
and Methyl Formate (CH$_3$OCHO) (28$_{2,16}$-27$_{12,15}$ E,  28$_{12,17}$-27$_{12,16}$ A, 
and 28$_{12,17}$-27$_{12,16}$ E). The integrated intensity maps 
of H$^{13}$CN/SO$_2$ are separated into blue- and red-shifted velocities and
plotted with blue and red contours, respectively.
CH$_3$OCHO and NS are too low in intensity and are integrated over the entire line profile
and plotted with green contours. The contours start at 3$\sigma$ and increase on 2$\sigma$ intervals.
 See Table 5 for more details
of the particular molecular transitions shown. The beams of the
continuum and molecular line data are shown in the lower right as black
and green ellipses, respectively. The continuum beams are $\sim$0\farcs11$\times$0\farcs10 
and the molecular line beams are $\sim$0\farcs15$\times$0\farcs14.
}
\label{additional-lines}
\end{figure}

We show the integrated intensity maps of the NS and methyl formate emission summed over the entire line(s)
for HOPS-108 in Figure \ref{additional-lines} and we show the blue and red-shifted integrated intensity maps for HOPS-370
in Figure \ref{additional-lines} as well. 
The H$^{13}$CN/SO$_2$ emission toward HOPS-108 appears to trace
an east-west velocity gradient, similar to the low excitation methanol (Figure \ref{methanol}), and
the H$^{13}$CN/SO$_2$ toward HOPS-370 also shows a rotation pattern across its disk similar to methanol.
The NS emission also appears to trace rotation toward HOPS-370, similar to the
methanol emission. However, toward HOPS-108 the NS total integrated intensity
emission is offset from the continuum source to the northwest, while the other molecular 
lines appear centered on the continuum source. However, the line widths and velocity centroids
of all the lines detected are consistent within the uncertainties of the measurements, meaning
that the lines could all be emitted from the same region.

\begin{figure}
\begin{center}
\includegraphics[scale=1.0,trim=0.5cm 15cm 7.5cm 0.5cm,clip=true]{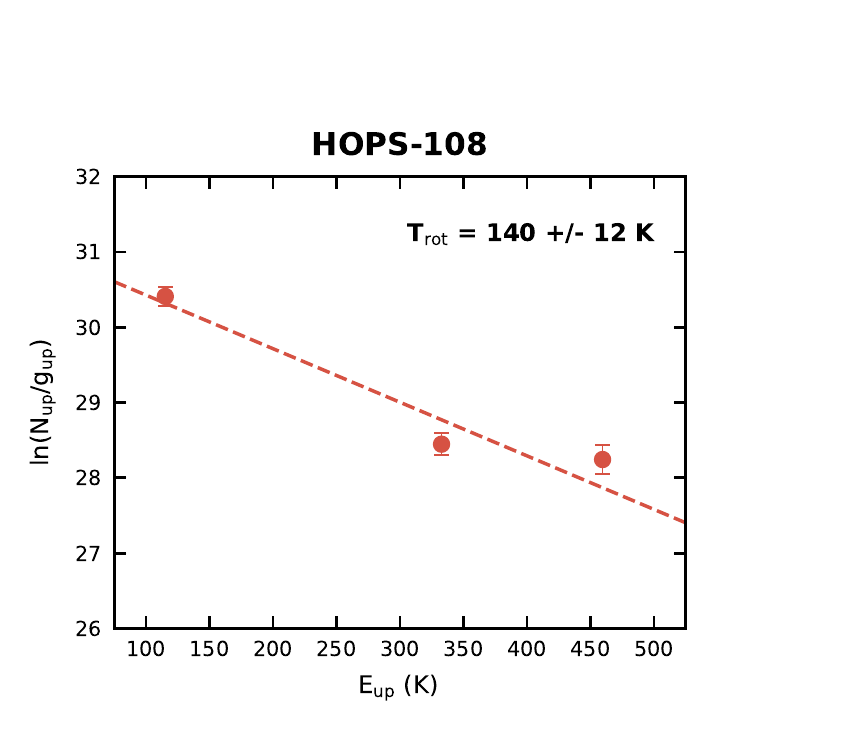}
\includegraphics[scale=1.0,trim=0.5cm 15cm 7.5cm 0.5cm,clip=true]{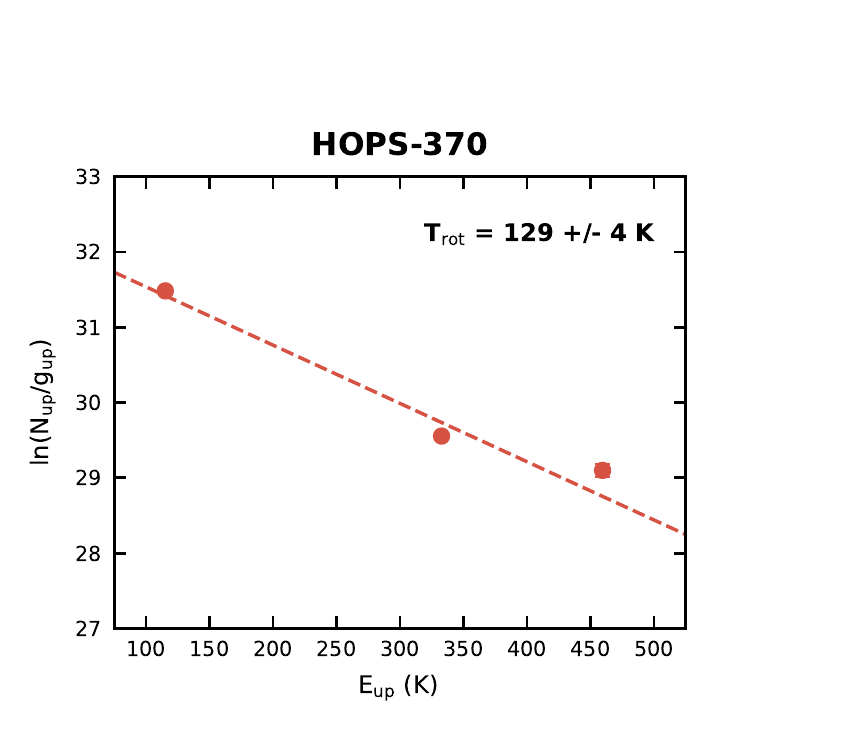}
\end{center}
\caption{Rotation diagrams of the three methanol transitions 
for HOPS-108 (left) and HOPS-370 (right) 
listed in Table 5 and
shown in Figures \ref{spectrum-H108} and \ref{spectrum-H370}.
The rotation temperatures of 140~K and 129~K for HOPS-108 and HOPS-370, respectively, are
indicative of the physical conditions required
to thermally evaporate methanol from the dust grains. 
The highest-excitation
transition appears to lie above the rotation temperature fit and could indicate a 
separate temperature component that is exciting the higher-energy transitions.
It is difficult, however, to be certain with a rotation diagram defined by only three transitions.
}
\label{rotdiagram}
\end{figure}

To characterize the excitation conditions of the methanol emission further in HOPS-370 and HOPS-108,
we used the four observed lines and their flux densities extracted
from 0\farcs5 (HOPS-108) and 0\farcs75 (HOPS-370) diameter apertures to derive their rotation temperatures.
Table 5 lists the measured line flux densities, and the uncertainties on the flux densities are
determined from the RMS flux density in regions devoid of emission.
We utilize the methodology outlined in 
\citet{goldsmith1999} to construct a rotation diagram from the three methanol transitions
shown in Figure \ref{rotdiagram}.
The lowest excitation line, the ($J=5_4\rightarrow6_3$) transition at 
$\sim$346.203~GHz is a blended transition of two lines, having the same upper level
excitation and Einstein-A coefficient. Thus, we divide the observed flux density by 2
and plot it as a single transition. From this analysis, we derive 
rotation temperatures of
140~K and 129~K for HOPS-108 and HOPS-370, respectively.
These temperatures are consistent with the conditions for
thermal evaporation of methanol from the dust grain surfaces. We note, however, 
that the rotation diagram analysis assumes that the line emission is arising from the same
physical structure and that the lines are optically thin. Thus, if the line emission
for the different transitions originates from different physical components of
the system (e.g., a rotating disk/inner envelope and/or the outflow) the derived rotation
temperature may not reflect the physical temperature of the gas around the protostar.
Furthermore, if any of the transitions are optically thick, then the column densities will
be inaccurate, making the rotation temperatures inaccurate as well.

The column density of methanol emission derived using the rotation diagram
indicates a methanol column density of 
4.3$\times$10$^{16}$~cm$^{-2}$ and 
1.4$\times$10$^{17}$~cm$^{-2}$ for HOPS-108 and HOPS-370, respectively. 
With these measurements of the methanol column density, and the disk masses from the
dust continuum we can estimate the fractional abundance of methanol.
We first convert the gas mass derived from the dust continuum
into a column density
by dividing the mass by the area defined by twice the ALMA
disk radius from Table 4 
and adopting a mean molecular weight of 2.8, 
finding 2.74$\times$10$^{24}$~cm$^{-2}$ and 1.64$\times$10$^{24}$~cm$^{-2}$
for HOPS-108 and HOPS-370, respectively.
We use 2$\times$ the HWHM disk radius from Table 4 because
it is determined from the HWHM and twice this value is a better representation of the full
extent of dust emission. Figure \ref{methanol} 
shows that the methanol emission is quite
coincident with the continuum emission, and this is a reasonable assumption 
for the total H$_2$ column density. 
We then find fractional 
abundances of methanol relative to H$_2$ to be
$\sim$1.6$\times$10$^{-8}$ and $\sim$8.5$\times$10$^{-8}$ for HOPS108 and HOP-370, respectively.
These values are lower than the observed methanol ice abundances toward low-mass protostars
\citep{boogert2015} ($\sim$10$^{-6}$ - 10$^{-5}$), but significantly higher
than the gas-phase methanol abundance of $\sim$10$^{-11}$ - 10$^{-12}$ found
within the disk of TW Hya. The disk of TW Hya is too cold to thermally
evaporate methanol throughout most of the disk and requires non-thermal desorption of methanol
to explain this low abundance \citep{walsh2016}.
Furthermore, other studies of the fractional abundance of methanol
toward high-mass star forming regions from
single-dish and interferometric studies employing different methodologies
also find fractional abundances of gas-phase methanol to similar 
our values \citep[e.g.,][]{gerner2014,feng2016}. 
Furthermore, we are calculating the fractional abundances relative to the total gas mass
derived from the dust continuum, which may over estimate the total mass from the
methanol emitting gas in the disk.

\citet{kama2010} also observed a large number of methanol lines 
using \textit{Herschel} HIFI toward OMC2-FIR4. They found that many of these lines originated
from a hot component with T$_{kin}$ = 145$\pm$12~K, comparable to our rotation
temperature. From the methanol fit alone they found a column density 
of 2.2$\times$10$^{14}$~cm$^{-2}$, and an LTE fit including an envelope and a hot
component with a size smaller than 760~AU indicates 
that the column density was $\sim$6$\times$10$^{16}$~cm$^{-2}$. Thus, the column density
inferred from the lower-angular resolution observations 
is also in agreement with our methanol column  densities.

In Section 4.3, we argued that HOPS-108 and HOPS-370
are consistent with being hot corinos. However,
a possible difference between HOPS-108 
and HOPS-370 and the hot corinos is the presence
of NS emission. IRAS16293-2422 is not known to have NS emission
within its spectrum despite a sensitive spectral survey and detections of
other sulfur-bearing species \citep{drozdovskaya2016}, and it
is not clear if the others exhibit NS emission either due to lack of spectral coverage. 
It was suggested by \citet{viti2001} that NS arises in shocked emission and 
that the ratio of NS to CS emission could be indicative of the
strength of that shock. However, the spatial
location of NS toward HOPS-108 and HOPS-370 being associated with the continuum source and not the 
\twco\ that overlaps with other shock tracers indicates that the NS is not likely tracing
shock-heated gas. The other molecules detected do not differentiate between a hot corino
or a hot core, but methanol must be formed on dust grains
through hydrogenation \citep{chuang2016} and must be released via thermal evaporation
to explain the quantities observed. Other molecules, however, such as HC$_3$N, CH$_3$CN could be
formed in the gas-phase and may have their formation catalyzed by high cosmic
ray flux \citep[Offner et al. submitted; ][]{fontani2017}.

\begin{small}
\bibliographystyle{apj}
\bibliography{ms}
\end{small}

\clearpage

\begin{deluxetable}{lllllll}
\tabletypesize{\scriptsize}
\tablewidth{0pt}
\tablecaption{ALMA and VLA Observation Summary}

\tablehead{\colhead{Fields} & \colhead{RA}       & \colhead{Dec.}     & \colhead{Date(s)} & \colhead{Max. Baseline} & \colhead{Antennas} & \colhead{PWV\tablenotemark{a}} \\
                            &  \colhead{(J2000)} & \colhead{(J2000)}  &                   & \colhead{(m)}           &                   & \colhead{(mm)} 
}
\startdata
ALMA 0.87~mm \\
(Band 7)\\
\hline
HOPS-66 & 05:35:26.843  & -05:09:24.58 & 2016 Sept 04, 05; 2017 Jul 19              & 2483, 2483, 3697 & 41, 41, 41 & 0.73, 0.53, 0.47 \\
HOPS-370 & 05:35:27.629 & -05:09:33.47 & 2016 Sept 06\tablenotemark{a}; 2017 Jul 19 & 2483, 2483, 3697 & 41, 41, 41 & 0.73, 0.53, 0.47 \\
HOPS-108 & 05:35:27.073 & -05:10:00.37 & 2016 Sept 04, 05; 2017 Jul 19              & 2483, 2483, 3697 & 39, 34, 42 & 0.42, 0.42, 0.42 \\
HOPS-369 & 05:35:26.972 & -05:10:17.14 & 2016 Sept 06\tablenotemark{a}; 2017 Jul 19 & 2483, 2483, 3697 & 39, 34, 42 & 0.42, 0.42, 0.42 \\
HOPS-368 & 05:35:24.725 & -05:10:30.21 & 2016 Sept 06\tablenotemark{a}; 2017 Jul 19 & 2483, 2483, 3697 & 39, 34, 42 & 0.42, 0.42, 0.42 \\
\hline
\hline
VLA 9.1~mm\\
(Ka-band)\\
\hline
HOPS-370 & 05:35:27.629 & -05:09:33.47 & 2016 Oct 26 & 36400 (A-config) & 26 & 8 \\
HOPS-108 & 05:35:27.073 & -05:10:00.37 & 2016 Dec 29 & 36400 (A-config) & 26 & 4\\
\enddata
\tablecomments{The ALMA observations of HOPS-370, HOPS-368, and HOPS-369 were observed as part of the
one ALMA scheduling block, while HOPS-66 and HOPS-108 were also observed together in another ALMA scheduling block.
The coordinates listed refer to the phase center of the observations.}
\tablenotetext{a}{Two executions were carried out on 2016 Sept 06.}
\end{deluxetable}

\begin{longrotatetable}
\begin{deluxetable}{llllllllll}
\tabletypesize{\tiny}
\tablewidth{0pt}
\tablecaption{ALMA 870~\micron\ Source Properties}
\tablehead{\colhead{Source} & \colhead{RA}      & \colhead{Dec.}      & \colhead{ALMA Field}     & \colhead{$\Delta\phi$}        & \colhead{F$_{\nu}$} & \colhead{Peak I$_{\nu}$} & \colhead{RMS} & \colhead{Decon. Size} & \colhead{Decon. PA}\\
                            & \colhead{(J2000)} & \colhead{(J2000)}   &              & \colhead{(\arcsec)}                        & \colhead{(mJy)}      & \colhead{(mJy~bm$^{-1}$)} & \colhead{(mJy~bm$^{-1}$)}      & \colhead{(\arcsec)}    & \colhead{(\degr)}         
}
\startdata
HOPS-66-B    & 05:35:26.927 & -05:09:22.43 & HOPS-66   & 2.5    & 34.62$\pm$1.10  & 12.56   & 0.28 (0.28) & 0.17$\times$0.12 & 165.8\\
HOPS-66-A    & 05:35:26.857 & -05:09:24.40 & HOPS-66   & 0.3    & 43.31$\pm$0.57  & 35.97   & 0.28 (0.28) & 0.05$\times$0.04 & 47.0\\
HOPS-370     & 05:35:27.634 & -05:09:34.42 & HOPS-370  & 1.0    & 533.28$\pm$10.05 & 109.89 & 0.39 (0.39) & 0.34$\times$0.11 & 109.7\\
MGM-2297     & 05:35:27.47  & -05:09:44.16 & HOPS-370  & 10.9   & 9.07$\pm$1.79   & 6.37    & 1.04 (0.36) & Unresolved & \nodata\\
HOPS-64      & 05:35:26.998 & -05:09:54.08 & HOPS-108  & 6.4    & 33.47$\pm$0.76  & 20.37   & 0.42 (0.28) & 0.10$\times$0.07 & 119.3\\
HOPS-108     & 05:35:27.086 & -05:10:00.06 & HOPS-108  & 0.4    & 62.63$\pm$0.98  & 27.67   & 0.31 (0.31) & 0.12$\times$0.12 & 105.7\\
VLA15        & 05:35:26.41  & -05:10:05.94 & HOPS-108  & 11.4   & 122.94$\pm$1.65 & 20.48   & 0.84 (0.26) & 0.41$\times$0.11 & 87.3\\
VLA16        & 05:35:26.824 & -05:10:05.62 & HOPS-108  & 6.4    & 6.66$\pm$0.81   & 4.24    & 0.37 (0.24) & 0.10$\times$0.08 & 164.8\\
OMC2-FIR4-ALMA1 & 05:35:26.785 & -05:10:08.83 & HOPS-369  & 8.8 & 5.57$\pm$0.73   & 3.30    & 0.59 (0.27) & 0.14$\times$0.04 & 110.5\\
HOPS-369     & 05:35:26.969 & -05:10:17.27 & HOPS-369  & 0.1    & 26.11$\pm$0.53  & 22.40   & 0.27 (0.27) & 0.05$\times$0.04 & 21.3\\
HOPS-368     & 05:35:24.725 & -05:10:30.08 & HOPS-368  & 0.1    & 135.91$\pm$2.62 & 60.96   & 0.25 (0.25) & 0.19$\times$0.08 & 105.2\\
\enddata
\tablecomments{Observied properties of the sources observed by ALMA at 0.87~mm. The column \textit{ALMA Field} corresponds to the
main target observed in a particular field and the column $\Delta\phi$ is the angular separation in arcseconds from the
phase center of the field. The source names VLA16 and VLA15 refer to sources
identified in \citet{osorio2017}, and MGM refers to \citet{megeath2012}.
The integrated flux densities and peak intensities are primary beam corrected; the RMS noise uncorrected for the primary beam
is given in paretheses.}
\end{deluxetable}
\end{longrotatetable}

\begin{longrotatetable}
\begin{deluxetable}{llllllllll}
\tabletypesize{\tiny}
\tablewidth{0pt}
\tablecaption{VLA 9~mm Source Properties}
\tablehead{\colhead{Source} & \colhead{RA}      & \colhead{Dec.}        &  \colhead{VLA Field} & \colhead{$\Delta\phi$}   & \colhead{F$_{\nu}$} & \colhead{Peak I$_{\nu}$} & \colhead{RMS} & \colhead{Decon. Size} & \colhead{Decon. PA}\\
                            & \colhead{(J2000)} & \colhead{(J2000)}     &                      & \colhead{(\arcsec)}       & \colhead{(mJy)}     & \colhead{(mJy~bm$^{-1}$)} &\colhead{($\mu$Jy~bm$^{-1}$)} & \colhead{(\arcsec)}    & \colhead{(\degr)}     
}
\startdata
HOPS-66-B  & 05:35:26.927 & -05:09:22.41 & HOPS-370  & 15.2  & 0.078$\pm$0.013 & 0.070  & 7.5 (6.8) & Unresolved & \nodata\\
HOPS-66-A  & 05:35:26.857 & -05:09:24.40 & HOPS-370  & 14.7  & 0.293$\pm$0.014 & 0.254  & 7.5 (6.8) & Unresolved & \nodata\\
HOPS-370   & 05:35:27.633 & -05:09:34.40 & HOPS-370  & 1.0   & 2.841$\pm$0.018 & 1.931  & 6.9 (6.9) & 0.08$\times$0.03 & 5.6\\
MGM-2297   & 05:35:27.474 & -05:09:44.16 & HOPS-370  & 10.9  & 0.146$\pm$0.018 & 0.102  & 7.6 (7.3) & 0.09$\times$0.03 & 10.9\\
HOPS-64    & 05:35:26.996 & -05:09:54.08 & HOPS-108  & 6.4   & 0.255$\pm$0.018 & 0.170  & 6.9 (6.8) & 0.06$\times$0.05 & 158.6\\
HOPS-108   & 05:35:27.084 & -05:10:00.06 & HOPS-108  & 0.4   & 0.099$\pm$0.019 & 0.068  & 6.8 (6.8) & 0.07$\times$0.06 & 156.2\\
VLA16      & 05:35:26.824 & -05:10:05.64 & HOPS-108  & 6.4   & 0.040$\pm$0.012 & 0.041  & 6.8 (6.8) & Unresolved & \nodata\\
VLA15      & 05:35:26.410 & -05:10:05.95 & HOPS-108  & 11.4  & 0.532$\pm$0.031 & 0.178  & 7.1 (6.8) & 0.18$\times$0.07 & 85.7\\
OMC2-FIR4-ALMA1 & 05:35:26.784 & -05:10:08.82 & HOPS-108  & 9.5 & 0.059$\pm$0.013 & 0.054  & 7.0 (6.7) & Unresolved & \nodata\\
HOPS-369   & 05:35:26.970 & -05:10:17.27 & HOPS-108  & 17.0  & 0.051$\pm$0.007 & 0.047  & 8.1 (7.2) & Unresolved & \nodata\\
HOPS-368   & 05:35:24.728 & -05:10:30.09 & HOPS-108  & 46.0  & 1.101$\pm$0.020 & 0.965  & 18.5 (7.2) & 0.06$\times$0.03 & 72.6\\
\enddata
\tablecomments{Observied properties of the sources observed by the VLA at 9~mm. The column \textit{VLA Field} corresponds to the
main target observed in a particular field and the column $\Delta\phi$ is the angular separation in arcseconds from the
phase center of the field. The source names VLA16 and VLA15 refer to sources
identified in \citet{osorio2017}, and MGM refers to \citet{megeath2012}.
The integrated flux densities and peak intensities are primary beam corrected; the RMS noise uncorrected for the primary beam
is given in paretheses.}
\end{deluxetable}
\end{longrotatetable}

\begin{longrotatetable}
\begin{deluxetable}{lllllllllll}
\tabletypesize{\scriptsize}
\tablewidth{0pt}
\tablecaption{ALMA and VLA Derived Parameters}
\tablehead{\colhead{Source} & \colhead{Other Names} & \colhead{L$_{bol}$}  & \colhead{T$_{bol}$} & \colhead{Class} & \colhead{HWHM$_{ALMA}$} & \colhead{HWHM$_{VLA}$} & \colhead{M$_{ALMA}$} & \colhead{M$_{VLA}$} & \colhead{Sp. Index} & \colhead{Sp. Index}\\
                            &                       & \colhead{(L$_{\sun}$)} & \colhead{(K)}     &                 & \colhead{(AU)}            & \colhead{(AU)}           & \colhead{(M$_{\sun}$)}  & \colhead{(M$_{\sun}$)}  &  \colhead{(0.87 - 9~mm)} &  \colhead{(8.1 - 10~mm)}   
}
\startdata
HOPS-66-B   & \nodata & 21.0 & 264.9 & Flat & 34.0$\pm$10.0 &  $\leq$10.0 & 0.0048$\pm$0.0002 & 0.015$\pm$0.003 & 2.6$\pm$0.09 & 0.5$\pm$1.40\\
HOPS-66-A   & \nodata & 21.0 & 264.9 & Flat & $\leq$10.0 &   $\leq$10.0 & 0.0060$\pm$0.0001 & 0.058$\pm$0.003 & 2.1$\pm$0.06 & 1.3$\pm$0.41\\
HOPS-370    & OMC2-FIR3 &  360.9 & 71.5 & I    & 67.0$\pm$10.0 &  16$\pm$10 & 0.0344$\pm$0.0006 & 0.276$\pm$0.002 & 2.2$\pm$0.06 & 0.7$\pm$0.05\\
MGM-2297    & \nodata & \nodata & \nodata & II & \nodata &  18$\pm$10 & 0.0030$\pm$0.0006 & 0.063$\pm$0.008 & 1.8$\pm$0.12 & 0.2$\pm$0.99\\
HOPS-64     & MGM 2293, V2457 Ori  & 15.3 & 29.7 & I    & 19.0$\pm$10.0 &  11$\pm$10 & 0.0050$\pm$0.0001 & 0.055$\pm$0.004 & 2.1$\pm$0.07 & 2.5$\pm$0.61\\
HOPS-108    & OMC2-FIR4 & 38.3 & 38.5 & 0    & 24.0$\pm$10.0 &  14$\pm$10 & 0.0073$\pm$0.0001 & 0.017$\pm$0.003 & 2.8$\pm$0.1 & 1.5$\pm$2.26\\
VLA16  & \nodata & \nodata & \nodata & \nodata & 19.0$\pm$10.0 &   $\leq$10.0 & 0.0022$\pm$0.0003 & 0.017$\pm$0.005 & 2.2$\pm$0.15 & 2.3$\pm$2.36\\
VLA15  & \nodata & \nodata & \nodata & \nodata & 81.0$\pm$10.0 &  35$\pm$ 10 & 0.0405$\pm$0.0005 & 0.228$\pm$0.013 & 2.3$\pm$0.07 & 1.7$\pm$0.5\\
OMC2-FIR4-ALMA1 & \nodata & \nodata & \nodata & \nodata & 28.0$\pm$10.0 &   $\leq$17.0 & 0.0018$\pm$0.0002 & 0.025$\pm$0.006 & 1.9$\pm$0.13 & 0.6$\pm$1.97\\
HOPS-369   & MGM 2282 & 35.3 & 379.2 & Flat & $\leq$10.0 & $\leq$10.0 & 0.0031$\pm$0.0001 & 0.009$\pm$0.001 & 2.7$\pm$0.08 & 2.2$\pm$1.03\\
HOPS-368   & MGM 2279  & 68.9 & 137.5 & I    & 37.0$\pm$10.0 &  11$\pm$10 & 0.0136$\pm$0.0003 & 0.162$\pm$0.003 & 2.1$\pm$0.06 & 0.0$\pm$0.11\\
\enddata
\tablecomments{The columns \textit{HWHM$_{ALMA}$ and HWHM$_{VLA}$} correspond to the half-width at half maximum radii
in AU, as a measure of the size of the continuum emission. The columns \textit{M$_{ALMA}$ and M$_{VLA}$} correspond to the
gas mass derived from the continuum flux density. The uncertainties on the masses are statistical only and do not
take into account the $\sim$10\% uncertainty in the absolute flux density scale.}
\end{deluxetable}
\end{longrotatetable}

\begin{longrotatetable}
\begin{deluxetable}{lccccccccccccc}
\tabletypesize{\tiny}
\tablewidth{0pt}
\tablecaption{Overview of the molecular line detections toward HOPS-108 and HOPS-370}
\tablehead{\colhead{Species} & \colhead{Transition} & \colhead{Frequency} & \colhead{V$_{lsr}$ (H108)} & \colhead{$\Delta$v (H108)} & \colhead{V$_{lsr}$ (H370)} & \colhead{$\Delta$v (H370)} & \colhead{$A_{\rm{ul}}$} & \colhead{$E_{\rm{up}}/k$} & \colhead{$g_{\rm{up}}$} & \colhead{$F_{\rm{peak}}$ (H108)} & \colhead{$F_{\rm{int}}$\tablenotemark{a} (H108)} & \colhead{$F_{\rm{peak}}$ (H370)} & \colhead{$F_{\rm{int}}$\tablenotemark{f} (H370)}\\     
    & & \colhead{(GHz)} & \colhead{(km~s$^{-1}$)} & \colhead{(km~s$^{-1}$)} & \colhead{(km~s$^{-1}$)} & \colhead{(km~s$^{-1}$)} & \colhead{(s$^{-1}$)} & \colhead{(K)} & & \colhead{(mJy beam$^{-1}$)} & \colhead{(Jy km s$^{-1}$)} & \colhead{(mJy beam$^{-1}$)} & \colhead{(Jy km s$^{-1}$)} \\
} 
\startdata
    $^{13}$CH$_3$OH	(A$^-$)	& $7_{-2,6}-6_{-2,5}$		& 330.535890 & 13.3$\pm$0.2) & 1.2$\pm$0.2 &  11.2$\pm$0.9  &  5.4$\pm$1.0   & 1.46$\times$10$^{-4}$ & 89 & 15  & 46$\pm$14.6 & 0.2$\pm$0.13 & 103$\pm$17 & 1.0$\pm$0.25\\
    $^{13}$CO & $J=3-2$                                 & 345.795990 & 13.2$\pm$0.6 & 1.1$\pm$0.6  &  \nodata  &  \nodata   & 2.19$\times$10$^{-6}$ & 32 & 7 & 75$\pm$14.6 &  0.99$\pm$0.16 & \nodata\tablenotemark{e} & \nodata\tablenotemark{e} \\
    CH$_3$CN            & $18_8-17_8$                   & 330.665206 & 11.8$\pm$0.8 & 0.5$\pm$1.5  &  11.3$\pm$1.2  &  3.5$\pm$1.2  & 1.74$\times$10$^{-3}$ & 608 & 37 & 67$\pm$14.6 & 0.2$\pm$0.11 & 78$\pm$17 & 0.5$\pm$0.2 \\
    H$^{13}$CN          & J=4-3                         & 345.339760 & 12.9$\pm$0.4\tablenotemark{b} & 3.3$\pm$0.4\tablenotemark{b}  &  12.5$\pm$0.5\tablenotemark{b}  &  5.4$\pm$0.6\tablenotemark{b}   & 1.74$\times$10$^{-1}$ & 41 & 9 & 117.5$\pm$18.8\tablenotemark{d} & 1.7$\pm$0.12\tablenotemark{d} & 223$\pm$13\tablenotemark{b} & 8.9$\pm$0.3\tablenotemark{b} \\
    SO$_2$              & $13_{2,12}-12_{1,11}$     & 345.338539 & 12.9$\pm$0.4\tablenotemark{b} & 3.3$\pm$0.4\tablenotemark{b}  &  12.5$\pm$0.5\tablenotemark{b}  &  5.4$\pm$0.6\tablenotemark{b}   & 2.38$\times$10$^{-4}$ & 93 & 27 & 117.5$\pm$18.8\tablenotemark{d} & 1.7$\pm$0.12\tablenotemark{d} & 223$\pm$13\tablenotemark{b} & 8.9$\pm$0.3\tablenotemark{b}\\
    CH$_3$OCHO   (A)    & 28$_{13,16}$-27$_{12,15}$ & 345.466962 & 12.4$\pm$0.3\tablenotemark{b} & 0.9$\pm$0.3\tablenotemark{b}  &  \nodata  &  \nodata   & 4.94$\times$10$^{-4}$ & 352 & 57 & 70.2$\pm$18.8\tablenotemark{b} & 0.34$\pm$0.1\tablenotemark{b} & 61$\pm$15 & $<$ \\
    CH$_3$OCHO   (A)    & 28$_{13,15}$-27$_{12,14}$ & 345.466962 & 12.4$\pm$0.3\tablenotemark{b} & 0.9$\pm$0.3\tablenotemark{b}  &  \nodata  &  \nodata   & 4.94$\times$10$^{-4}$ & 352 & 57 &  70.2$\pm$18.8\tablenotemark{b} & 0.34$\pm$0.1\tablenotemark{b} & $<$ & $<$ \\
    CH$_3$OCHO   (E)    & 28$_{13,16}$-27$_{12,15}$ & 345.466962 & 12.4$\pm$0.3\tablenotemark{b} & 0.9$\pm$0.3\tablenotemark{b}  &  \nodata  &  \nodata   & 4.94$\times$10$^{-4}$ & 352 & 57 &  70.2$\pm$18.8\tablenotemark{b} & 0.34$\pm$0.1\tablenotemark{b} & $<$ & $<$\\
    HC$_3$N\tablenotemark{g}             & $J=38-37$                 & 345.609010  & \nodata & \nodata  &  11.6$\pm$0.4\tablenotemark{c}  &  4.2$\pm$0.5\tablenotemark{c}   & 3.29$\times$10$^{-3}$ & 323 & 77 &   66$\pm$18.8 & 0.24$\pm$0.1 & 158$\pm$14\tablenotemark{c} & 3.0$\pm$0.2\tablenotemark{c}\\
    CH$_3$OCHO\tablenotemark{g}   (E)    & 14$_{13,2}$-14$_{12,3}$   & 345.613535  & \nodata & \nodata  &  11.6$\pm$0.4\tablenotemark{c}  &  4.2$\pm$0.5\tablenotemark{c}   &  1.5$\times$10$^{-5}$ & 174 & 29 &  66$\pm$18.8\tablenotemark{c} & 0.24$\pm$0.1\tablenotemark{c} & 158$\pm$14\tablenotemark{c} & 3.0$\pm$0.2\tablenotemark{c}\\
    CO & $J=3-2$                                    & 345.795990 & \nodata & \nodata  &  \nodata  & \nodata  & 2.5$\times$10$^{-6}$ & 33 & 7 & \nodata  &  \nodata\tablenotemark{e}  & \nodata  &  \nodata\tablenotemark{e}\\
    CH$_3$OH	(A$^-$)	& $16_1-15_2$	            & 345.903916  & 12.7$\pm$0.3 & 1.7$\pm$0.3  &  10.7$\pm$0.3  &  4.0$\pm$0.3   & 8.8$\times$10$^{-5}$ & 333 & 33 & 133.4$\pm$18.8 & 0.69$\pm$0.10 & 171$\pm$15 & 4.7$\pm$0.23 \\
    CH$_3$OH	(E2)		& $18_3-17_4$	        & 345.919260 & 12.6$\pm$0.4 & 1.5$\pm$0.4  &  11.4$\pm$0.5  & 4.8$\pm$0.5   & 7.1$\times$10$^{-5}$ & 459 & 37 & \hspace{0.15cm}99.1$\pm$18.8 & 0.52$\pm$0.10 & 138$\pm$14 & 2.7$\pm$0.23\\
    CH$_3$OCHO   (E)    & 28$_{12,16}$-27$_{12,15}$ & 345.974664 & 13.7$\pm$0.7 & 1.3$\pm$0.7  &  \nodata  &  \nodata   & 5.16$\times$10$^{-4}$ & 335  & 57 & 73.6$\pm$18.8\tablenotemark{b} & 1.1$\pm$0.2\tablenotemark{b} & 84$\pm$15\tablenotemark{b} & $<$  \\
    CH$_3$OCHO   (A)    & 28$_{12,17}$-27$_{12,16}$ & 345.985381 & 13.1$\pm$0.8\tablenotemark{b} & 2.1$\pm$0.8\tablenotemark{b}  &  \nodata  &  \nodata   & 5.16$\times$10$^{-4}$ & 335  & 57 &  73.6$\pm$18.8\tablenotemark{b} & 0.95$\pm$0.2\tablenotemark{b} & 84$\pm$15\tablenotemark{b} & $<$  \\
    CH$_3$OCHO   (A)    & 28$_{12,16}$-27$_{12,15}$ & 345.985381 & 13.1$\pm$0.8\tablenotemark{b} & 2.1$\pm$0.8\tablenotemark{b}  &  \nodata  &  \nodata   & 5.16$\times$10$^{-4}$ & 335  & 57 &  73.6$\pm$18.8\tablenotemark{b} & 0.95$\pm$0.2\tablenotemark{b}  & 84$\pm$15\tablenotemark{b} & $<$ \\
    CH$_3$OCHO   (E)    & 28$_{12,17}$-27$_{12,16}$ & 346.001616 & 12.1$\pm$0.4 & 1.0$\pm$0.4  & \nodata  &  \nodata   & 5.16$\times$10$^{-4}$ & 335  & 57 &  73.6$\pm$18.8\tablenotemark{b} & 0.95$\pm$0.2\tablenotemark{b} & $<$ & $<$ \\
    CH$_3$OH	(A$^-$)	& $5_4-6_3$		            & 346.202719 & 12.0$\pm$0.4\tablenotemark{b} & 2.3$\pm$0.4\tablenotemark{b}  &  10.5$\pm$0.4\tablenotemark{b} & 5.0$\pm$0.5\tablenotemark{b} & 2.1$\times$10$^{-5}$ & 115 & 11 & \hspace{0.15cm}113.3 $\pm$ 18.8\tablenotemark{b} & \hspace{0.15cm}0.78 $\pm$ 0.11\tablenotemark{b} & 182$\pm$15\tablenotemark{b} & 5.2$\pm$0.25\tablenotemark{b}  \\
    CH$_3$OH	(A$^+$)	& $5_4-6_3$		            & 346.204271 & 12.0$\pm$0.4\tablenotemark{b} & 2.3$\pm$0.4\tablenotemark{b}  &  10.5$\pm$0.4\tablenotemark{b} & 5.0$\pm$0.5\tablenotemark{b} & 2.1$\times$10$^{-5}$ & 115 & 11 & \hspace{0.15cm}113.3 $\pm$ 18.8\tablenotemark{b} & \hspace{0.15cm}0.78 $\pm$ 0.11\tablenotemark{b} & 182$\pm$15\tablenotemark{b} & 5.2$\pm$0.25\tablenotemark{b}  \\
    NS                  & J=$\frac{15}{2}-\frac{13}{2}$, F=$\frac{17}{2}-\frac{15}{2}$         & 346.220137 & 12.6$\pm$0.3\tablenotemark{b} & 1.2$\pm$0.3\tablenotemark{b} & 10.3$\pm$0.6\tablenotemark{b} & 5.7$\pm$0.9\tablenotemark{b} & 7.38$\times$10$^{-4}$ & 71 & 31 & 79.6$\pm$18.8\tablenotemark{b} & 0.36$\pm$0.1\tablenotemark{b} & 102$\pm$14\tablenotemark{b} & 1.9$\pm$0.2\tablenotemark{b}\\
    NS                  & J=$\frac{15}{2}-\frac{13}{2}$, F=$\frac{15}{2}-\frac{13}{2}$         & 346.221163 & 12.6$\pm$0.3\tablenotemark{b} & 1.2$\pm$0.3\tablenotemark{b} & 10.3$\pm$0.6\tablenotemark{b} & 5.7$\pm$0.9\tablenotemark{b} & 7.38$\times$10$^{-4}$ & 71 & 31 & 79.6$\pm$18.8\tablenotemark{b} & 0.36$\pm$0.1\tablenotemark{b} & 102$\pm$14\tablenotemark{b} & 1.9$\pm$0.2\tablenotemark{b} \\
    NS                  & J=$\frac{15}{2}-\frac{13}{2}$, F=$\frac{13}{2}-\frac{11}{2}$        & 346.221163 & 12.6$\pm$0.3\tablenotemark{b} & 1.2$\pm$0.3\tablenotemark{b} & 10.3$\pm$0.6\tablenotemark{b} & 5.7$\pm$0.9\tablenotemark{b} & 7.38$\times$10$^{-4}$ & 71 & 31 & 79.6$\pm$18.8\tablenotemark{b} & 0.36$\pm$0.1\tablenotemark{b} & 102$\pm$14\tablenotemark{b} & 1.9$\pm$0.2\tablenotemark{b} \\
\enddata
\tablenotetext{a}{Within a 0.5$\arcsec$ diameter aperture.}  
\tablenotetext{b}{Both/all lines combined.}
\tablenotetext{c}{HC$_3$N and CH$_3$OCHO combined.}
\tablenotetext{d}{H$^{13}$CN and SO$_2$ combined.}
\tablenotetext{e}{Measurements not performed for CO/$^{13}$CO due to significant spatial filtering.}
\tablenotetext{f}{Within a 0.75$\arcsec$ diameter aperture.} 
\tablenotetext{g}{CH$_3$COOH (acetic acid) and other methyl formate (CH$_3$OCHO) transitions could also contribute to the total line flux.}
\end{deluxetable}
\end{longrotatetable}


\end{document}